\documentclass{jfm}
\newcommand{\ppA}{p_{\rule{0pt}{1.6ex} \mkern-3mu A}}
\newcommand{\qA}{q_{\rule{0pt}{1.6ex} \mkern-3mu A}}
\newcommand{\rA}{r_{\rule{0pt}{1.6ex} \mkern-4mu A}}
\newcommand{\gSb}{\boldsymbol{\dot{\gamma}}^*_{\rule{0pt}{-2ex} \mkern-2mu S}}
\newcommand{\gWb}{\boldsymbol{\dot{\gamma}}^*_{\rule{0pt}{-2ex} \mkern-2mu W}}
\newcommand{\gS}{\dot{\gamma}^2_{\rule{0pt}{-2ex} \mkern-2mu S}}
\newcommand{\gW}{\dot{\gamma}^2_{\rule{0pt}{-2ex} \mkern-2mu W}}
\newcommand{\gsq}{\dot{\gamma}^2}
\newcommand{\esq}{\dot{\epsilon}^2}
\newcommand{\psq}{\dot{\varphi}^2}
\newcommand{\src}{\boldsymbol{\dot{\varphi}} \mkern-5mu : \mkern-5mu \boldsymbol{\dot{\gamma}}}
\newcommand{\Aavg}[1]{\big\langle {#1} \big\rangle_{\mkern-3mu A^2}}
\usepackage{graphicx}
\usepackage{newtxtext}
\usepackage{newtxmath}
\usepackage{natbib}
\usepackage{xpatch}
\usepackage[colorlinks=true, urlcolor=blue, allcolors=blue]{hyperref}
\hypersetup{
    colorlinks = true,
    urlcolor   = blue,
    citecolor  = blue,
}
\usepackage{orcidlink}
\usepackage{tikz}
\usepackage{xcolor}
\usepackage{colortbl}
\usepackage{comment}
\usepackage{soul}% for highlighting with \hl
\usepackage[nameinlink,noabbrev]{cleveref}% for references
\newcommand{\copyedit}[2][green]{{\colorlet{foo}{#1}\sethlcolor{foo}\hl{#2}}}% copy-editor changes
\newcommand{\rahul}[2][yellow]{{\colorlet{foo}{#1}\sethlcolor{foo}\hl{#2}}}% Rahul changes
\renewcommand\copyedit[1]{#1}
\renewcommand\rahul[1]{#1}
\newcommand{\RomanNumeralCaps}[1]
\linenumbers
\makeatletter
\xpatchcmd\NAT@citex
 {%
  \@citea\NAT@hyper@{%
    \NAT@nmfmt{\NAT@nm}%
    \hyper@natlinkbreak{\NAT@aysep\NAT@spacechar}{\@citeb\@extra@b@citeb}%
    \NAT@date
  }%
 }
 {%
  \@citea
  \NAT@nmfmt{\NAT@nm}%
  \NAT@aysep\NAT@spacechar
  \NAT@hyper@{\NAT@date}%
 }
 {}{}
\xpatchcmd\NAT@citex
 {%
  \@citea\NAT@hyper@{%
    \NAT@nmfmt{\NAT@nm}%
    \hyper@natlinkbreak{\NAT@spacechar\NAT@@open\if*#1*\else#1\NAT@spacechar\fi}%
    {\@citeb\@extra@b@citeb}%
    \NAT@date
  }%
 }
 {
  \@citea
    \NAT@nmfmt{\NAT@nm}%
    \NAT@spacechar\NAT@@open\if*#1*\else#1\NAT@spacechar\fi
    \NAT@hyper@{\NAT@date}%
 }
 {}{}
\makeatother
\title{Velocity gradient analysis of a head-on vortex ring collision}

\shorttitle{Velocity gradient analysis of a vortex ring collision}
\shortauthor{R. Arun and T. Colonius}

\author{Rahul Arun\aff{1}
  \corresp{\email{\href{mailto:rarun@caltech.edu}{rarun@caltech.edu}}} \and Tim Colonius\aff{2}}

\affiliation{\aff{1}Graduate Aerospace Laboratories, California Institute of Technology, Pasadena, CA 91125, USA
\aff{2}Department of Mechanical and Civil Engineering, California Institute of Technology, Pasadena, CA 91125, USA}
\begin{document}
\maketitle
%%%%%%%%%%%%%%%%%%%%%%%%%%%%%%%%%%%%%%%%%%

% abstract
%%%%%%%%%%%%%%%%%%%%%%%%%%%%%%%%%%%%%%%%%%

\begin{abstract}
We simulate the head-on collision between vortex rings \rahul{with circulation Reynolds numbers of 4000} using an adaptive, multiresolution solver based on the lattice Green's function. The simulation fidelity is established with integral metrics representing symmetries and discretization errors. Using the velocity gradient tensor and structural features of local streamlines, we characterize the evolution of the flow with a particular focus on its transition and turbulent decay. Transition is excited by the development of the elliptic instability, which grows during the mutual interaction of the rings as they expand radially at the collision plane. The development of antiparallel secondary vortex filaments along the circumference mediates the proliferation of small-scale turbulence. During turbulent decay, the partitioning of the velocity gradients approaches an equilibrium that is dominated by shearing and agrees well with previous results for forced isotropic turbulence. We also introduce new phase spaces for the velocity gradients that reflect the interplay between shearing and rigid rotation and highlight geometric features of local streamlines. In conjunction with our other analyses, these phase spaces suggest that, while the elliptic instability is the predominant mechanism driving the initial transition, its interplay with other mechanisms, e.g. the Crow instability, becomes more important during turbulent decay. Our analysis also suggests that the geometry-based phase space may be promising for identifying the effects of the elliptic instability and other mechanisms using the structure of local streamlines. Moving forward, characterizing the organization of these mechanisms within vortices and universal features of velocity gradients may aid in \copyedit{modelling} turbulent flows.
\end{abstract}
%%%%%%%%%%%%%%%%%%%%%%%%%%%%%%%%%%%%%%%%%%

% keywords
%%%%%%%%%%%%%%%%%%%%%%%%%%%%%%%%%%%%%%%%%%

%{\bf MSC Codes }  {\it(Optional)} Please enter %your MSC Codes here

% \newpage% new page after abstract
%%%%%%%%%%%%%%%%%%%%%%%%%%%%%%%%%%%%%%%%%%

%%%%%%%%%%%%%%%%%%%%%%%%%%%%%%%%%%%%%%%%%%
% main sections
%%%%%%%%%%%%%%%%%%%%%%%%%%%%%%%%%%%%%%%%%%

%%%%%%%%%%%%%%%%%%%%%%%%%%%%%%%%%%%%%%%%%%
\section{Introduction}\label{sec:introduction}
%%%%%%%%%%%%%%%%%%%%%%%%%%%%%%%%%%%%%%%%%%

%%%%%%%%%%%%%%%%%%%%%%%%%%%%%%%%%%%%%%%%%%
\subsection{Vortex rings}\label{sec:intro:vortex_rings}
%%%%%%%%%%%%%%%%%%%%%%%%%%%%%%%%%%%%%%%%%%

Vortex rings are ubiquitous flow phenomena in both applied and theoretical settings, with applications including sound generation, \rahul{transport,} \copyedit{mixing} and vortex interactions \citep{Sha1992}. In geophysical settings, vortex rings can be used to model entrainment and dispersion in particle clouds \citep{Bus2003}. They play important roles in the initial jets of volcanic eruptions \citep{Tad2015} and the transport of contaminated sediments disposed \copyedit{of} in open-water settings \citep{Rug2000}. In biomechanical settings, vortex rings have been observed in the motions of blood in the human heart \citep{Arv2016} and in the propulsive motion of oblate medusan jellyfish \citep{Dab2005}. Remarkably, separated vortex rings augment dandelion seed dispersal by prolonging flight through drag enhancement \citep{Cum2018}. In aerodynamic settings, vortex rings are responsible for the so-called vortex ring state, which negatively impacts lift in helicopters \citep{Joh2005} and the performance of offshore wind turbines \citep{Kyl2020}. In experimental and numerical settings, the formation and pinch-off of vortex rings are of particular interest in jet flows involving nozzles and orifices \citep{Gha1998,Moh2001,Kru2003,OFa2014,Lim2021}.

Vortex rings are also associated with complex instabilities and dynamics that relate more generally to the sustenance of turbulence. Flow instabilities in vortex rings depend primarily on the core vorticity distribution, the circulation Reynolds number $\left( Re_\mathit{\Gamma} = \mathit{\Gamma} / \nu \right)$ \copyedit{and} the slenderness ratio $\left( \delta = a / R \right)$ \citep{Bal2020}. Here, $\mathit{\Gamma}$ is the circulation, $\nu$ is the kinematic viscosity, $a$ is the core \copyedit{radius} and $R$ is the ring radius. We focus on the evolution of thin-cored vortex rings with Gaussian core vorticity profiles, no \copyedit{swirl} and centroids ($Z$) that propagate along the $z$-axis. In cylindrical coordinates ($r, \theta, z$), this initial vorticity profile is written as
\begin{equation}\label{eq:intro:Gaussian}
    \omega_\theta (r, z; t = 0) = \pm \frac{\mathit{\Gamma}_0}{\pi a_0^2} {\rm exp} \left( - \frac{(z - Z_0)^2 + (r - R_0)^2}{a_0^2} \right),
\end{equation}
where subscripts \copyedit{$( \, \cdot \, )_0$} denote parameter values at $t = 0$ and the sign of $\omega_\theta$ dictates the propagation direction. Since Gaussian vortex rings only satisfy the governing equations with infinitesimal core thickness, they initially undergo a rapid period of equilibration in which vorticity is redistributed throughout the core \citep{Sha1994,Arc2008,Bal2020}. Following instability growth, transition is often marked by the development of secondary vorticity in a halo around the core vorticity \citep{Daz2006,Ber2007,Arc2008}. During turbulent decay, the shedding of secondary vortex structures to the wake can result in a stepwise decay in circulation \citep{Wei1994,Ber2007}.

Stability analyses of thin vortex rings are often (classically) formulated in terms of asymptotic expansions in $\delta$ \citep{Wid1974,Wid1977,Fuk2005}. Infinitesimally thin vortex rings ($\delta \to 0$) are neutrally stable \citep{Sha1992}. For rings with finite thickness ($\delta > 0$), the curvature instability occurs at first order in $\delta$ and the elliptic instability occurs at second order in $\delta$. The curvature and elliptic instabilities occur at short wavelengths and arise due to parametric resonance between Kelvin waves with core azimuthal wavenumbers separated by one and two, respectively \citep{Fuk2005,Hat2019}. The curvature instability is attributed to a dipole field produced by the vortex ring curvature \citep{Fuk2005,Bla2015,Bla2017}. By contrast, the elliptic instability is attributed to a quadrupole field generated by straining induced by the ring or some external source \citep{Fuk2005,Bla2015,Bla2016}.

This elliptic instability acts to break up elliptic streamlines and is key to the development of three-dimensional transitional and turbulent flows \citep{Ker2002}. In the context of vortex rings (or, more generally, strained vortices), it is sometimes called the \copyedit{Moore--Saffman--Tsai--Widnall} (MSTW) instability \citep{Fuk2005,Cha2021} based on the initial investigations of \citet{Moo1975} and \citet{Tsa1976}. The elliptic instability dominates the curvature instability for thin Gaussian vortex rings without swirl. However, the curvature instability becomes increasingly important for vortex rings with increasing $Re_\mathit{\Gamma}$ and decreasing $\delta$, as well as in vortex rings with swirl \citep{Bla2017, Hat2019}.

While interesting in their own right, thin vortex rings often form canonical building blocks of more complex turbulent flows. Modified vortex geometries, such as elliptic vortex rings \citep{Che2016,Che2019} and trefoil knots \citep{Zha2021,Yao2021}, provide alternative means of probing vortex dynamics and interactions. Collisions between vortex rings and other vortex rings, \copyedit{walls} and free surfaces are also commonly studied to investigate mechanisms underlying the turbulent cascade and the generation of small scales (see \citet{Mis2021} for a review). These mechanisms can be characterized using a variety of collision geometries, including head-on collisions \citep{Che2018,McK2018,McK2020,Mis2021}, inclined collisions \citep{Kid1991,Yao2020,Yao2020b} \copyedit{and} axis-offset collisions \citep{Zaw1991,Smi1994,Ngu2021}, among others. Boundary layers play an important role in \copyedit{vortex--wall} interactions (e.g. by causing rebounding events) \citep{Wal1987} and interactions with free surfaces can often be understood in terms of image vortices \citep{Arc2010}.

Here, we focus on head-on collisions between identical vortex rings of opposite circulation, which have been noted for their rapid enstrophy production \citep{Lu2008,Aya2017,Kan2020}. They have been classically studied in the contexts of the formation of smaller rings through vortex reconnection and the formation of turbulent clouds at high $Re_\mathit{\Gamma}$ \citep{Osh1978,Lim1992,Chu1995}. Many recent investigations have focused particularly on the mechanisms (e.g. instabilities) underlying these transitional and turbulent processes \citep{McK2018,McK2020,Mis2021}.

For the head-on vortex ring collisions under consideration, the elliptic instability competes and interacts with the longer-wavelength Crow instability. The Crow instability \citep{Cro1970} is associated with the mutual interaction of perturbed counter-rotating vortices, which, in the linear regime, locally displaces the vortices without modifying their core structures \citep{Lew2016}. \citet{Mis2021} provides a focused review of vortex ring collisions in the context of the these instabilities. For collisions at relatively low Reynolds numbers, the Crow instability can lead to the pinch-off of secondary vortex rings via local reconnections. At higher Reynolds numbers, the elliptic instability \copyedit{favours} rapid disintegration of the vortex rings into a turbulent cloud.

\citet{McK2020} proposed that iterative elliptic instabilities between successive generations of antiparallel vortices can mediate the turbulent cascade in head-on vortex ring collisions. \citet{Mis2021} also observed that the elliptic instability tends to dominate at high $Re_\mathit{\Gamma}$, although this \copyedit{behaviour} is also sensitive to the slenderness ratio and vorticity distribution. In a different configuration involving symmetrically perturbed antiparallel vortices, \citet{Yao2020a} attributed the turbulent cascade at high $Re_\mathit{\Gamma}$ to an avalanche of successive vortex reconnections. In general, \citet{Ost2021} found that collisions between counter-rotating vortices are indeed highly sensitive to the geometry of their configuration. They particularly found that the mechanisms mediating the cascade bear resemblance to the reconnection scenario \citep{Yao2020a} when the vortices are nearly perpendicular, whereas they are more reminiscent of the iterative elliptic instability scenario \citep{McK2020} when the vortices are more acutely aligned. These recent works share two common themes, (i) that the mode of transition and the formation of a cascade are sensitive to the details of the initial flow configuration and (ii) that the interplay between relevant instabilities is simultaneously important to the flow physics and difficult to capture.

%%%%%%%%%%%%%%%%%%%%%%%%%%%%%%%%%%%%%%%%%%
\subsection{Velocity gradients and vortices}\label{sec:intro:vortex_flow}
%%%%%%%%%%%%%%%%%%%%%%%%%%%%%%%%%%%%%%%%%%

The elliptic instability, which typically dominates head-on collisions between the vortex rings of interest at high $Re_\mathit{\Gamma}$ \citep{McK2020,Mis2021}, is associated with elliptic streamlines \citep{Ker2002}. This generic feature of strained vortical flows can be used to characterize the elliptic instability, which is typically difficult to discern in the complex interactions of multiscale vortices \citep{Mis2021,Ost2021}. Given the inherent complexity of turbulent flows, the geometry of local streamlines provides a relatively simple and interpretable means for characterizing flow features (e.g. vortices).

The instantaneous trajectory of a \copyedit{materially advecting} fluid particle follows the streamlines, which are \copyedit{frame dependent}. At a critical point, e.g. in a frame advecting with the particle, the velocity gradient tensor (VGT), \rahul{$\mathsfbi{A} = \boldsymbol{\nabla} \boldsymbol{u}$}, determines, to linear order, the local structure of streamlines \citep{Per1975, Per1987, Cho1990}. \rahul{The scale-invariant shape of local streamlines is captured by normalizing the VGT as $\mathsfbi{\tilde{A}} = \mathsfbi{A}/A$} \citep{Gir1995,Das2019}, \rahul{where $A = \left\Vert \mathsfbi{A} \right\Vert_{\rm F} = {\rm tr}( \mathsfbi{A}^{\rm T} \mathsfbi{A} )^{1/2}$ is the Frobenius norm of the VGT,} \copyedit{$( \, \cdot \, )^{\rm T}$} \rahul{represents the transpose and, unless otherwise stated, non-bold versions of bold tensor quantities represent their Frobenius norms.} This normalized VGT has been used to investigate the scalings, \copyedit{forcings} and non-local features of the VGT dynamics \citep{Das2019, Das2020a, Das2022} and a similar analysis of vorticity gradients has been used to classify the geometry of local vortex lines \citep{Sha2021}.

The principal invariants of \rahul{$\mathsfbi{\tilde{A}}$} instantaneously characterize local streamline topologies and geometries \citep{Cho1990,Das2019,Das2020a,Das2020}. They are given by
\begin{equation}\label{eq:intro:invariants}
\ppA = -{\rm tr}\left( \mathsfbi{\tilde{A}} \right), \quad \qA = \frac{1}{2}\left( {\rm tr}\left( \mathsfbi{\tilde{A}} \right){\vphantom{\mathsfbi{\tilde{A}}}}^{2} - \, {\rm tr}\left( \mathsfbi{\tilde{A}}{\vphantom{\mathsfbi{\tilde{A}}}}^{2} \right) \right), \quad \rA = -{\rm det}\left( \mathsfbi{\tilde{A}} \right),
\end{equation}
where \copyedit{${\rm tr}( \, \cdot \, )$} and \copyedit{${\rm det}( \, \cdot \, )$} represent the trace and determinant, respectively. For incompressible flows ($\ppA = 0$), four classes of local streamline topologies are separated by degenerate geometries in the $\qA-\,\rA$ plane. Using the invariants of \rahul{$\mathsfbi{\tilde{A}}$} is advantageous \copyedit{compared with} using the invariants of \rahul{$\mathsfbi{A}$} since the $\qA-\,\rA$ plane is a bounded phase space and it provides a more complete representation of streamline geometries \citep{Das2019, Das2020a, Das2020}. For example, the aspect ratio of purely elliptic local streamlines ($\qA > 0$ and $\rA = 0$) is completely characterized by $\qA$, but not by $Q = A^2 \qA$. However, while the $\qA-\,\rA$ plane efficiently characterizes local streamline geometries at critical points, additional parameters are required to fully describe all geometries \citep{Das2020a}.

Following \citet{Das2020}, we consider \copyedit{the} local streamline geometry in the context of the modes of deformation of a fluid parcel: extensional straining, (symmetric and antisymmetric) \copyedit{shearing} and rigid rotation. The well-known \copyedit{Cauchy--Stokes} decomposition of the VGT, \rahul{$\mathsfbi{\tilde{A}} = \mathsfbi{\tilde{S}} + \mathsfbi{\tilde{W}}$}, disambiguates contributions from the symmetric strain rate tensor, \rahul{$\mathsfbi{\tilde{S}} = ( \mathsfbi{\tilde{A}} + \mathsfbi{\tilde{A}}^{\rm T} )/2$}, and the antisymmetric vorticity tensor, \rahul{$\mathsfbi{\tilde{W}} = ( \mathsfbi{\tilde{A}} - \mathsfbi{\tilde{A}}^{\rm T} )/2$}. It has enabled insightful characterizations of \copyedit{the} VGT dynamics from the perspective of the strain rate eigenframe \citep{Tom2021}. However, it does not disambiguate symmetric shearing from extensional straining in \rahul{$\mathsfbi{\tilde{S}}$} or antisymmetric shearing from rigid rotation in \rahul{$\mathsfbi{\tilde{W}}$}. This limitation motivated the development of the triple decomposition of the VGT \citep{Kol2007}, which disambiguates all three fundamental modes of deformation.

\citet{Kol2004, Kol2007} originally formulated the triple decomposition of the VGT by identifying a `basic' reference frame in which motions associated with elongation, rigid \copyedit{rotation} and pure shearing can be isolated. However, identifying a basic reference frame requires a challenging pointwise optimization problem, the solution of which is typically approximated over a finite number of frames \citep{Kol2007, Nag2020}. More recently, \citet{Gao2018, Gao2019} introduced a unique triple decomposition, based on a related vorticity tensor decomposition \citep{Liu2018, Gao2019a}, that is more computationally practical than that of \citet{Kol2004, Kol2007}. This triple decomposition is formally performed in a local `principal' coordinate system ($x^*, y^*, z^*$), which is related to the global coordinate system ($x, y, z$) by an orthogonal transformation. In this principal frame, denoted by \copyedit{$( \, \cdot \, )^*$}, the triple decomposition is given in normalized form by
\begin{equation}\label{eq:intro:triple}
    \mathsfbi{\tilde{A}}^* = \underbrace{\begin{bmatrix} \dot{\epsilon}_{x^*} & 0 & 0 \\ 0 & \dot{\epsilon}_{y^*} & 0 \\ 0 & 0 & \dot{\epsilon}_{z^*} \end{bmatrix}}_{\textstyle \boldsymbol{\dot{\epsilon}}^*} + \underbrace{\begin{bmatrix} 0 & 0 & 0 \\ \dot{\gamma}_{z^*} & 0 & 0 \\ \dot{\gamma}_{y^*} & \dot{\gamma}_{x^*} & 0 \end{bmatrix}}_{\textstyle \boldsymbol{\dot{\gamma}}^*} + \underbrace{\begin{bmatrix} 0 & -\dot{\varphi_{z^*}} & 0 \\ \dot{\varphi_{z^*}} & 0 & 0 \\ 0 & 0 & 0 \end{bmatrix}}_{\textstyle \boldsymbol{\dot{\varphi}}^*}.
\end{equation}
Here, $\boldsymbol{\dot{\epsilon}}^*$, $\boldsymbol{\dot{\gamma}}^*$ \copyedit{and} $\boldsymbol{\dot{\varphi}}^*$ represent the normal straining, pure \copyedit{shearing} and rigid body rotation tensors, respectively. Their constituents can be directly identified from the components of the VGT in the principal frame \citep{Gao2018, Gao2019, Das2020}. Their representations in the global coordinates ($\boldsymbol{\dot{\epsilon}}$, $\boldsymbol{\dot{\gamma}}$ \copyedit{and} $\boldsymbol{\dot{\varphi}}$) can subsequently be recovered  by inverting (i.e. transposing) the original orthogonal transformation \citep{Gao2019a}.

The components of the normal straining tensor represent the real parts of the eigenvalues of \rahul{$\mathsfbi{\tilde{A}}$}, which are identical to those of \rahul{$\mathsfbi{\tilde{A}}^*$}. For points with rotational local streamlines, \rahul{$\mathsfbi{\tilde{A}}$} has a pair of complex eigenvalues and the real eigenvector defines the local rotation axis. In this case, the transformation to the principal frame is identified by (i) using a real Schur decomposition to align the $z^*$-axis with the real eigenvector of the VGT and (ii) orienting the $x^*-y^*$ plane to minimize the local rotational speed \citep{Liu2018, Das2020}. One advantage of (\ref{eq:intro:triple}) is that it provides representations of the strength ($2 \dot{\varphi}_{z^*}$) and the axis ($z^*$) of rigid rotation that are Galilean invariant \citep{Wan2018}. Unlike the rotational case, the VGT has only real eigenvalues when the local streamline geometry is non-rotational ($\boldsymbol{\dot{\varphi}}^* = \boldsymbol{0}$). In this case, the principal frame is identified by using a Schur decomposition to transform the VGT into a triangular tensor. The modes of deformation are then isolated by decomposing this transformed tensor into a normal, diagonal tensor representing normal straining and a non-normal, strictly triangular tensor representing pure shearing \citep{Key2018, Das2020}.

The triple decomposition enables refined analyses of the influences of fundamental constituents of the VGT. For example, the original triple decomposition \citep{Kol2007} has been used to show that lifetimes of fundamental flow structures at macroscopic scales (where viscosity can be neglected) can be related to stability of rigid rotation, linear instability of pure \copyedit{shearing} and exponential instability of irrotational straining \citep{Hof2021}. At small scales, the more recent triple decomposition \citep{Gao2018, Gao2019} has been used to show that pure shearing is typically the dominant contributor to energy dissipation \citep{Wu2020} and intermittency \citep{Das2020} in turbulent flows. Further, the symmetric and antisymmetric components of $\boldsymbol{\dot{\gamma}}^*$ are given by $\gSb = \left( \boldsymbol{\dot{\gamma}}^* + \, \boldsymbol{\dot{\gamma}}^*{}^{\rm T} \right)/2$ and $\gWb = \left( \boldsymbol{\dot{\gamma}}^* - \, \boldsymbol{\dot{\gamma}}^*{}^{\rm T} \right)/2$, respectively. In this manner, the triple decomposition is more refined than the \copyedit{Cauchy--Stokes} decomposition since \rahul{$\mathsfbi{\tilde{S}}^* = \boldsymbol{\dot{\epsilon}}^* + \, \gSb$} and \rahul{$\mathsfbi{\tilde{W}}^* = \boldsymbol{\dot{\varphi}}^* + \, \gWb$} \citep{Gao2019, Das2020}. As described in detail by \citet{Das2020}, the triple decomposition also enables a natural characterization of local streamline topologies and geometries. Similar topological analyses of vortical flow features have also been proposed \citep{Nak2017}, but we focus on the triple decomposition for the advantages outlined herein.

The ability of the triple decomposition to capture local streamline structure in terms of fundamental modes of deformation has guided efforts to define improved vortex criteria. There are an abundance of criteria to identify vortices that are based on various features (e.g. eigenvalues) of the VGT and that adopt various philosophies of what constitutes a vortex \citep{Cha2005, Epp2017, Gun2018, Liu2019b, Hal2021}. Debates surrounding these criteria primarily involve their (i) philosophical underpinnings, (ii) threshold \copyedit{sensitivities} and (iii) observational invariances.

Regarding (i), the \copyedit{Cauchy--Stokes} decomposition underlies many common symmetry-based vortex criteria, including the $Q$ \citep{Hun1988} and $\lambda_2$ \citep{Jeo1995} criteria. Local streamline topology underlies many common geometry-based vortex criteria, including the $\Delta$ \citep{Cho1990} and $\lambda_{\rm ci}$ \citep{Cha2005} criteria. Like the geometry-based methods, and unlike the symmetry-based methods, the rigid vorticity criterion ($\dot{\varphi}_{z^*} > 0$) \citep{Tia2018} captures all rotational local streamline geometries under the assumption that rigid rotation is an essential ingredient of a vortex \citep{Liu2019b, Das2020}. The philosophical distinction between symmetry-based and geometry-based criteria also underlies the so-called `disappearing vortex problem' in which, fixing the VGT configuration and strain rate, increasing only the vorticity magnitude can remove a geometry-based vortex from the flow \citep{Cha2005, Kol2020, Kol2022}. However, we here adopt the geometry-based viewpoint since, unlike vorticity, rigid rotation persistently underlies rotational local streamline topologies in all inertial frames. This interpretation in terms of local streamline topology has the potential to elucidate connections to related (e.g. elliptic) instabilities.

Regarding (ii), the Omega ($\Omega$) class of vortex criteria \citep{Liu2016,Don2018,Don2019,Liu2019} is advantageous since it uses quantities that are bounded and less \copyedit{threshold sensitive} than the aforementioned methods. Regarding (iii), whereas most common vortex criteria are Galilean invariant, they are typically not objective since they are not preserved in rotating reference frames \citep{Epp2017, Gun2018}. \copyedit{However,} the objectivized $\dot{\varphi}_{z^*}$ \citep{Liu2019a} and objectivized $\Omega$ \citep{Liu2019c} criteria, which are formulated by replacing \rahul{$\mathsfbi{\tilde{W}}$} with its deviation from its global spatial mean, remain invariant in these reference frames. Moreover, they are among the only compatible (i.e. self-consistent) objectivized vortex criteria out of the modifications commonly associated with the vortex criteria we have discussed \citep{Hal2021}. This advantage enhances the experimental verifiability and clarifies the physical significance of visualizations of the corresponding vortex structures.

Synthesizing the advantages of the geometry-based vortex definitions and the $\Omega$ class of vortex criteria, we identify vortices using the $\Omega_r$ method in the present investigation. This criterion is formulated in terms of the quantity
\begin{equation}
    \Omega_r = \frac{\left( \boldsymbol{\omega} \boldsymbol{\cdot} \boldsymbol{e}_{z^*} \right)^2}{2 \left( \boldsymbol{\omega} \boldsymbol{\cdot} \boldsymbol{e}_{z^*} \right)^2 - 4 \lambda_{\rm ci}{\vphantom{\left( \boldsymbol{\omega} \boldsymbol{\cdot} \boldsymbol{e}_{z^*} \right)}}^{\mkern-14mu 2} + 4 \varepsilon_{\rm vort}},
\end{equation}
where $\lambda_{\rm ci}$ is the imaginary part of the complex eigenvalues of \rahul{$\mathsfbi{A}$}, $\boldsymbol{e}_{z^*}$ is the unit vector along the $z^*$-\copyedit{axis} and $\varepsilon_{\rm vort}$ is a numerical threshold used to prevent division by zero. Vortices are theoretically identified as spatially connected regions satisfying $\Omega_r > 0.5$ when $\varepsilon_{\rm vort} = 0$. In practice, however, vortices are identified using a small $\varepsilon_{\rm vort} > 0$ and $\Omega_r \geq 0.52$ \citep{Liu2019b} \copyedit{to} e.g. remove weak vortices.

%%%%%%%%%%%%%%%%%%%%%%%%%%%%%%%%%%%%%%%%%%
\subsection{Contributions}\label{sec:intro:contributions}
%%%%%%%%%%%%%%%%%%%%%%%%%%%%%%%%%%%%%%%%%%

In this paper, we utilize the advantageous properties of geometry-based analyses of the VGT to efficiently characterize turbulence initiated by a vortex ring collision. We use the adaptive, multiresolution computational techniques discussed in \textsection \ref{sec:methods} to perform a direct numerical simulation of this flow at $Re_{\mathit{\Gamma}_0} = 4000$. In \textsection \ref{sec:res:evol_metrics}, we establish the fidelity of our simulation and we visualize and discuss the various regimes of its evolution. In \textsection \ref{sec:res:partition}, we \copyedit{analyse} the partitioning of the velocity gradients to characterize these regimes in terms of the modes of deformation. In \textsection \ref{sec:res:phase}, we introduce a geometry-based phase space that characterizes the action of the elliptic instability and its interplay with other mechanisms driving the turbulent flow. Our analyses reveal statistical features of the VGT that are similar to those of previous simulations. They also provide tools with the potential to help disentangle mechanisms underlying vortex interactions during transition and turbulent decay. Finally, we summarize our results in the context of previous works and highlight promising future research prospects in \textsection \ref{sec:conclusions}.

%%%%%%%%%%%%%%%%%%%%%%%%%%%%%%%%%%%%%%%%%%
\section{Methods}\label{sec:methods}
%%%%%%%%%%%%%%%%%%%%%%%%%%%%%%%%%%%%%%%%%%

%%%%%%%%%%%%%%%%%%%%%%%%%%%%%%%%%%%%%%%%%%
\subsection{Computational method}\label{sec:meth:comp}
%%%%%%%%%%%%%%%%%%%%%%%%%%%%%%%%%%%%%%%%%%

To efficiently simulate a turbulent vortex ring collision, we adopt a \copyedit{recently developed} multiresolution solver for viscous, incompressible flows on unbounded domains \citep{Lis2016, Dor2020, Yu2021, Yu2022}. \citet{Yu2022} \copyedit{provide} a detailed discussion of the formulation, \copyedit{properties} and performance of the method. We summarize the key advantages of the solver here and expound the computational formulation in \autoref{sec:app:comp_meth}. The advantages we discuss allow us to simulate a relatively high Reynolds number vortex ring collision at a relatively low computational cost.

The Navier-Stokes equations (NSE) are spatially discretized onto a staggered Cartesian grid using a second-order-accurate finite-volume scheme that endows discrete operators with useful properties (i) \citep{Lis2016}. Discrete differential operators are constructed to mimic the symmetry, \copyedit{orthogonality} and integration properties of their continuous counterparts. They also commute with the Laplacian and integrating \copyedit{factor} operators, as defined in \autoref{sec:app:comp_meth}. Furthermore, the discretization of the nonlinear term in the momentum equations preserves relevant (e.g. energy) conservation properties in the absence of viscosity. Together, the mimesis, \copyedit{commutativity} and conservation properties of the discretization scheme facilitate fast, stable, high-fidelity simulations of turbulent flows.

The computational methods we employ also have high parallel efficiency (ii) and linear algorithmic complexity (iii). The computational efficiency of the flow solver is primarily \copyedit{centred} around solving the discrete pressure Poisson equation on a formally unbounded grid using the lattice Green's function (LGF) \citep{Lis2014, Lis2016, Lis2016a}. The Poisson equation is obtained by taking the divergence of the NSE in rotational form, such that the source term is \copyedit{$\boldsymbol{\nabla} \boldsymbol{\cdot} \boldsymbol{r}$}, where $\boldsymbol{r} = \boldsymbol{u} \times \boldsymbol{\omega}$ is the Lamb vector. By considering flows with at least exponentially decaying far-field vorticity, the approximate support of this source field can be captured using a finite computational domain. Given a source cutoff threshold, the finite domain is adaptively truncated to capture only the regions relevant to the Poisson problem. Solving the Poisson problem over this domain involves the convolution of the LGF with the source field. The flow solver achieves (ii) and (iii) by efficiently evaluating this convolution via a fast multipole method \citep{Lis2014} that compresses the kernel using polynomial interpolation. This method is accelerated by exploiting the efficiency of fast Fourier transforms on a block-structured Cartesian grid.

In addition to spatially adapting the extent of the computational domain, adaptive multiresolution discretization (iv) is achieved by using adaptive mesh refinement (AMR) to reduce the number of degrees of freedom required for solutions. As discussed previously \citep{Dor2020, Yu2022}, the present AMR framework is carefully constructed to preserve the desirable operator properties (i) and augment the efficiencies (ii, iii) associated with the uniform-grid framework \citep{Lis2014, Lis2016}. In the AMR framework, the computational grid is partitioned into multiple levels, each with double the resolution in each direction as the previous level. The spatial regions associated with each level are \copyedit{non-overlapping,} except for extended regions that are used to compute a combined source term that includes a correction induced by the difference between the coarse-grid and fine-grid partial solutions. As formulated in \autoref{sec:app:comp_meth}, a region is refined when its combined source exceeds a threshold and it is coarsened when its combined source falls below a smaller threshold. As shown in \textsection \ref{sec:res:metrics} (see \cref{tab:res:regimes}), this AMR formulation drastically reduces the number of computational cells required to capture a head-on vortex ring collision compared \copyedit{with} a fixed-resolution scheme.

%%%%%%%%%%%%%%%%%%%%%%%%%%%%%%%%%%%%%%%%%%
\subsection{Vortex ring collision simulation}\label{sec:simulation}
%%%%%%%%%%%%%%%%%%%%%%%%%%%%%%%%%%%%%%%%%%

\begin{figure}
    \centering
    \includegraphics[width=\textwidth,trim=30 10 30 10,clip]{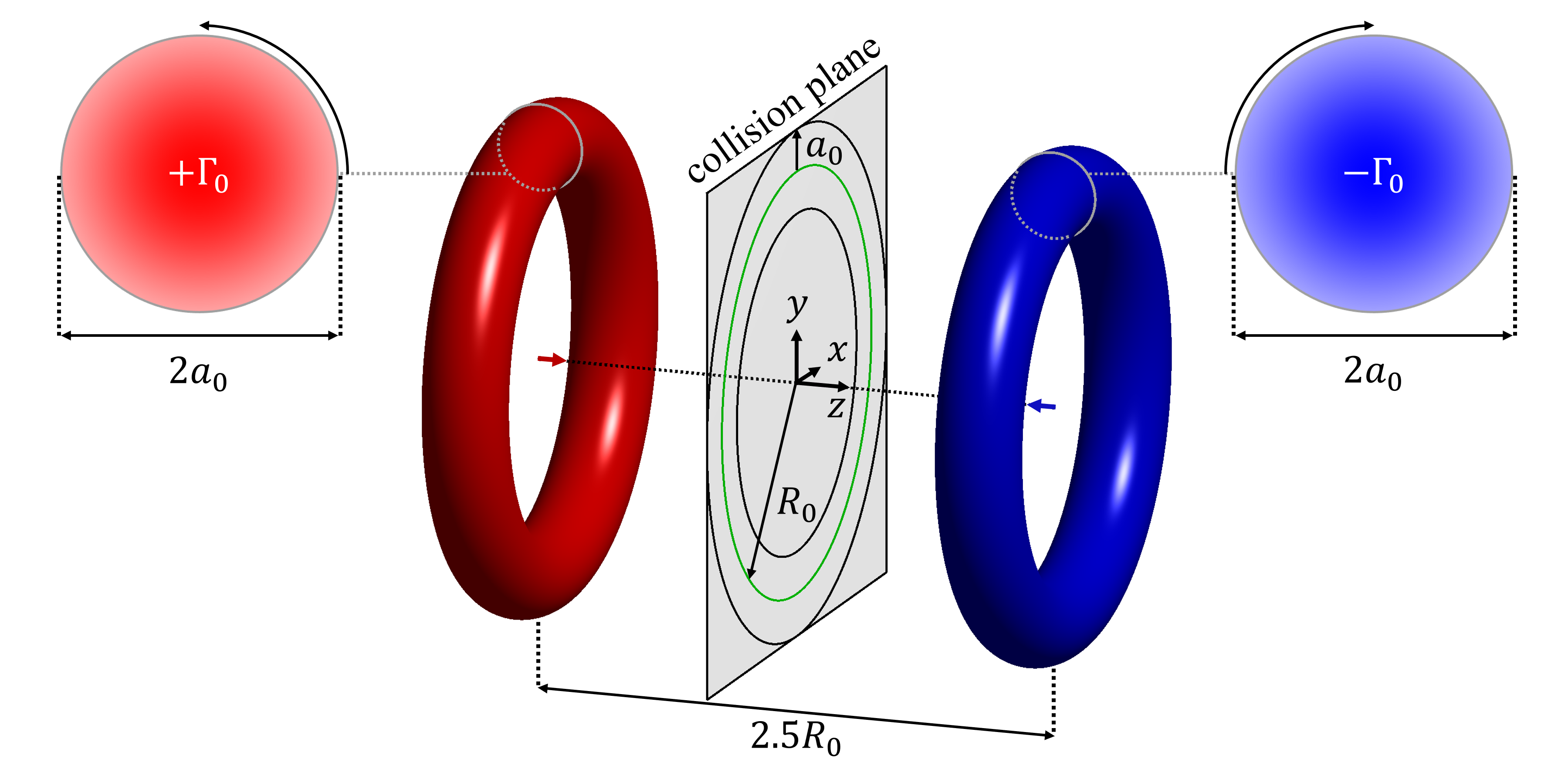}
    \caption{Initial geometry of the flow configuration used to simulate the head-on collision between vortex rings. The shading of the vortex cores reflects their Gaussian vorticity profiles.}
    \label{fig:meth:config}
\end{figure}

As depicted in \cref{fig:meth:config}, we consider a flow configuration in which the vortex rings are initialized with opposing circulations such that they propagate toward one another along the $z$-axis and meet at the collision plane at $z = 0$. The rings are initialized a distance $L_z = 2.5 R_0$ apart, which is sufficiently large to mitigate their mutual influence during the most vigorous period of equilibration. Both rings are initialized with Gaussian vorticity distributions (\ref{eq:intro:Gaussian}) such that $Re_{\mathit{\Gamma}_0} = 4000$ and $\delta_0 = 0.2$. Unless otherwise stated, we use the initial circulation, $\mathit{\Gamma}_0 = 1$, and radius, $R_0 = 1$, of each ring to non-dimensionalize all variables. To excite transition, we randomly perturb the radii of the vortex rings using the first 32 Fourier modes in $\theta$, which are prescribed random phases and uniform magnitudes, $R_{\rm pert} = 5 \times 10^{-4}$. Consistent with previous tests \citep{Yu2022}, these initial perturbations are sufficiently large to dominate perturbations incurred by discretization errors.

The computational mesh we use has $N_{\rm level} = 2$ levels of refinement beyond the base level such that the ratio of the \copyedit{coarsest-grid} spacing to the \copyedit{finest-grid} spacing is $\Delta x_{\rm base} / \Delta x_{\rm fine} = 4$. Based on preliminary simulations of turbulent vortex rings \citep{Lis2016} and vortex ring collisions \citep{Yu2022}, we select $a_0/\Delta x_{\rm base} = 5$ and $\Delta t/ \Delta x_{\rm fine} = 0.35$ to ensure the flow is \copyedit{well resolved} throughout the simulation. Finally, parameters controlling the spatial and mesh refinement thresholds are chosen as discussed in \autoref{sec:app:comp_meth}.

%%%%%%%%%%%%%%%%%%%%%%%%%%%%%%%%%%%%%%%%%%
\subsection{Simulation integral metrics}\label{sec:meth:metrics}
%%%%%%%%%%%%%%%%%%%%%%%%%%%%%%%%%%%%%%%%%%

We track the evolution and fidelity of the simulation using integral metrics associated with incompressible flows \citep{Lis2016}. Particularly, we compute the hydrodynamic impulse, helicity, vortical kinetic \copyedit{energy} and enstrophy of the flow, which are denoted by $\boldsymbol{I}_{V}$, $H$, $K_V$ \copyedit{and} $E$, respectively. These integrals are formally evaluated on an unbounded domain, but we evaluate them using the finite AMR grid as
\begin{equation}\label{eq:metrics}
\begin{alignedat}{14}
    &\;\boldsymbol{I}_{V}&&(t) &&= &&\int_{V(t)} &&\,(&&\boldsymbol{x} \times \boldsymbol{\omega})  &&dV, \quad &&H(t) &&= &&\int_{V(t)} &&( &&\boldsymbol{u} &&\boldsymbol{\cdot} \, \boldsymbol{\omega} ) &&dV, \\
     &K_{V}&&(t) &&= &&\int_{V(t)} \boldsymbol{u} \, \boldsymbol{\cdot} &&\,( &&\boldsymbol{x} \times \boldsymbol{\omega} ) &&dV, \quad &&E(t) &&= \frac{1}{2} &&\int_{V(t)} &&\lvert &&\boldsymbol{\omega} &&\rvert^2 \; &&dV,
\end{alignedat}
\end{equation}
where $V(t)$ is the time-varying AMR grid. The impulse is the appropriate measure of momentum since it converges for flows on unbounded domains with compact vorticity. The vortical kinetic energy can also be expressed as  $K_V = K + K_{\partial V}$, where $K$ represents the kinetic energy and $K_{\partial V}$ is a correction term based on the flow at the boundary of the grid, $\partial V(t)$. These metrics can be expressed as
\begin{equation}
    K(t) = \frac{1}{2} \int_{V(t)} \lvert \boldsymbol{u} \rvert^2 \; dV, \quad K_{\partial V}(t) = \int_{\partial V(t)} \boldsymbol{x} \boldsymbol{\cdot} \left( ( \boldsymbol{u} \boldsymbol{u} ) \boldsymbol{\cdot} \boldsymbol{n} - \frac{1}{2} \lvert \boldsymbol{u} \rvert^2 \boldsymbol{n} \right) dS, \label{eq:quality:kindefs}
\end{equation}
where $\boldsymbol{n}$ is the normal vector of $\partial V$ \citep{Wu2015}. For vanishing far-field velocity, $K_{\partial V}$ vanishes on unbounded domains and, for the present grid, we make use of the smallness of $K_{\partial V}$ when \copyedit{analysing} dissipation.

In the absence of non-conservative external body forces, the hydrodynamic impulse is conserved for incompressible flows on unbounded domains \citep{Saf1993}. The helicity would also be conserved in the absence of viscosity, and it is useful for assessing simulation fidelity as the vortex rings initially approach the collision plane since the evolution of the flow is dominated by inviscid effects. \rahul{These integral metrics initially evaluate to $\boldsymbol{I}_{V}(0) = \boldsymbol{0}$ and $H(0) = 0$ due to the spatial symmetries of the initial flow configuration. These initial symmetries hold to the extent that the vorticity is} \copyedit{well captured} \rahul{and the contributions of the random perturbations used to excite instability growth are negligible.} As the flow evolves, subsequent deviations from these initial values reflect the degree to which the corresponding initial symmetries are broken.

The enstrophy and the kinetic energy provide a more detailed picture of simulation fidelity during transition and turbulent decay, when viscous dissipation at small scales becomes relevant. For unsteady, incompressible flows on unbounded domains, the dissipation governs the decay rate of kinetic energy and can be expressed in terms of the enstrophy. Comparing these integrals is useful for characterizing (i) the degree to which small-scale features are resolved during peak dissipation and (ii) the flux of kinetic energy out of the finite computational domain \citep{Arc2008}. We therefore introduce effective Reynolds numbers, which are given by
\begin{equation}\label{eq:quality:Re}
    \frac{Re^{\rm eff}_S (t)}{Re_{\mathit{\Gamma}_0}} = -\frac{\Phi_S (t)}{dK/dt}, \quad\quad \frac{Re^{\rm eff}_W (t)}{Re_{\mathit{\Gamma}_0}} = -\frac{\Phi_W (t)}{dK/dt},
\end{equation}
where $\Phi_S$ is the volume-integrated dissipation and $\Phi_W = 2 E/Re_{\mathit{\Gamma}_0}$ is its enstrophy-based counterpart \citep{Ser1959}. Here, we differentiate $K$ instead of $K_V$ when computing the effective Reynolds numbers to prevent amplification of the noise associated with adaptations in the computational domain, to which $K_V$ is more sensitive. This is justified since $K$ and $K_V$ are nearly identical throughout the present simulations (see \cref{fig:metrics_v2}). The ratio $Re^{\rm eff}_S$ is useful for assessing spatial resolution since the dissipation can vary significantly during transition and turbulent decay. The corresponding Kolmogorov scale, $\eta = (\nu^3/\Phi_S)^{1/4}$, can also be used to validate the selected grid spacings. The difference between $Re^{\rm eff}_S$ and $Re^{\rm eff}_W$ reflects the relative significance of the acceleration of the flow on $\partial V$ through the boundary integral in the \copyedit{Bobyleff--Forsyth} formula \citep{Ser1959}. Together, the error metrics defined in this section comprehensively characterize the fidelity of the simulation as its flow structures evolve and the computational domain adapts accordingly.

%%%%%%%%%%%%%%%%%%%%%%%%%%%%%%%%%%%%%%%%%%
\section{Evolution of integral metrics and vortical structures}\label{sec:res:evol_metrics}
%%%%%%%%%%%%%%%%%%%%%%%%%%%%%%%%%%%%%%%%%%

%%%%%%%%%%%%%%%%%%%%%%%%%%%%%%%%%%%%%%%%%%
\subsection{Evolution of integral metrics}\label{sec:res:metrics}
%%%%%%%%%%%%%%%%%%%%%%%%%%%%%%%%%%%%%%%%%%

\begin{figure}
    \centering
    \includegraphics[width=\textwidth,trim=150 160 150 70,clip]{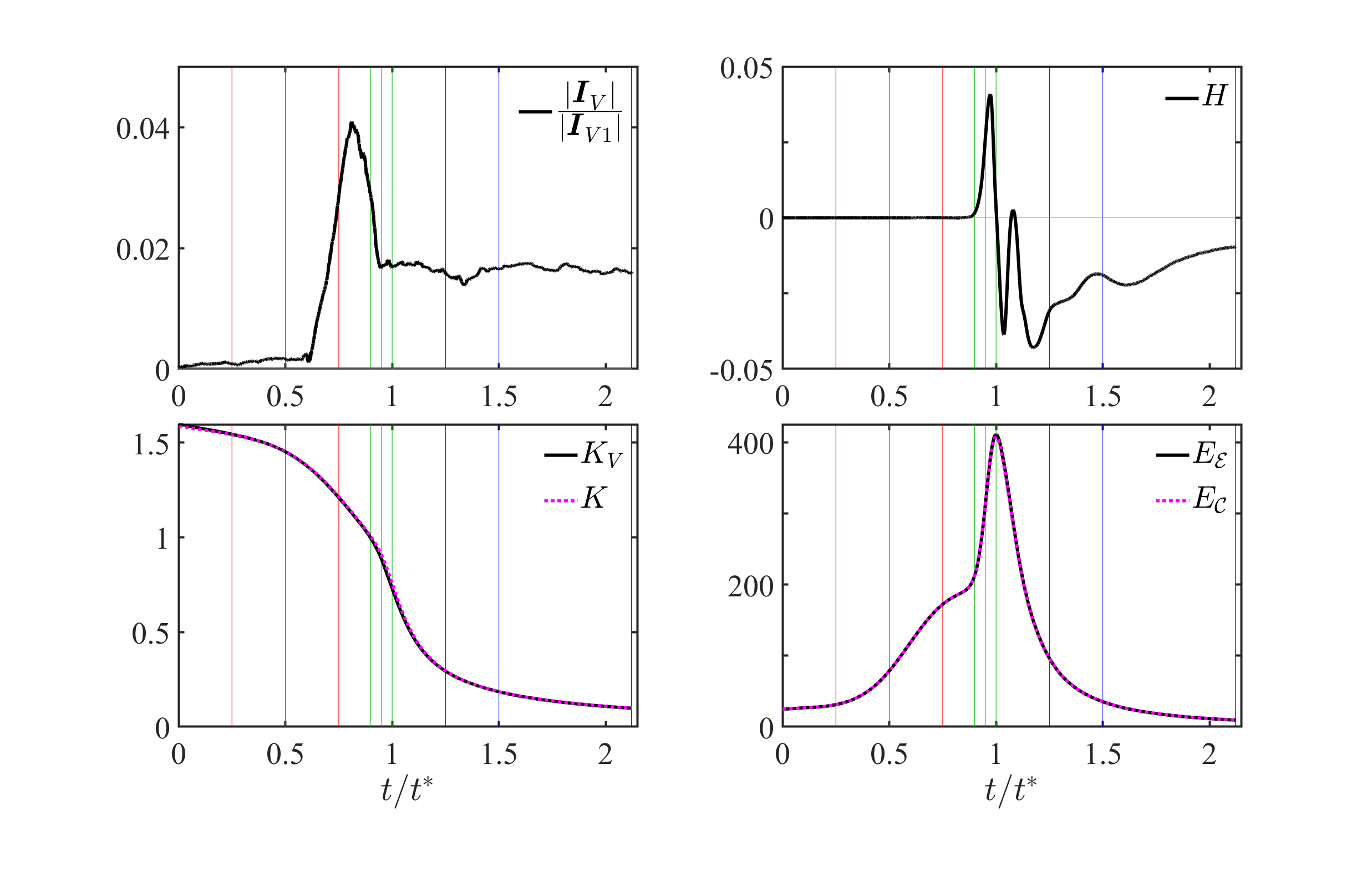}
    \includegraphics[width=\textwidth,trim=150 90 150 88,clip]{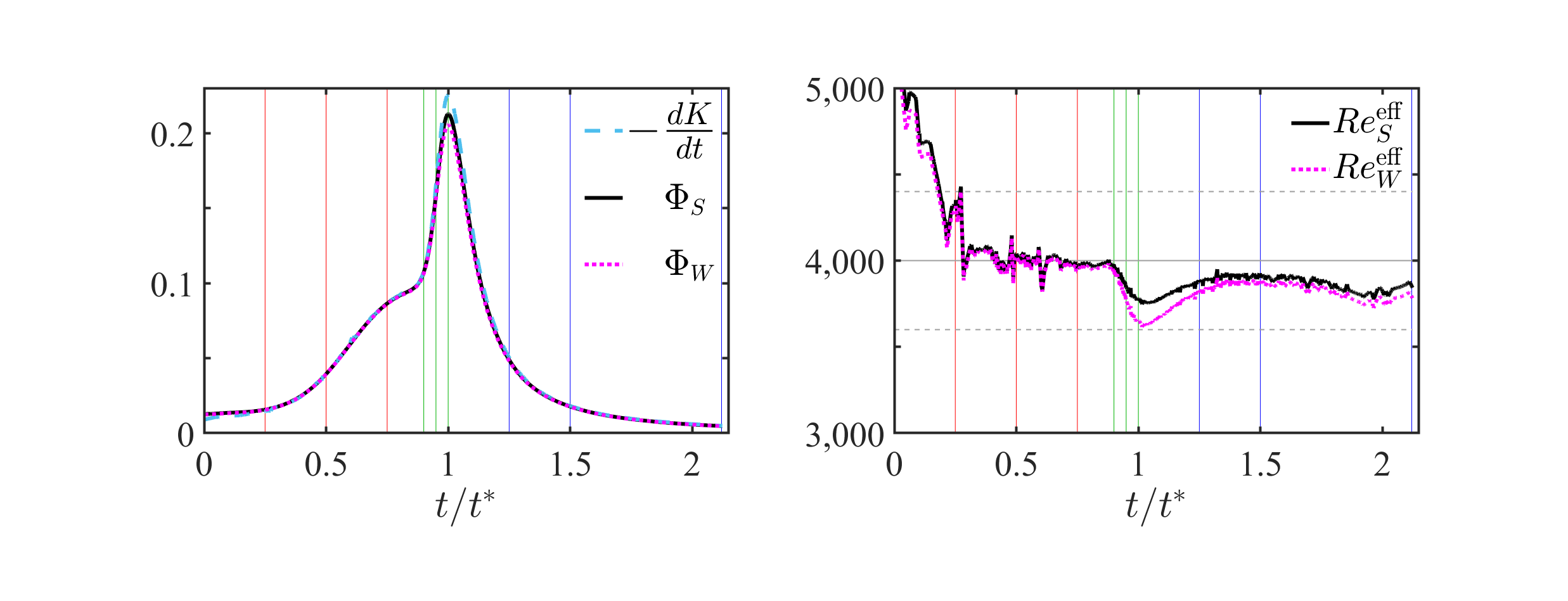}
    \caption{Temporal evolution of the integral metrics defined in \textsection \ref{sec:meth:metrics} over the course of the simulation. The vertical lines correspond to the reference times in \cref{tab:res:regimes} and they are \copyedit{coloured} accordingly. The horizontal lines in the $Re^{\rm eff}$ panel represent $Re_{\mathit{\Gamma}_0}$ (solid) with 10\% margins (dashed). The impulse magnitude is normalized by that of each vortex ring in isolation, $\lvert \boldsymbol{I}_{V1} \rvert \approx 1.02\pi \approx 3.204$. The enstrophies $E_{\mathcal{E}}$ and $E_{\mathcal{C}}$ are computed using vorticities located at the edges and \copyedit{centres}, respectively, of the computational cells.}
    \label{fig:metrics_v2}
\end{figure}

\autoref{fig:metrics_v2} shows the evolution of the integral metrics from \textsection \ref{sec:meth:metrics} over the course of the simulation. In the subsequent analysis, we reference the various regimes of flow development with respect to the time, $t^* = 14.77$, at which maximum dissipation is attained. \autoref{tab:res:regimes} qualitatively characterizes the state of the simulation at each reference time we consider for the initial, \copyedit{transitional} and turbulent regimes of the simulation.

\definecolor{dr}{rgb}{0.957,0.738,0.699}% dark red
\definecolor{dg}{rgb}{0.700,0.950,0.700}% dark green
\definecolor{db}{rgb}{0.630,0.790,0.950}% dark blue
\colorlet{lr}{dr!70}% light red
\colorlet{lg}{dg!70}% light green
\colorlet{lb}{db!70}% light blue

\begin{table}
    \centering
    \renewcommand*{\arraystretch}{1.15}
    \begin{tabular}{|r|c|c|c|ccc|}
    \hline
    Regime of evolution & $\;\,t/t^*$ & $\bar{R}_p$ & $V$ & $\phi_0$ (\%) & $\phi_1$ (\%) & $\phi_2$ (\%) \\
    \hline
    \cellcolor{lr} Post-equilibration vortex rings & \cellcolor{lr} 0.25 & \cellcolor{lr} 1.18 & \cellcolor{lr} 201 & \cellcolor{lr} 80.9 & \cellcolor{lr} 13.8 & \cellcolor{lr} 5.3 \\
    \cellcolor{lr} Vortex boundary merger & \cellcolor{lr} 0.50 & \cellcolor{lr} 2.01 & \cellcolor{lr} 243 & \cellcolor{lr} 76.6 & \cellcolor{lr} 18.5 & \cellcolor{lr} 4.9 \\
    \cellcolor{lr} Enhanced radial expansion & \cellcolor{lr} 0.75 & \cellcolor{lr} 3.58 & \cellcolor{lr} 356 & \cellcolor{lr} 80.6 & \cellcolor{lr} 15.7 & \cellcolor{lr} 3.7 \\
    \hline
    \cellcolor{lg} Formation of secondary vortices & \cellcolor{lg} 0.90 & \cellcolor{lg} 4.51 & \cellcolor{lg} 408 & \cellcolor{lg} 76.6 & \cellcolor{lg} 20.0 & \cellcolor{lg} 3.4 \\
    \cellcolor{lg} Interaction of secondary vortices & \cellcolor{lg} 0.95 & \cellcolor{lg} 4.77 & \cellcolor{lg} 421 & \cellcolor{lg} 72.0 & \cellcolor{lg} 24.1 & \cellcolor{lg} 3.9 \\
    \cellcolor{lg} Proliferation of small scales & \cellcolor{lg} 1.00 & \cellcolor{lg} 4.97 & \cellcolor{lg} 445 & \cellcolor{lg} 63.8 & \cellcolor{lg} 30.9 & \cellcolor{lg} 5.3 \\
    \hline
    \cellcolor{lb} Early turbulent decay & \cellcolor{lb} 1.25 & \cellcolor{lb} 5.38 & \cellcolor{lb} 507 & \cellcolor{lb} 61.4 & \cellcolor{lb} 30.8 & \cellcolor{lb} 7.8 \\
    \cellcolor{lb} Intermediate turbulent decay & \cellcolor{lb} 1.50 & \cellcolor{lb} 5.60 & \cellcolor{lb} 531 & \cellcolor{lb} 66.7 & \cellcolor{lb} 26.7 & \cellcolor{lb} 6.6 \\
    \cellcolor{lb} Late turbulent decay & \cellcolor{lb} 2.12 & \cellcolor{lb} 5.82 & \cellcolor{lb} 564 & \cellcolor{lb} 72.1 & \cellcolor{lb} 26.8 & \cellcolor{lb} 1.1 \\
    \hline
    \end{tabular}
    \caption{Reference times used to \copyedit{analyse} the initial (red), transitional (green) \copyedit{and} turbulent (blue) regimes of the present vortex ring collision, where $t^* = 14.77$ is the time of maximum dissipation. Here, $\bar{R}_p$ represents the mean vortex ring radius (see \autoref{sec:app:elliptic}), $V$ is the volume of the computational \copyedit{domain} and $\phi_k$ is the fraction of $V$ occupied by level $k$ of the AMR grid, which has grid spacing $\Delta x_k = \Delta x_{\rm base}/2^{k}$ with $\Delta x_{\rm base} = 0.04$.}
    \label{tab:res:regimes}
\end{table}

The initial evolution of the flow involves a rapid period of equilibration ($t \lesssim 0.25 t^*$) and the propagation of the equilibrated rings towards the collision plane ($0.25 t^* \lesssim t \lesssim 0.50 t^*$). \rahul{The interaction of the rings accelerates their radial expansion ($0.50 t^* \lesssim t \lesssim 0.75 t^*$) and the elliptic instability eventually emerges along the expanding rings ($0.75 t^* \lesssim t \lesssim 0.90 t^*$).} \autoref{sec:app:elliptic} \rahul{supports the importance of the elliptic instability during the early stages of transition.} Subsequently, the flow transitions to turbulence ($0.90 t^* \lesssim t \lesssim t^*$) and rapidly produces small-scale flow structures. Following transition, the flow undergoes turbulent decay for the remainder of the simulation (i.e. for $t \gtrsim t^*$). See \textsection \ref{sec:res:vortex} for visualizations of the flow at the reference times from \cref{tab:res:regimes} associated with each of these regimes of evolution.

As the vortex rings initially propagate towards the collision plane ($t \lesssim 0.50 t^*$), the kinetic energy decays slowly and the enstrophy and dissipation are relatively small. The effective Reynolds numbers rapidly adjust to the value of $Re_{\mathit{\Gamma}_0}$ during the initial equilibration period ($t \lesssim 0.25 t^*$) and remain roughly constant as the equilibrated rings approach the collision plane ($0.25 t^* \lesssim t \lesssim 0.50 t^*$). The helicity is \copyedit{well conserved} in this regime since the flow evolves in a nearly inviscid fashion. Further, the impulse is initially small and grows relatively slowly during this period. These results suggest that the symmetries associated with the handedness and momentum distribution of the flow are \copyedit{well preserved} in the initial regime of evolution.

As the rings expand radially at the collision plane ($0.50 t^* \lesssim t \lesssim 0.75 t^*$) and the elliptic instability emerges ($0.75 t^* \lesssim t \lesssim 0.90 t^*$), the kinetic energy decays more rapidly and the dissipation grows. During these periods, the helicity symmetry remains \copyedit{well preserved} and the effective Reynolds numbers remain relatively constant near $Re_{\mathit{\Gamma}_0}$, suggesting that the flow is \copyedit{well resolved}. However, the impulse magnitude varies more rapidly in time due to the rapid radial expansion of the rings. In following the expanding vortical flow at the collision plane, the adaptations of the domain break the symmetry associated with impulse integral more significantly than during the initial evolution of the rings. The resulting growth in $\lvert \boldsymbol{I}_V \rvert$ is primarily attributed to its component in the $z$ direction, along which the domain is compressed as the flow concentrates about the collision plane. 

As the flow transitions to turbulence ($0.90 t^* \lesssim t \lesssim t^*$), the kinetic energy decays even more rapidly and the dissipation approaches its maximum value. Due to the proliferation of small-scale flow structures during this period, the effective Reynolds numbers drop to their minimum values at $t \approx t^*$, when the flow is most difficult to resolve. The increased difference between $Re^{\rm eff}_S$ and $Re^{\rm eff}_W$ reflects that the acceleration of the flow near $\partial V$ is more relevant at this time. The rapid generation of small-scale flow structures also implies that viscosity plays a more important role in this regime. Correspondingly, the helicity begins to vary in time in this regime, reaching its maximum rate of change at the time of peak dissipation. Its variations reflect that vortex lines in the flow undergo rapid topological changes during transition. By contrast, the impulse magnitude decays to a roughly constant value as the radial expansion of the rings slows in this transitional regime.

During the turbulent decay of the flow ($t \gtrsim t^*$), the kinetic energy becomes small and the dissipation decays rapidly, eventually falling below its initial value (at $t \approx 1.62 t^*$). The dissipation matches the kinetic energy decay rate more closely for this regime than for transition. Further, as the turbulence develops, the effective Reynolds numbers agree well with one another and, to a lesser extent, with $Re_{\mathit{\Gamma}_0}$. These features reflect, respectively, that the acceleration near $\partial V$ is less significant and that the \copyedit{small scales are relatively well resolved}, especially with respect to the transitional period. The helicity variations in this turbulent regime also eventually slow relative to those observed during transition. Similarly, the impulse remains roughly constant around its value at $t = t^*$. Whereas the $z$ component dominates the impulse magnitude during the radial expansion of the rings, all impulse components have similar magnitudes in this turbulent regime.

The evolution of the integral metrics characterizes the various regimes of the flow and supports the fidelity of our simulation. For example, the helicity is \copyedit{well conserved during the nearly inviscid} evolution of the vortex rings and its subsequent variations are relatively small in magnitude. Further, the variations in the impulse magnitude throughout the simulation remain less than 5\% of the impulse associated with each vortex ring in isolation. These results indicate that the symmetries associated with the handedness and momentum distribution of the flow also remain \copyedit{well preserved} in the appropriate regimes of the simulation.

Our dissipation analysis also suggests that the small-scale flow structures remain reasonably \copyedit{well resolved} throughout the entire simulation. After equilibration, the maximum relative deviation in $Re^{\rm eff}_S$ from $Re_{\mathit{\Gamma}_0}$ is roughly 6.5\% and it occurs around $t \approx t^*$. This relative deviation is similar to that of a previous simulation of a single vortex ring at $Re_{\mathit{\Gamma}_0} = 7500$ using a finite computational grid \citep{Arc2008}. Moreover, it is considerably smaller during the approach and radial expansion of the rings and, to a lesser extent, during turbulent decay. Even during peak dissipation, when the Kolmogorov scale is smallest, the finest grid has acceptable resolution since $\Delta x_{\rm fine} / \eta \approx 3.42$. Altogether, these results suggest that our \rahul{simulation is} \copyedit{well resolved} and support our analysis of the mechanisms underlying transition and turbulent decay.

%%%%%%%%%%%%%%%%%%%%%%%%%%%%%%%%%%%%%%%%%%
\subsection{Evolution of vortical flow structures}\label{sec:res:vortex}
%%%%%%%%%%%%%%%%%%%%%%%%%%%%%%%%%%%%%%%%%%

\begin{figure}
    \centering
    \includegraphics[width=0.5\textwidth,trim=320 0 320 0,clip]{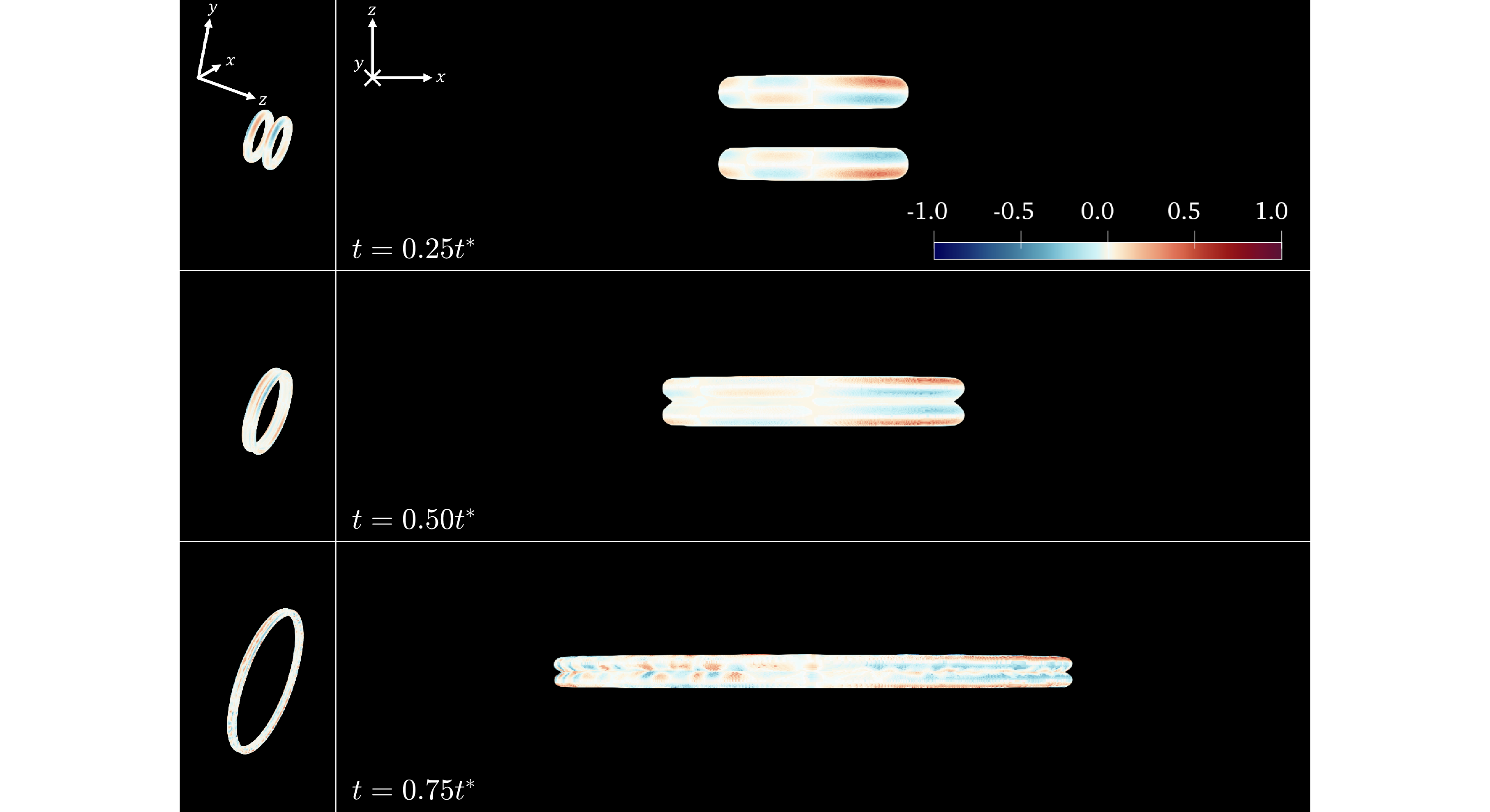}\includegraphics[width=0.5\textwidth,trim=320 0 320 0,clip]{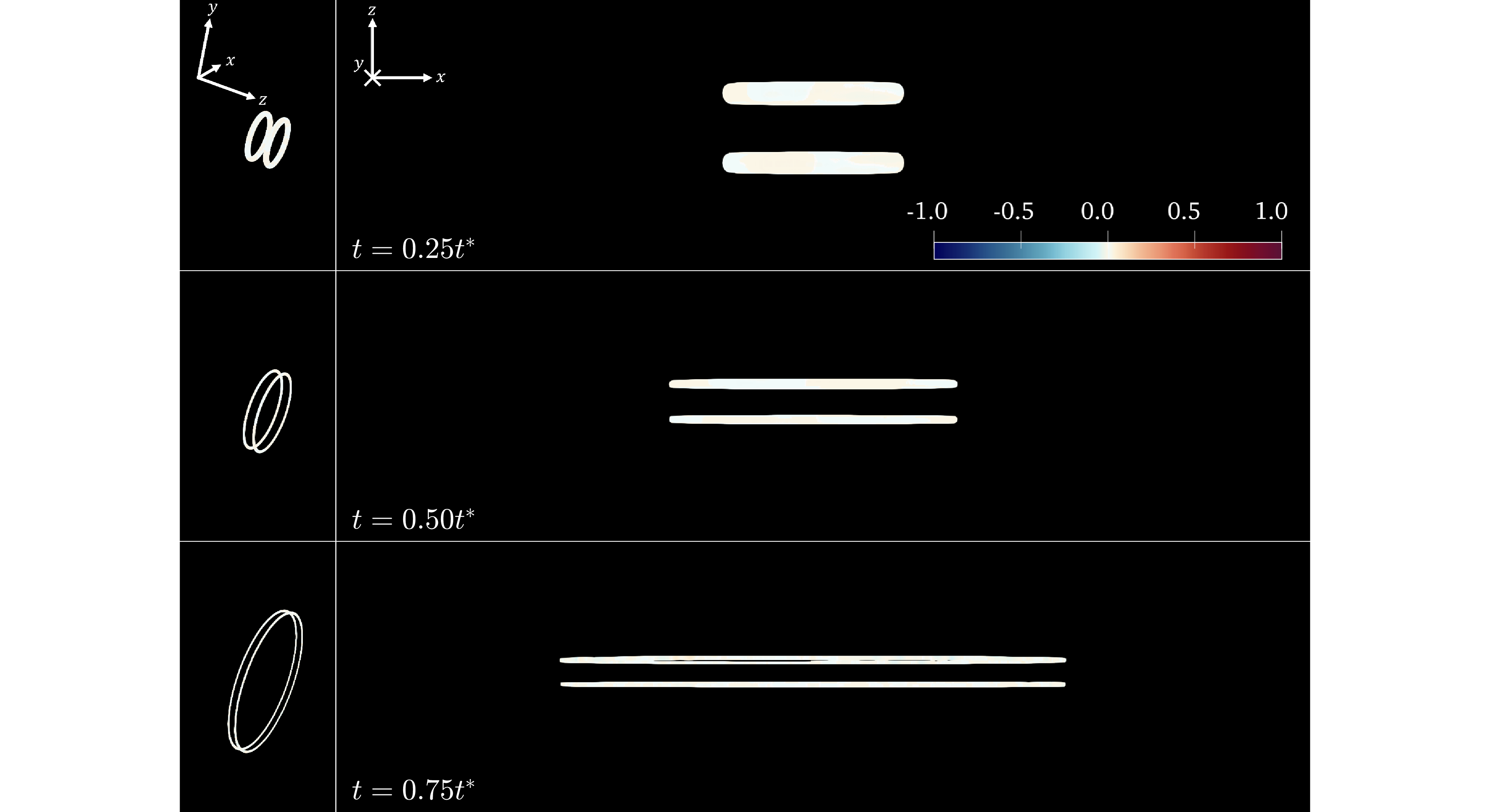}
    \includegraphics[width=0.5\textwidth,trim=320 0 320 0,clip]{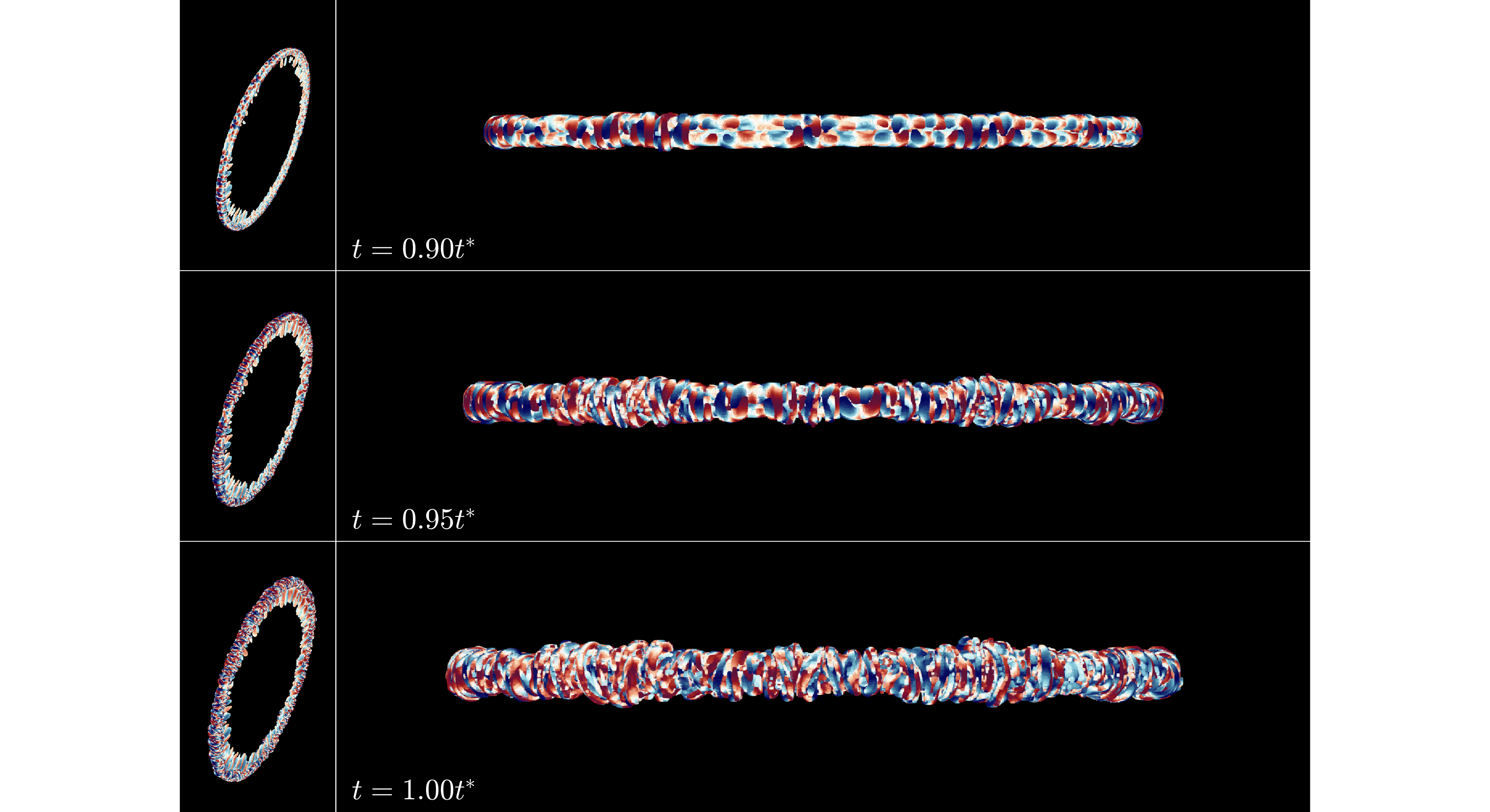}\includegraphics[width=0.5\textwidth,trim=320 0 320 0,clip]{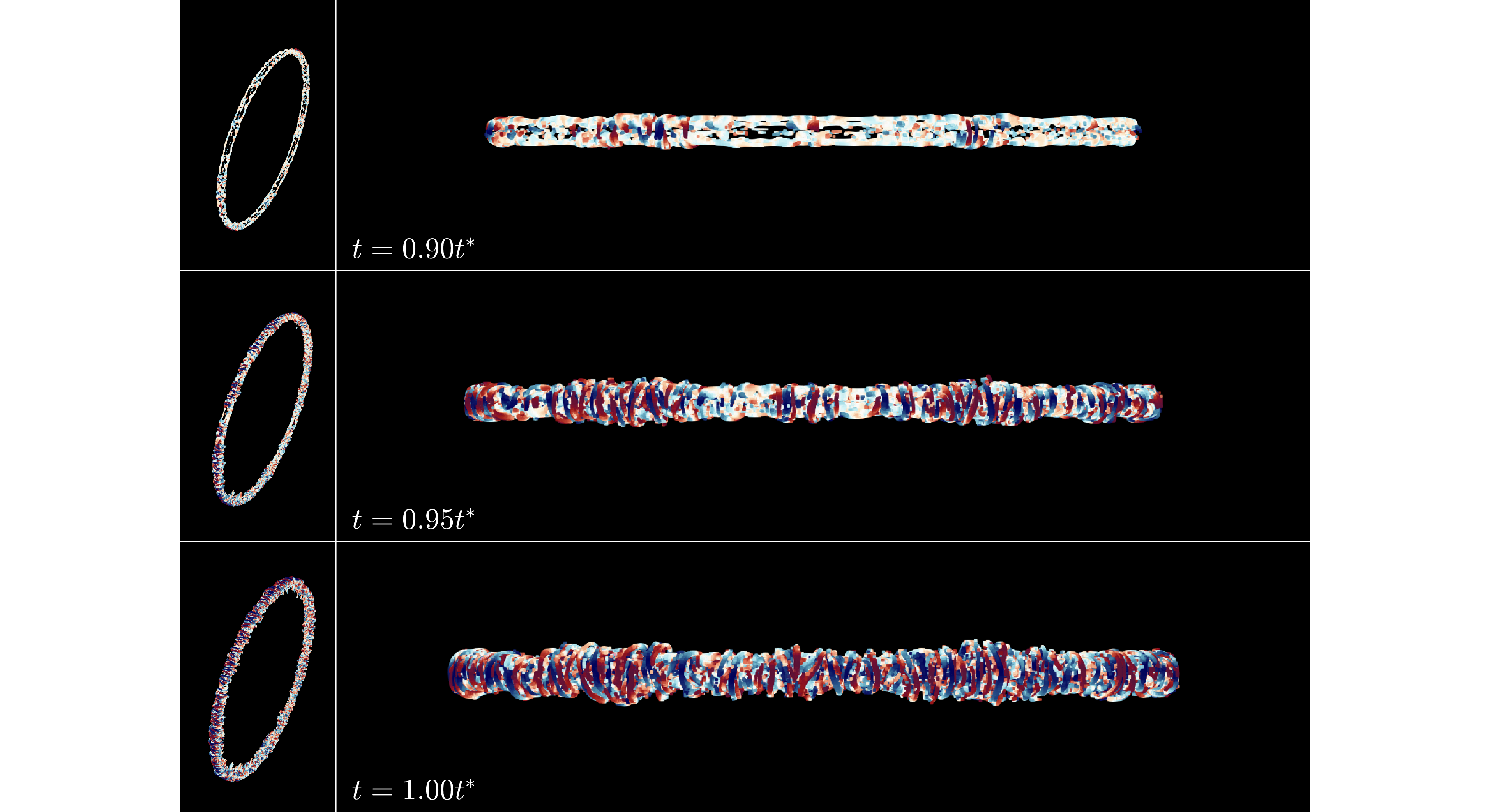}
    \includegraphics[width=0.5\textwidth,trim=320 0 320 0,clip]{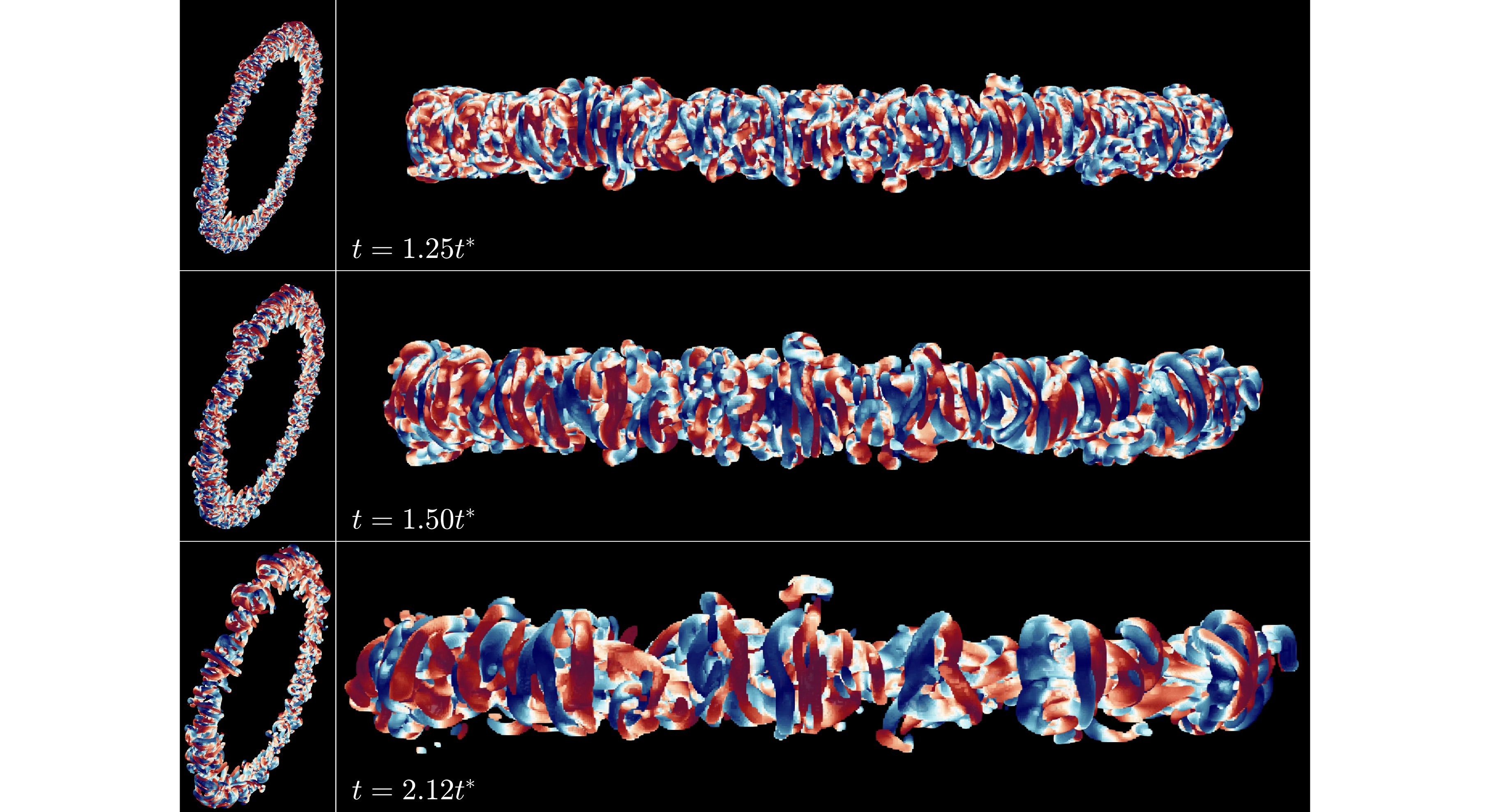}\includegraphics[width=0.5\textwidth,trim=320 0 320 0,clip]{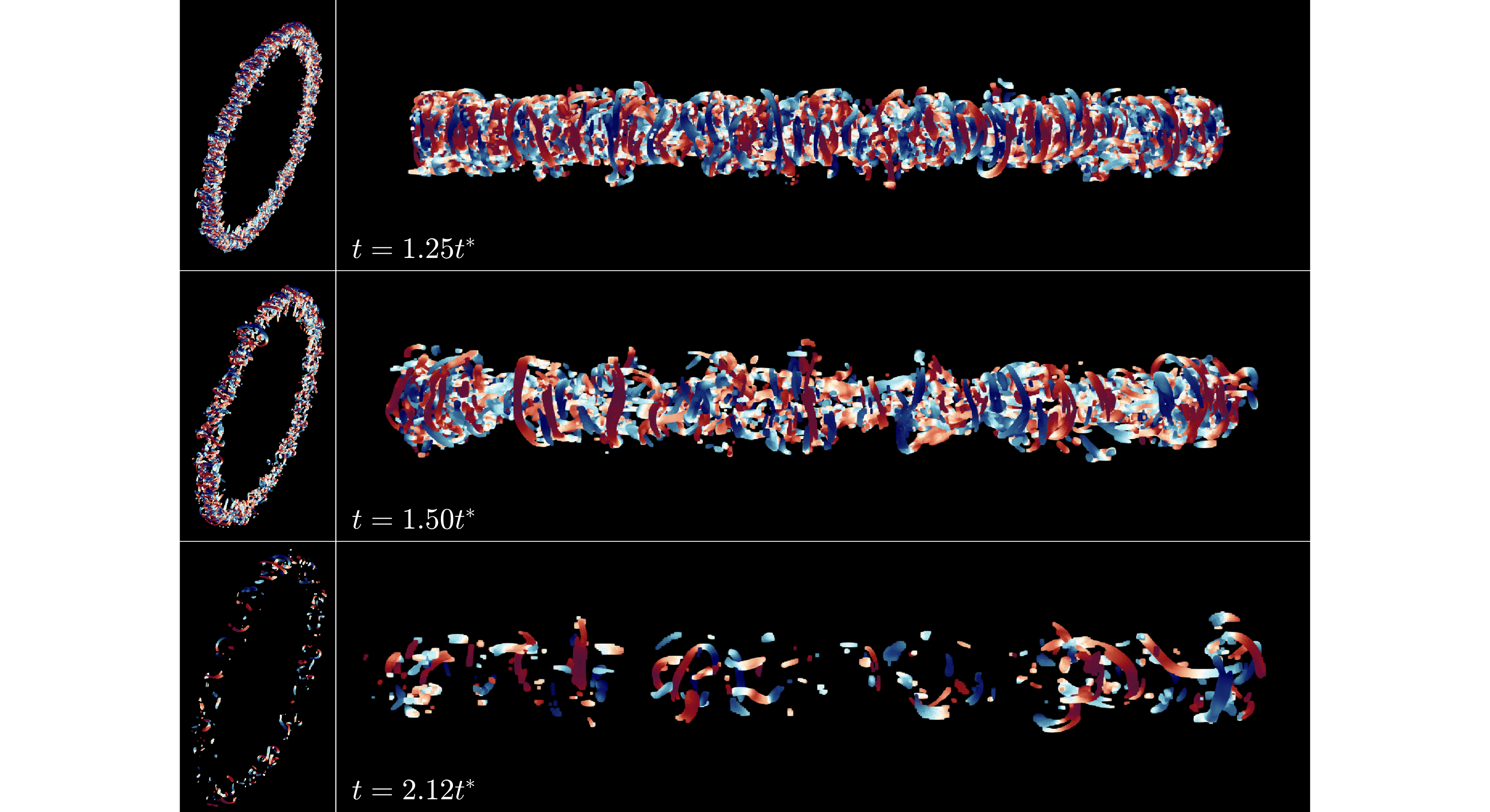}
    \caption{Visualizations of the vortex boundaries ($\Omega_r = 0.52$, left side) and vortex cores ($\Omega_r = 0.93$, right side), \copyedit{coloured} by ${\rm cos} \, \theta^*$, for each reference time from \cref{tab:res:regimes}. A movie depicting the evolution of the vortex boundaries from the auxiliary viewpoint (leftmost column) is provided as supplementary material \copyedit{available at} \href{https://doi.org/10.1017/jfm.2024.90}{https://doi.org/10.1017/jfm.2024.90}.}
    \label{fig:tim_try_lvec2}
\end{figure}

For the present simulation, we identify vortices using the $\Omega_r$ criterion \citep{Liu2019b} with a numerical threshold of $\varepsilon_{\rm vort}$ = 0.04. This criterion provides connections to the triple decomposition of the VGT and the structure of local streamlines. Due to the  well-preserved symmetries of the flow, the global spatial mean of the vorticity tensor is nearly zero and, hence, the $\Omega_r$ criterion is nearly objective \citep{Liu2019c} for the present simulation. We specifically visualize the flow using $\Omega_r = 0.52$ and $\Omega_r = 0.93$ to investigate the structures of the vortex boundaries and the vortex cores, respectively. We \copyedit{colour} these structures using ${\rm cos} \, \theta^*$, where $\theta^*$ is the angle between the $z^*$-axis and the $z$-axis. This \copyedit{colour} scheme enables the identification of antiparallel vortices along the $z$-axis, which play an important role in mediating transition and generating small-scale flow structures in the present vortex ring collision. In \cref{fig:tim_try_lvec2}, we visualize the vortical structures in the flow at each reference time from \cref{tab:res:regimes}.

During the initial evolution of the flow, the equilibrated vortex rings approach the collision plane and expand radially due to their mutual interaction. In this regime, the thinning of the vortex boundaries and cores illustrates the mechanisms driving the shift from a rigid-rotation-dominated regime to a shearing-dominated regime. Further, the visualizations at $t = 0.75 t^*$ depict the emergence of the \copyedit{short-wave} elliptic instability, which is consistent with previous vortex ring collision simulations in similar parameter regimes \citep{McK2020,Mis2021}.

The transitional regime of the flow is marked by the development of secondary vortex filaments and the subsequent generation of small-scale vortical flow structures. At $t = 0.90 t^*$, the visualizations show the development of secondary vortical structures around the circumference of the collision. These structures consist of antiparallel vortex filament pairs that arise in regions where the elliptic instability drives local interactions between the rings. This \copyedit{behaviour} supports the notion that the elliptic instability mediates the initial transition of the rings, leading to the development of secondary vortical structures. The antiparallel secondary filaments become increasingly densely-packed as transition progresses and they mediate the proliferation of small-scale vortical flow structures, e.g. as observed at $t = t^*$.

In \autoref{sec:app:elliptic}, we decompose the flow into azimuthal Fourier modes to characterize the wavenumbers of the perturbations that dominate transition. Our analysis confirms that the \copyedit{short-wave} elliptic instability, with wavelength \copyedit{of} the order of the core radius, mediates the initial stages of transition. We further show that, at $t = t^*$, the most prominent antiparallel vortices occur at the second harmonic of an originally dominant perturbation. Taken together, our visualizations and perturbation analysis are consistent with the initial stages of the iterative elliptic instability pathway, which is driven by subsequent generations of antiparallel vortex filaments \citep{McK2020}.

During the turbulent decay of the flow, the geometric features of the vortex boundaries remain similar at each reference time. However, as energy is dissipated, the smallest-scale vortices are progressively destroyed and the vortical flow structures grow larger in time. The structures of the vortex cores and boundaries reinforce the importance of the interactions between the secondary vortex filaments in mediating the evolution of the turbulent flow. The vortex boundaries also show the formation and ejection of vortex rings from the turbulent cloud resulting from the collision. These ejections, which are a hallmark of the Crow instability, often occur in regions where antiparallel vortex filaments interact and are of similar size to those filaments. This observation provides further evidence of the interplay between the elliptic and Crow instabilities driven by interacting vortex filaments around the turbulent cloud \citep{Mis2021,Ost2021}. In what follows, we develop machinery to probe these mechanisms in the context of features of the velocity gradients, with a particular emphasis on characterizing the action of the elliptic instability among other mechanisms.

%%%%%%%%%%%%%%%%%%%%%%%%%%%%%%%%%%%%%%%%%%
\section{Partitioning of velocity gradients}\label{sec:res:partition}
%%%%%%%%%%%%%%%%%%%%%%%%%%%%%%%%%%%%%%%%%%

\begin{figure}
    \centering
    \includegraphics[width=0.631\textwidth,trim=0 120 0 140,clip]{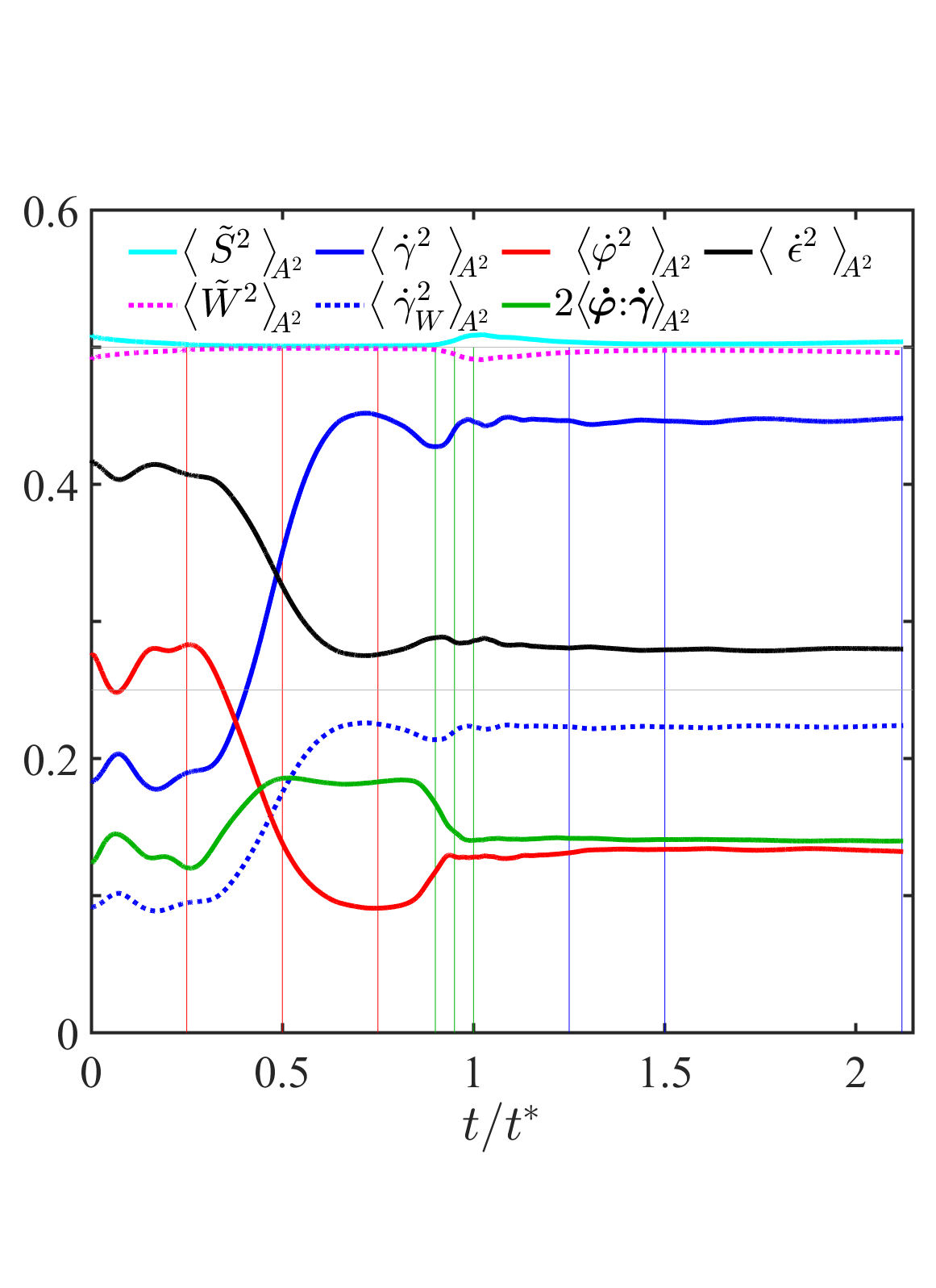}
    \caption{Relative contributions of the constituents of the \copyedit{Cauchy--Stokes} decomposition and the triple decomposition to $A^2$. The vertical lines correspond to the reference times in \cref{tab:res:regimes} and they are \copyedit{coloured} accordingly.}
    \label{fig:phase_space_v1_fig1}
\end{figure}

Here, we investigate the partitioning of the velocity gradients to characterize the evolution of the flow in the context of the fundamental modes of deformation. We first consider volumetric weighted averages of the relative contributions of various constituents of \rahul{$\mathsfbi{\tilde{A}}$} to the strength of the velocity gradients. These averages may be expressed as
\begin{equation}
    \Aavg{\xi} = \frac{\int_{V(t)} A^2 \xi dV}{\int_{V(t)} A^2 \;\, dV},
\end{equation}
where $\xi \in \{\tilde{S}^2, \tilde{W}^2\}$ for the \copyedit{Cauchy--Stokes} decomposition and $\xi \in \{ \esq, \gsq, \psq, 2 \src \}$ for the triple decomposition. Here, we have used that the Frobenius inner product, denoted by \rahul{$:$}, of a symmetric tensor with an antisymmetric tensor is zero and that $\gW = \gS = \gsq/2$. The \copyedit{shear--rotation} correlation term, $2 \src = 2 \dot{\varphi}_{z^*} \dot{\gamma}_{z^*} > 0$, reflects the presence of shearing in the plane of rigid rotation. All of the relative contributions discussed are unitarily invariant and, thus, they apply to both the principal coordinates and the global coordinates. \autoref{fig:phase_space_v1_fig1} shows how these relative contributions evolve during the simulation for both decompositions of \rahul{$\mathsfbi{\tilde{A}}$}. 

Consistent with the equivalence of $\Phi_S$ and $\Phi_W$ for incompressible flows on unbounded domains \citep{Ser1959}, $\Aavg{\tilde{S}^2} \approx \Aavg{\tilde{W}^2} \approx 0.50$ for the present simulations. The largest deviations from this balance occur during equilibration ($t \lesssim 0.25 t^*$) and around the time of peak dissipation ($t \approx t^*$). These deviations are consistent with the \copyedit{behaviour} of the effective Reynolds numbers in \cref{fig:metrics_v2} and their smallness further validates the ability of the finite computational grid to approximate a formally unbounded flow. However, since $\Aavg{\tilde{S}^2}$ and $\Aavg{\tilde{W}^2}$ remain relatively constant throughout the simulation, they provide limited information about the nature of the velocity gradients as the flow progresses through its initial, \copyedit{transitional} and turbulent regimes.

Compared \copyedit{with} the constituents of the \copyedit{Cauchy--Stokes} decomposition, the constituents of the triple decomposition show more pronounced variations associated with the different regimes of evolution. For the initial Gaussian vorticity profiles, the contribution of rigid rotation to the enstrophy dominates the contribution of antisymmetric shearing. During equilibration ($t \lesssim 0.25 t^*$), the fluctuations in all contributions of the triple decomposition constituents reflect the redistribution of velocity gradients in the cores of the vortex rings. As the equilibrated rings approach the collision plane and spread ($0.25 t^* \lesssim t \lesssim 0.75 t^*$), $\Aavg{\esq}$ and $\Aavg{\psq}$ decrease and $\Aavg{\gsq}$ and $2\Aavg{\src}$ increase. As the elliptic instability emerges, these contributions level off in a regime where antisymmetric shearing dominates rigid rotation and \copyedit{shear--rotation} correlations are enhanced. The subsequent development of the elliptic instability ($0.75 t^* \lesssim t \lesssim 0.90t^*$) is marked by slight rebounds in the contributions of $\Aavg{\psq}$ and $\Aavg{\gsq}$. These rebounds are associated with the emergence of secondary vortex filaments and, hence, the nonlinear evolution of the elliptic instability. The transition to turbulence ($0.90 t^* \lesssim t \lesssim t^*$), which is associated with the generation of small scales and enhanced dissipation, is marked by a decrease in the contribution of $2\Aavg{\src}$.

Remarkably, even though the flow is not stationary during turbulent decay, the relative contributions of the constituents of the triple decomposition to the strength of the velocity gradients remain roughly constant after transition. Further, as summarized in \cref{tab:res:eqpart}, these `equilibrium' relative contributions are similar to those computed by \citet{Das2020} for forced isotropic turbulence at high Taylor-scale Reynolds numbers. This agreement suggests that the velocity gradient partitioning may encode a relatively common balance in unbounded, incompressible turbulence with appropriate symmetries. In this balance, shearing \copyedit{makes} the largest contribution to the velocity gradients and rigid rotation \copyedit{makes} the smallest contribution.

\begin{table}
    \centering
    \renewcommand*{\arraystretch}{1.2}
    \begin{tabular}{|c|c|c|c|c|}
    \hline
        Constituent & $\Aavg{\esq}$ & $\Aavg{\gsq}$ & $\Aavg{\psq}$ & $2 \Aavg{\src}$ \\
        \hline
        Present vortex ring collision & $\;$28.0$\,$\% & $\;$44.6$\,$\% & $\;$13.3$\,$\% & $\;$14.1$\,$\% \\
        \hline
        Forced isotropic turbulence \citep{Das2020} & $\;$24$\;\;\;$\% & $\;$52$\;\;\;$\% & $\;$11$\;\;\;$\% & $\;$13$\;\;\;$\% \\
        \hline
    \end{tabular}
    \caption{Comparison of the equilibrium partitioning of the velocity gradients for the present vortex ring collision with the partitioning computed for forced isotropic turbulence \citep{Das2020}. Here, the equilibrium partitioning is computed as the mean over the turbulent decay regime ($t \gtrsim t^*$) and it is insensitive to the length of the averaging interval.}
    \label{tab:res:eqpart}
\end{table}

Beyond the strength of velocity gradients, it is also useful to examine the interplay between the modes of deformation in the context of vortical flow structures. Here, we introduce a new phase space defined by the relative contributions of $\Aavg{\gW}$, $\Aavg{\psq}$ \copyedit{and} (implicitly) $2\Aavg{\src}$ to $\Aavg{\tilde{W}^2}$. The upper bound of this phase space is found by maximizing $\Aavg{\psq} / \Aavg{\tilde{W}^2}$, which occurs when $2 \Aavg{\src} / \Aavg{\tilde{W}^2} = 0$. The bottom boundary is found by minimizing $\Aavg{\psq} / \Aavg{\tilde{W}^2}$, which occurs when \rahul{$2 \Aavg{\src} / \Aavg{\tilde{W}^2} = 2 \sqrt{\Aavg{\psq} \Aavg{\gW}} / \Aavg{\tilde{W}^2}$}. Correspondingly, these boundaries may be expressed as
\begin{alignat}{8}
    \Bigg( 1 &- &&\sqrt{\frac{\Aavg{\gW}}{\Aavg{\tilde{W}^2}} } &&\Bigg)^2 &&\leq &&\;\,\frac{\Aavg{\psq}}{\Aavg{\tilde{W}^2}}&& &&\leq 1 &&- \frac{\Aavg{\gW}}{\Aavg{\tilde{W}^2}}, \label{eq:shear_rot_bounds_1}\\
    & && && &&\Updownarrow &&\;\;\;\; && &&\Updownarrow && \nonumber \\
    -2 \Bigg( \frac{\Aavg{\gW}}{\Aavg{\tilde{W}^2}} &- &&\sqrt{\frac{\Aavg{\gW}}{\Aavg{\tilde{W}^2}}} &&\Bigg) &&\geq &&\frac{2 \Aavg{\src}}{\Aavg{\tilde{W}^2}}&& &&\geq &&\;0, \label{eq:shear_rot_bounds_2}
\end{alignat}
highlighting that lower and upper bounds of the rigid rotation contribution correspond to the upper and lower bounds, respectively, of the \copyedit{shear--rotation} correlation contribution. The maximum value of $2 \Aavg{\src}/\Aavg{\tilde{W}^2}$ varies along the lower boundary of this phase space to ensure that the relative contributions sum to unity. The global maximum occurs when $2 \src / \tilde{W}^2 = 0.5$ or, equivalently, $\psq / \tilde{W}^2 = \gW / \tilde{W}^2 = 0.25$ at all points, which corresponds to the (pointwise) maximum of $2 \src = ( \sqrt{2} + 1 )^{-1}$ reported by \citet{Das2020}. This maximum corresponds to local streamlines in the principal frame for which shearing occurs exclusively in the plane of rigid rotation.

\begin{figure}
    \centering
    \includegraphics[width=0.83\textwidth,trim=0 50 0 50,clip]{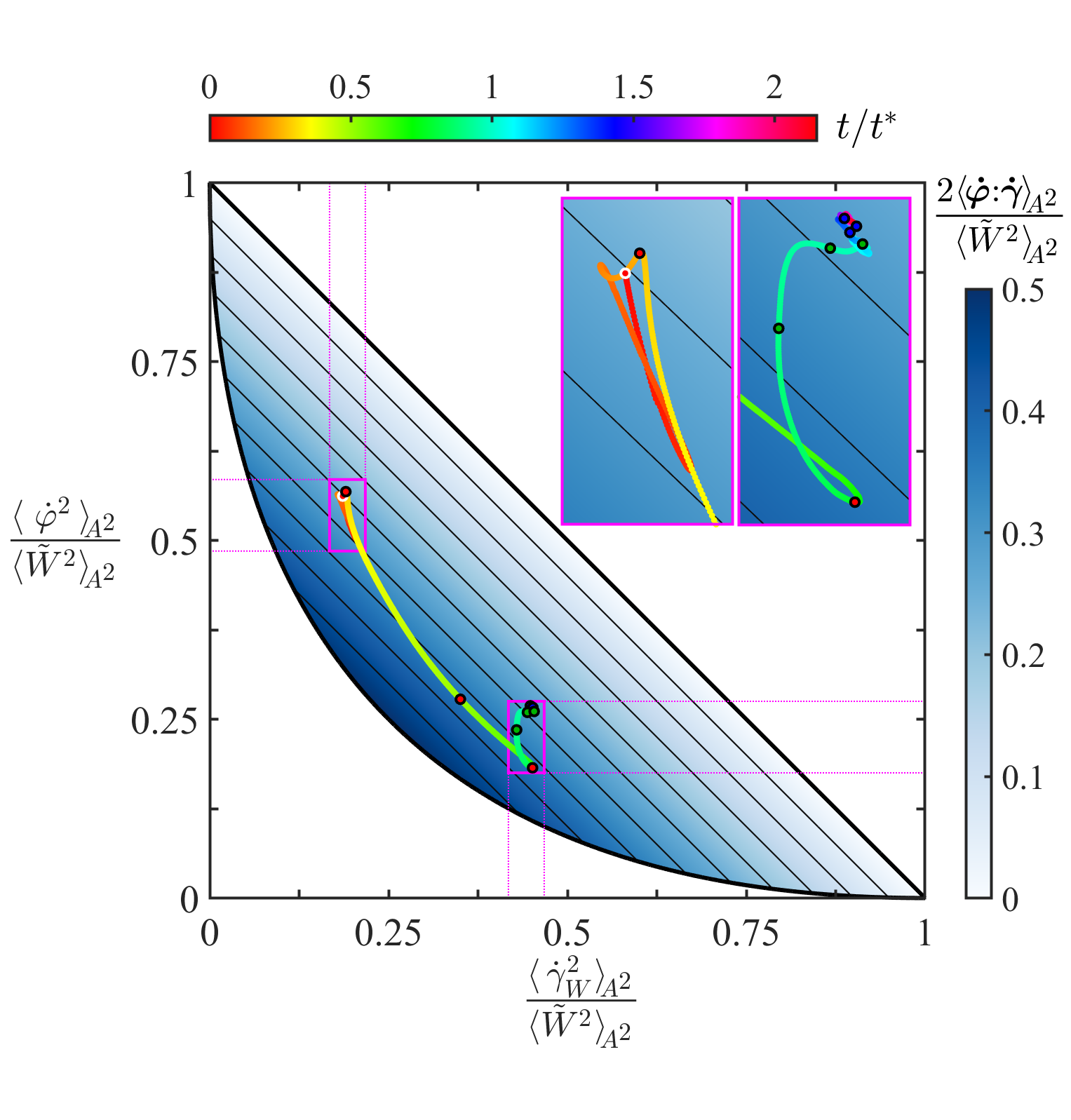}
    \caption{\copyedit{Shear--rotation} phase space trajectory of the flow, highlighting the evolution during equilibration (left inset) and transition and turbulent decay (right inset). The contours represent increments of 0.05 in the \copyedit{colour} scale. The white circle marks the initial condition and the fill \copyedit{colours} of the black circles correspond to the reference times they represent from \cref{tab:res:regimes}.}
    \label{fig:phase_space_v1_fig2}
\end{figure}

\autoref{fig:phase_space_v1_fig2} depicts the trajectory of the flow in this \copyedit{shear--rotation} phase space and elucidates how the relationships between the constituents of enstrophy associated with the triple decomposition evolve in time. Following rapid variations during equilibration, the trajectory returns to a position in phase space similar to that of the initial condition at $t \approx 0.25 t^*$. The trajectory then undergoes a shift across the phase space as the equilibrated rings approach the collision plane and spread radially ($0.25 t^* \lesssim t \lesssim 0.75 t^*$). This shift from a rigid-rotation-dominated regime to a shearing-dominated regime is associated with an enhanced contribution from $2\Aavg{\src}/\Aavg{\tilde{W}^2}$. The development of the elliptic instability ($0.75 t^* \lesssim t \lesssim 0.90 t^*$) and transition ($0.90 t^* \lesssim t \lesssim t^*$) are associated with a shift in the direction of the trajectory towards smaller contributions of $2\Aavg{\src}/\Aavg{\tilde{W}^2}$. During turbulent decay ($t \gtrsim t^*$), the trajectory remains roughly fixed in the phase space and very close to its position at $t = t^*$.

Considering this phase space trajectory in the context of the dissipation (see \cref{fig:metrics_v2}) reveals that, while the initial growth in dissipation is associated with enhanced \copyedit{shear--rotation} correlations, its subsequent enhancement during transition is associated with a reduction in \copyedit{shear--rotation} correlations. In \autoref{sec:app:src_vfs}, we reexamine the visualizations from \cref{fig:tim_try_lvec2} in terms of the \copyedit{shear--rotation} correlations in the flow to highlight their relationship to the vortical flow structures. In \textsection \ref{sec:res:phase}, we interpret the effects of these \copyedit{shear--rotation} correlations using a new, related phase space, based on local streamline geometry, to characterize the elliptic instability and other mechanisms.

%%%%%%%%%%%%%%%%%%%%%%%%%%%%%%%%%%%%%%%%%
\section{Statistical geometry of local streamlines}\label{sec:res:phase}
%%%%%%%%%%%%%%%%%%%%%%%%%%%%%%%%%%%%%%%%%

%%%%%%%%%%%%%%%%%%%%%%%%%%%%%%%%%%%%%%%%%
\subsection{Phase space transformations}\label{sec:res:phase_sub1}
%%%%%%%%%%%%%%%%%%%%%%%%%%%%%%%%%%%%%%%%%

The elliptic instability is associated with the resonance of the vortical flow with the underlying strain field and acts to break up elliptic streamlines. Consistent with this picture, we introduce a new geometry-based phase space that captures local flow features that (i) are conducive to the elliptic instability and (ii) characterize its action.

To address (i), we consider the angle, $\theta_\omega$, between the vorticity vectors associated with antisymmetric shearing and rigid rotation, which is given by
\begin{equation}\label{eq:newphase_forward_p1}
 \theta_\omega = {\rm cos}^{-1}\left( \sqrt{ \frac{\left(\src\right)^2}{\psq \gW} } \right).
\end{equation}
Our focus on \copyedit{shear} straining is consistent with the classical models of strained vortices used to characterize the elliptic instability \citep{Ker2002}. Since decreasing $\theta_\omega$ corresponds to increasing the alignment between shearing and rigid rotation, it can be associated with conditions conducive to the elliptic instability.

To address (ii), we consider the aspect ratio, $\zeta$, of the elliptic component of rotational local streamlines in the plane of rigid rotation, which is given by
\begin{equation}\label{eq:newphase_forward_p2}
\zeta = \sqrt{1-e^2} = \sqrt{\frac{\psq}{\psq + 2 \src}},
\end{equation}
where $e$ represents the eccentricity of an ellipse with aspect ratio $\zeta$. This aspect ratio characterizes the scale-invariant geometry of the local streamlines in the plane of rigid rotation. As such, it can be used alongside $\theta_\omega$ to characterize the action of the elliptic instability by identifying how alignment between shearing and rigid rotation affects local streamline geometry.

The present \copyedit{shear--rotation} and geometry-based phase spaces can be understood through nonlinear transformations of the $\qA-\,\rA$ phase space. The transformations we derive express the relative contributions of the triple decomposition to the velocity gradients using $\qA$ and $\rA$, and they represent the inverse transformations to those presented by \citet{Das2020}. However, an additional parameter, $\theta_\omega$, is generally required to evaluate our transformations. This extra parameter demonstrates that the invariants of \rahul{$\mathsfbi{\tilde{A}}$} alone cannot generally characterize the relative contributions of the constituents of the triple decomposition to the velocity gradients.

For rotational local streamlines, the transformations are given by
\begin{equation}\label{eq:newphase_inverse_rot}
\begin{split}
   \esq &= \frac{\left( -3 \times 2^{2/3} \qA + 3^{2/3} \left( 2 \sqrt{3 \Delta} + 9 \lvert \rA \rvert \right)^{2/3} \right)^2}{6^{5/3} \left( 2 \sqrt{3 \Delta} + 9 \lvert \rA \rvert \right)^{2/3}}, \quad\;\; \gsq = 2 \gW = 1 - 2 \esq - 2 \qA \;\;, \\
   \psq &= \left( -\sqrt{\gW} {\rm cos} \theta_{\rm \omega} + \sqrt{1 - \esq - \gW \left( 2 - {\rm cos}^2 \theta_{\omega} \right)}  \right)^2, \quad 2 \src = 2 \sqrt{\psq \gW} {\rm cos} \theta_{\omega},
\end{split}
\end{equation}
where $\Delta = \qA{\mkern-6mu}^3 + \frac{27}{4} \rA{\mkern-6mu}^2$ is proportional to the discriminant of the characteristic equation of \rahul{$\mathsfbi{\tilde{A}}$}. In this case, $\esq$ and $\gsq$ can be determined directly from $\qA$ and $\rA$ and $\psq$, $2 \src$ \copyedit{and} $\zeta$ can be determined if $\theta_\omega$ is known. For non-rotational local streamlines, the transformations are given by
\begin{equation}\label{eq:newphase_transform_nonrot}
    \esq = -2 \qA, \quad \gsq  = 1- \esq, \quad \psq = 0,
\end{equation}
which can be determined directly from a single parameter, $\qA$. The rotational and non-rotational transformations are continuous with one another at their boundary ($\Delta = 0$) and they are both symmetric about the $\qA$-axis. However, the aspect ratio is only \copyedit{well defined} for $\Delta > 0$, consistent with our focus on rotational local streamlines.

\begin{figure}
    \centering
    \includegraphics[width=\textwidth,trim=80 90 60 30,clip]{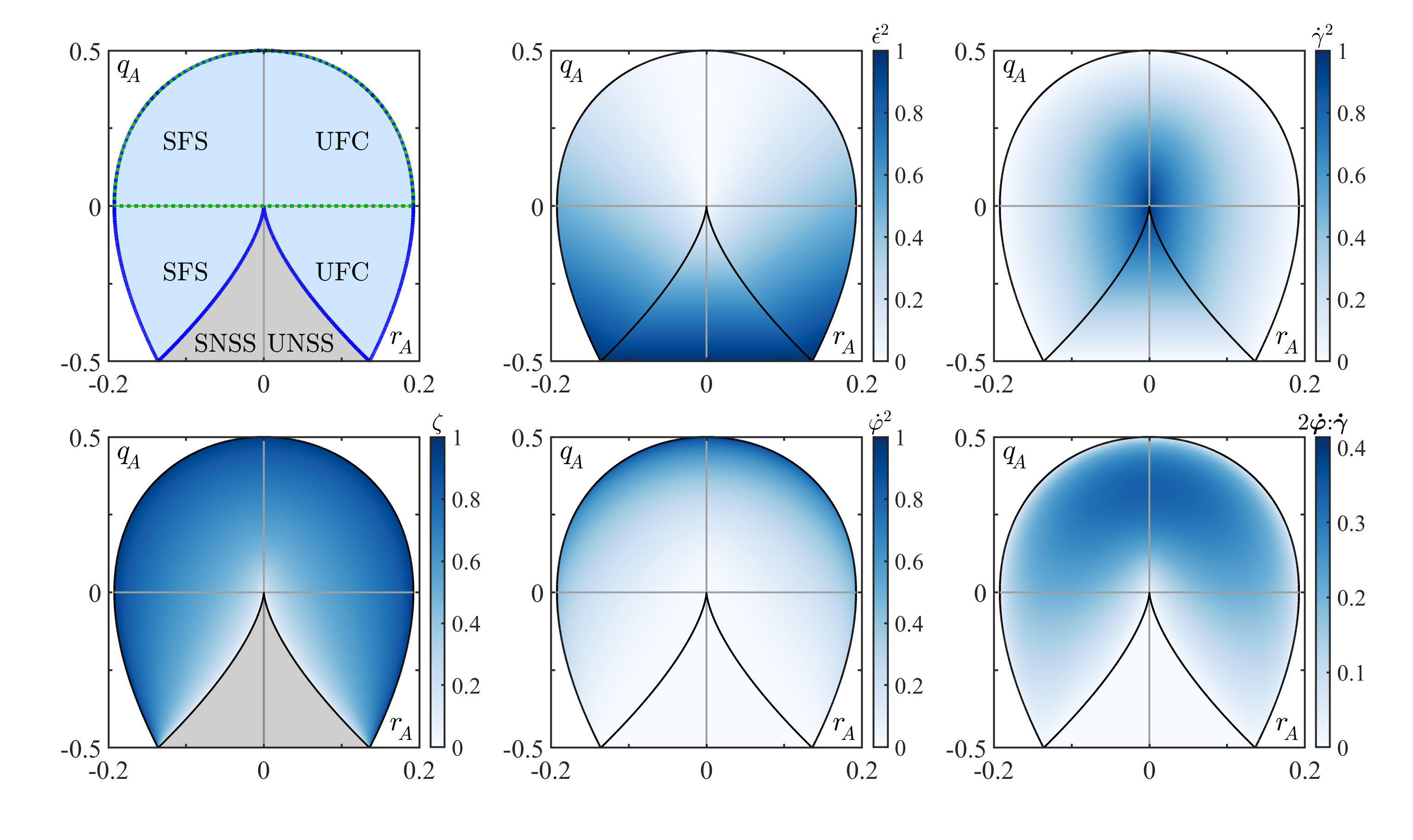}
    \caption{The $\qA -\, \rA$ phase space (top left) and the variations in $\esq$, $\gsq$, $\psq$, $2\src$ \copyedit{and} $\zeta$ in this space when $\theta_\omega = 43.57^\circ$ is held constant. The top left panel shows the boundaries corresponding to the symmetry-based (dotted green) and the geometry-based (solid blue) vortex criteria (see \textsection \ref{sec:intro:vortex_flow}). It also labels the four classes of non-degenerate local streamline topologies: \copyedit{stable-focus--stretching} (SFS), \copyedit{unstable-focus--compression} (UFC), \copyedit{stable-node--saddle-saddle} (SNSS) \copyedit{and} \copyedit{unstable-node--saddle-saddle} (UNSS).}
    \label{fig:qr_and_nrh_v1}
\end{figure}

In \cref{fig:qr_and_nrh_v1}, we illustrate how $\esq$, $\gsq$, $\psq$, $2\src$ \copyedit{and} $\zeta$ vary within the $\qA -\, \rA$ phase space. In this phase space, the rotational geometries are externally bounded by $3\sqrt{3}\lvert \rA \rvert = ( 1 + \qA ) ( 1 - 2 \qA )^{1/2}$ and the non-rotational geometries are externally bounded by $\qA = -\frac{1}{2}$. For the rotational geometries, we display $\psq$, $2 \src$ \copyedit{and} $\zeta$ for $\theta_\omega = 43.57^\circ$, which corresponds to the mean value at $t = 1.25 t^*$ (see \cref{fig:eccentricity2}) and approximates the equilibrium value during turbulent decay. We document how each of these quantities varies with $\theta_\omega$ in the $\qA -\, \rA$ phase space in \autoref{sec:app:shear_rot}.

\autoref{fig:qr_and_nrh_v1} shows that, generically, the contribution of pure shearing ($\gsq$) tends to dominate the velocity gradients near the origin of the $\qA -\, \rA$ plane. The contribution of normal straining ($\esq$) grows large when the velocity gradients are dominated by the strain rate tensor. By contrast, for rotational geometries, the contribution of rigid rotation ($\psq$) grows large near the external boundary in regions where the vorticity tensor dominates. The contribution of \copyedit{shear--rotation} correlations ($2\src$) grows largest in the intermediate region of the phase space and it decays near the discriminant line and the external boundary. The aspect ratio $\zeta$ is unity along the external boundary and it decays to zero at the discriminant line.

The elliptic instability, which is relevant in strained vortical flows, is expected to be most active for an intermediate range of aspect ratios. Further, vortex stretching and squeezing, which are known to play important roles in turbulent flows, can most readily be associated with the SFS and UFC streamline topologies, respectively \citep{Lop1993}. In the corresponding regions in the $\qA -\, \rA$ plane, the transformations in \cref{fig:qr_and_nrh_v1} suggest that the interplay between shearing and rotation is pertinent to these fundamental turbulent processes. Within this context, the transformations have the potential to characterize the role of the elliptic instability, among other mechanisms, in mediating such processes. In what follows, we investigate this premise by \copyedit{analysing} the evolution of the velocity gradient distributions in the $\qA -\, \rA$, \copyedit{shear--rotation} and $\zeta - \theta_\omega$ phase spaces.

%%%%%%%%%%%%%%%%%%%%%%%%%%%%%%%%%%%%%%%%%
\subsection{Phase space distributions}\label{sec:res:phase_sub2}
%%%%%%%%%%%%%%%%%%%%%%%%%%%%%%%%%%%%%%%%%

The joint probability density functions (\copyedit{p.d.f.s}) of the normalized velocity gradients in the phase spaces we investigate encode information about the local streamline geometries. Although these phase spaces are related to one another, the choice of phase space plays an important role in interpreting the statistical distributions of the flow. In the $\qA -\, \rA$ phase space, incompressible turbulent flows often follow a near-universal teardrop-like distribution about the origin \citep{Das2019,Das2020a}. The \copyedit{shear--rotation} phase space highlights the distribution of rotational streamline geometries and it characterizes the interplay of rigid rotation and antisymmetric shearing. The $\zeta - \theta_\omega$ phase space also considers rotational streamline geometries and it characterizes the flow in terms of geometric features of local streamlines that are associated with the elliptic instability. 

\begin{figure}
    \centering
    \includegraphics[width=0.88\textwidth,trim=0 30 70 10,clip]{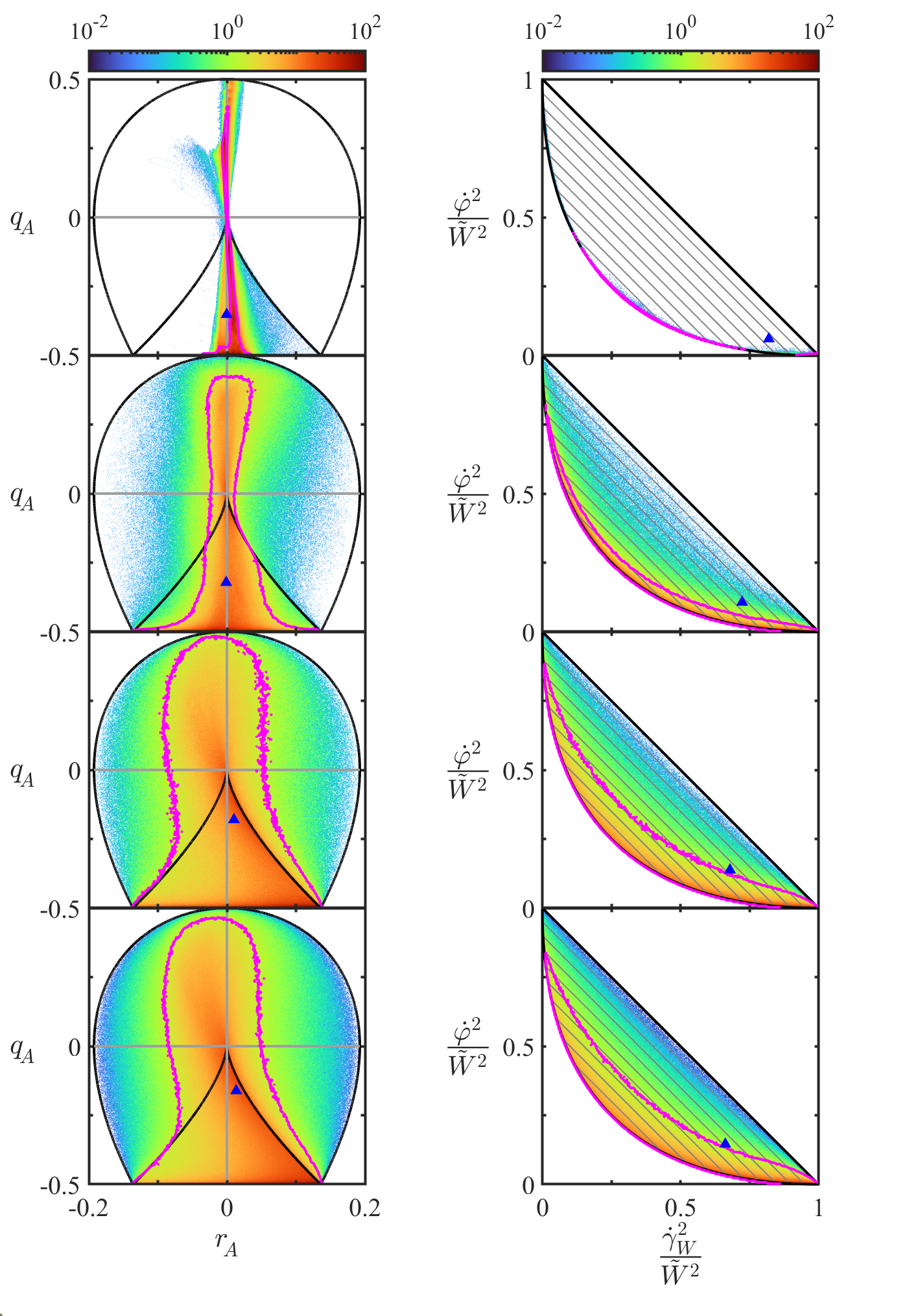}
    \caption{Joint \copyedit{p.d.f.s} of the velocity gradients satisfying $A^2/{\rm max}(A^2) \geq 0.1\%$ in the $\qA -\, \rA$ phase space (left) and $W^2/{\rm max}(W^2) \geq 0.1\%$ in the \copyedit{shear--rotation} phase space (right) at times $t = 0.75 t^*$, $0.90 t^*$, $1.00 t^*$ \copyedit{and} $1.25 t^*$ (from top to bottom). The blue triangles represent the centroids of the distributions and the magenta contours represent the \copyedit{p.d.f.} levels for which 90\% of the flow (by volume) resides at higher \copyedit{p.d.f.} levels. These contours are smoothed by using coarser \copyedit{p.d.f.} bins to ensure that they roughly enclose the regions with higher \copyedit{p.d.f.} levels.}
    \label{fig:main_scatter}
\end{figure}

\autoref{fig:main_scatter} shows the $\qA -\, \rA$ and the \copyedit{shear--rotation} phase space distributions at reference times, selected from \cref{tab:res:regimes}, that pertain to the development of the elliptic instability, \copyedit{transition} and turbulent decay. As the elliptic instability emerges ($t \approx 0.75 t^*$), the velocity gradients are concentrated near the $\qA$-axis. Since $2\Aavg{\src}$ is relatively large around this time, the rotational regions of the flow are concentrated near the bottom boundary of the \copyedit{shear--rotation} phase space, particularly near the location where $2 \Aavg{\src}$ is maximized.

As the flow transitions ($t \approx 0.90t^*$), the distribution remains \copyedit{centred} about the $\qA$-axis but it expands towards larger $\lvert \rA \rvert$ as it begins to fill the $\qA -\, \rA$ phase space. Similarly, the \copyedit{p.d.f. in the shear--rotation} phase space begins extending away from the bottom boundary of the phase space, although its bulk remains concentrated at the boundary. The broadening \copyedit{p.d.f.s} represent the generation of more diverse local streamline topologies (see \cref{fig:qr_and_nrh_v1}), which is consistent with the formation of more complex vortical structures (see \cref{fig:tim_try_lvec2}).

At the time of peak dissipation ($t \approx t^*$), the distributions populate nearly all of the area in both the phase spaces. During the subsequent turbulent decay (as shown for $t \approx 1.25 t^*$), the flow approaches its equilibrium distributions in both phase spaces, which remain similar to the distributions for $t \approx t^*$. In the $\qA -\, \rA$ phase space, the equilibrium \copyedit{p.d.f.} above the $\rA$-axis is concentrated slightly left of the $\qA$-axis. This slight preference of the SFS topology is consistent with the typical presence of positive vortex stretching in regions of turbulent flows with rotational geometries. Below the $\rA$-axis, the \copyedit{p.d.f.} is concentrated along the discriminant line for $\rA > 0$. The equilibrium distribution of our vortex ring collision in this phase space is similar to the near-universal teardrop-like shapes reported previously for forced isotropic turbulence \citep{Das2019,Das2020a}. This similarity suggests that, in addition to the velocity gradient partitioning (see \cref{tab:res:eqpart}), the teardrop-like distribution may be more broadly applicable to incompressible flows with appropriate symmetries.

The \copyedit{p.d.f.s in the shear--rotation} phase space evolve similarly to those in the $\qA -\, \rA$ phase space (e.g. by broadening) and specifically highlight rotational geometries. However, the difference between vortex stretching and squeezing, which is encoded in the sign of the real eigenvalue of the VGT, cannot be distinguished in this phase space since its constituents are non-negative. The high concentration of the \copyedit{p.d.f.s} near the lower corner of this phase space, representing the origin of the $\qA -\, \rA$ phase space, shifts the centroids accordingly and highlights the importance of shearing in the generation and evolution of turbulent flows. The equilibrium distribution remains relatively concentrated about the lower boundary of this phase space, including regions where 2$\Aavg{\src}$ is relatively large. This \copyedit{behaviour} highlights the potential for the elliptic instability to be active during turbulent decay, but the breadth of the distribution suggests it may not be a completely dominant mechanism driving turbulent flow in rotational regions.

\begin{figure}
    \centering
    \includegraphics[width=\textwidth,trim=0 70 30 30, clip]{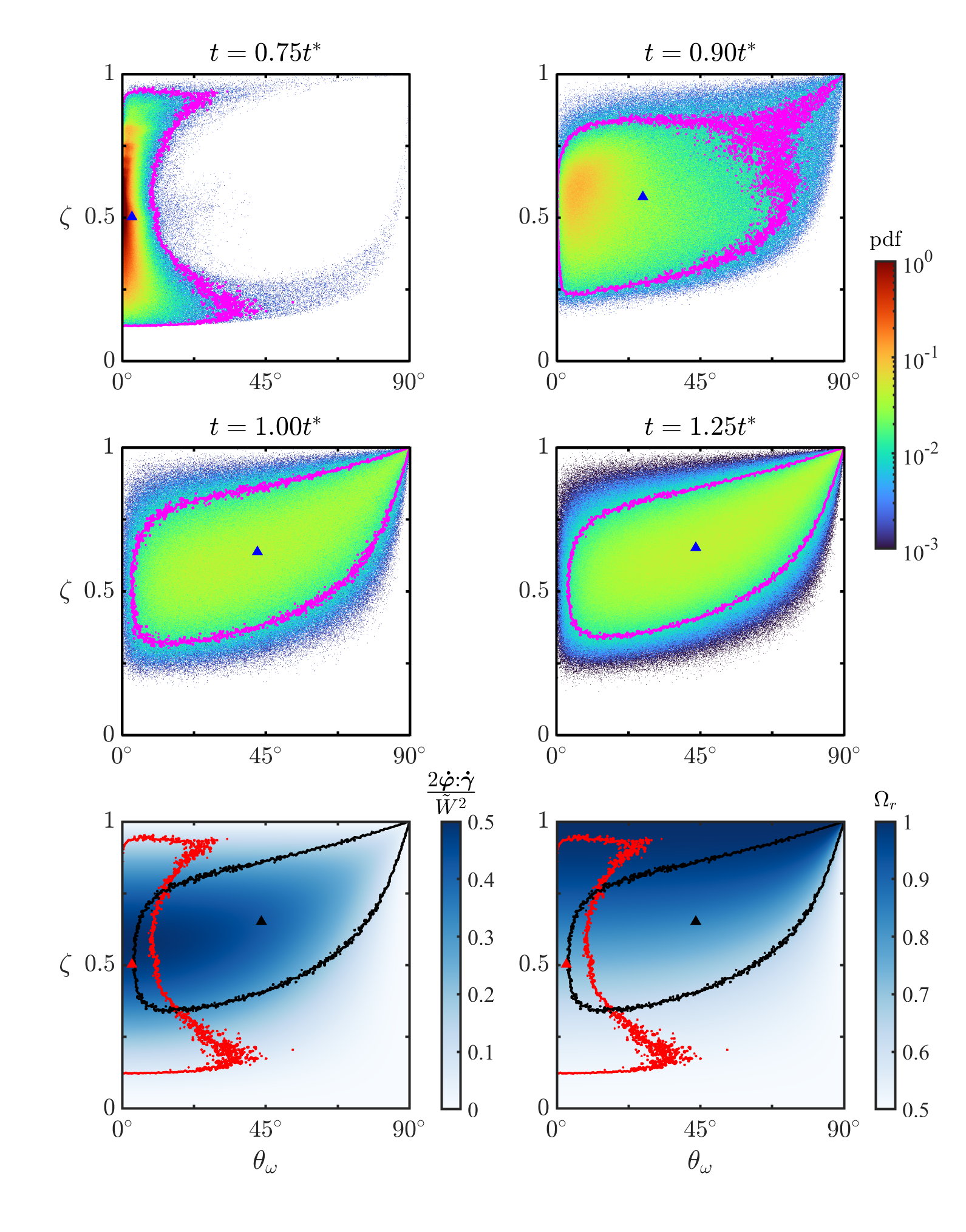}
    \caption{Joint \copyedit{p.d.f.s} (top and middle) of the velocity gradients in the $\zeta - \theta_\omega$ phase space at the same times, at the same \copyedit{points} and in the same style as those in \cref{fig:main_scatter}. The bottom plots superimpose the 90\% contours and centroids for $t = 0.75 t^*$ (red) and $t = 1.25 t^*$ (black) on $2 \src / \tilde{W}^2$ (left) and on $\Omega_r$ (right), as given by (\ref{eq:geo_transforms}).}
    \label{fig:eccentricity2}
\end{figure}

To further investigate the evolution of the flow in the context of the elliptic instability, we again consider the relationship between the $\zeta - \theta_\omega$ and \copyedit{shear--rotation} phase spaces. Using $\theta_\omega$ and $\zeta$, the constituents of the \copyedit{shear--rotation} phase space and the corresponding \copyedit{shear--rotation} correlation term and vortex identification criterion are given by
\begin{equation}\label{eq:geo_transforms}
\begin{split}
    \frac{\psq}{\tilde{W}^2} &= \frac{4 \zeta^4 {\rm cos}^2 \theta_\omega}{1 + \left( 
    -2 + 4{\rm cos}^2 \theta_\omega \right)\zeta^2 + \zeta^4}, \quad \frac{\gW}{\tilde{W}^2} = \frac{\left( 1 - \zeta^2 \right)^2}{1 + \left(-2 + 4{\rm cos}^2 \theta_\omega \right) \zeta^2 + \zeta^4},\\
    \frac{2 \src}{\tilde{W}^2} &= \frac{4 \zeta^2 {\rm cos}^2 \theta_\omega \left( 1 - \zeta^2 \right)}{1 + \left(-2 + 4{\rm cos}^2 \theta_\omega \right) \zeta^2 + \zeta^4}, \quad \; \Omega_r = \left( 1 + \left( 1 - \sqrt{\frac{\psq}{\tilde{W}^2}} \right)^2 \right)^{-1}.
\end{split}
\end{equation}
\autoref{fig:eccentricity2} shows the \copyedit{p.d.f.s} of the velocity gradients alongside the distributions of $2 \src/\tilde{W}^2$ and $\Omega_r$ in the geometry-based phase space at the same times as those in \cref{fig:main_scatter}.

As the elliptic instability emerges ($t \approx 0.75 t^*$), the \copyedit{p.d.f.} is highly concentrated at small values of $\theta_\omega$ over a broad range of $\zeta$. This distribution is consistent with the enhancement of $2 \Aavg{\src}$ around this time (see \cref{fig:phase_space_v1_fig1}) and reflects conditions conducive to the elliptic instability. The centroid of the distribution is located around $\zeta \approx 0.5$, which suggests that rotational streamlines with this aspect ratio may be particularly susceptible to these conditions.

During transition ($t \approx 0.90 t^*$), the distribution broadens to a much larger range of $\theta_\omega$ and, at the high end of this range, $\zeta$ becomes increasingly correlated with $\theta_\omega$. This broadening reflects the diversification of rotational local streamline topologies to include those for which shearing and rigid rotation are not \copyedit{well aligned}. However, despite this broadening, the \copyedit{p.d.f.} is still concentrated at small $\theta_\omega$. This \copyedit{behaviour} is consistent with the importance of the elliptic instability during transition (see \autoref{sec:app:elliptic}).

During turbulent decay ($t \gtrsim t^*$), the features of the \copyedit{p.d.f.} are relatively invariant in time. In the equilibrium distribution (approximated at $t \approx 1.25 t^*$), the centroid is located around $\theta_\omega \approx 44^\circ$, which is considerably larger than its value ($\theta_\omega \approx 3^\circ$) at $t = 0.75 t^*$. Further, the 90\% contour of the \copyedit{p.d.f.} spans nearly the entire range of $\theta_\omega$ and highlights a well-defined sharpening in the correlation of $\zeta$ with $\theta_\omega$ with increasing $\theta_\omega$. The upper limit $(\theta_\omega,\zeta) = (90^\circ,1)$ corresponds to circular local streamlines in the plane of rigid rotation subject to out-of-plane shearing.

The shifts in the centroid and the 90\% contour support the hypothesis that, while the elliptic instability still plays a role in the turbulent regime, other mechanisms also contribute significantly to the local flow structure. Specifically, they suggest that, in addition to the breakup of elliptic streamlines via the elliptic instability, the deformation of vortices, e.g. through the action of the Crow instability, may play an important role during turbulent decay. As discussed in \autoref{sec:app:src_vfs}, the partitioning of \copyedit{shear--rotation} correlations between the cores and boundaries of vortices in the flow (see \cref{fig:tim_try_corr2}) provides an interesting opportunity for \copyedit{analysing} these mechanisms. Given current challenges in disentangling the elliptic and Crow instabilities in turbulent flows \citep{Mis2021,Ost2021}, the present geometry-based ($\zeta - \theta_\omega$) phase space has the potential to help distinguish flow features associated with these ubiquitous mechanisms.

Altogether, the results in this section reinforce the notion that the elliptic instability is the dominant mechanism mediating the transition of the present vortex ring collision. They also support the notion that, after transition, the elliptic instability is no longer a strictly dominant mechanism underlying the turbulent decay of the flow. The results point to increased contributions from rotational geometries with out-of-plane shearing, which may reflect interactions associated with mechanisms like the Crow instability. The results also highlight the ability of the new \copyedit{shear--rotation} and geometry-based phases spaces to characterize the relative contributions of different modes of deformation and rotational features of local streamlines, respectively.

%%%%%%%%%%%%%%%%%%%%%%%%%%%%%%%%%%%%%%%%%%
\section{Conclusions}\label{sec:conclusions}
%%%%%%%%%%%%%%%%%%%%%%%%%%%%%%%%%%%%%%%%%%

We use a \copyedit{recently developed} adaptive, multiresolution numerical scheme based on the \copyedit{LGF} to efficiently simulate the head-on collision between two vortex rings at a relatively high Reynolds number ($Re_{\mathit{\Gamma}_0} = 4000$). The fidelity of this simulation is confirmed using various integral metrics that reflect the symmetries, conservation \copyedit{properties} and discretization errors of the flow. We provide a detailed analysis of the initial evolution, \copyedit{transition} and turbulent decay of the flow to elucidate flow features that are pertinent to the mechanisms driving its evolution, e.g. the elliptic instability.

Our visualizations of vortex structures enable qualitative characterizations of the various regimes through which the flow evolves. They depict the \copyedit{short-wave} elliptic instability as the mechanism driving the initial transition of the rings as they merge at the collision plane. Consistent with previous studies \citep{McK2020,Mis2021}, late transition and (to a lesser extent) turbulent decay are mediated by antiparallel secondary vortex filaments that arise from local interactions associated with the elliptic instability. We confirm that the elliptic instability dominates transition by \copyedit{analysing} the scales of dominant \copyedit{wave-like} perturbations in that regime. During turbulent decay, we observe local ejections of vortex rings in regions where antiparallel vortex filaments interact. This observation supports the notion that interplay between the elliptic and Crow instabilities can impact vortex interactions, consistent with previous findings \citep{Mis2021,Ost2021}.

Our analysis of the flow \copyedit{centres} around using the triple decomposition of the \copyedit{VGT} to characterize the contributions of axial straining, shearing, rigid rotation \copyedit{and shear--rotation} correlations to the velocity gradients. The mutual interaction of the rings is marked by the development of shearing-dominated vorticity and enhanced \copyedit{shear--rotation} correlations, reflecting conditions conducive to the elliptic instability. These conditions are consistent with the initial elliptic instability observed in our visualizations and previously in similar configurations \citep{McK2020,Mis2021}. During turbulent decay, the relative contributions of the different modes of deformation to the velocity gradient strength (which is not stationary) are roughly invariant in time, suggesting an equilibrium partitioning of the VGT. This equilibrium partitioning is remarkably similar to the partitioning observed for forced isotropic turbulence \citep{Das2020}, suggesting that it may provide a broadly applicable avenue for \copyedit{modelling} incompressible flows with appropriate symmetries.

During the transition and turbulent decay of the flow, we also consider instantaneous distributions of the velocity gradients in various phase spaces. The broadening of the phase space distributions in these regimes reflects the generation of more diverse local streamline topologies. The distributions in the $\qA -\, \rA$ phase space show that the present vortex ring collision produces velocity gradients that follow the near-universal teardrop-like distribution observed previously for forced isotropic turbulence \citep{Das2019,Das2020a}. In addition to the $\qA -\, \rA$ phase space, we introduce the \copyedit{shear--rotation} phase space to characterize the interplay of shearing and rigid rotation in rotational settings and highlight the role of their correlations during transition and turbulent decay. 

Finally, we introduce a geometry-based ($\zeta - \theta_\omega$) phase space to further characterize the action of the elliptic instability (and other mechanisms) during transition and turbulent decay. As the rings interact, the emergence of the elliptic instability spurring transition is associated with the alignment of shearing and rigid rotation ($\theta_\omega \approx 3^\circ$). In this regime, the elliptic local streamlines in the plane of rigid rotation have aspect ratios \copyedit{centred} about $\zeta \approx 0.5$. During late transition and turbulent decay, the generation and interaction of secondary vortical structures broadens the distribution to include larger $\theta_\omega$, and the equilibrium distribution is ultimately \copyedit{centred} near $\theta \approx 44^\circ$. In this regime, regions with high $\theta_\omega$ and high $\zeta$ become increasingly correlated as they approach ($\theta_\omega,\zeta$) = ($90^\circ,1$). In conjunction with our visualizations, these results suggest that proximity to vortex cores and boundaries may be a useful tool for \copyedit{modelling} the interplay between mechanisms \copyedit{such as} the elliptic and Crow instabilities. As a whole, the geometry-based phase space we introduce has the potential to help distinguish effects associated with the elliptic instability (small $\theta_\omega$) and other mechanisms, which is an ongoing challenge for turbulent flows driven by interacting vortex filaments \citep{Mis2021,Ost2021}. 

Moving forward, the VGT phase spaces we introduce may provide a useful setting for \copyedit{analysing} a broad class of turbulent flows. For vortex ring collisions, \copyedit{analysing} regimes where the Crow instability dominates the elliptic instability would clarify the extent to which the phase spaces can disambiguate these mechanisms. More generally, it would also be useful to identify the conditions under which (i) the equilibrium partitioning of the VGT \citep{Das2020} and (ii) the teardrop-like distribution in the $\qA -\, \rA$ phase space \citep{Das2019,Das2020a} are applicable. The present VGT analyses are limited by the local, instantaneous nature of the streamline geometries under consideration. It would be interesting to generalize these analyses to capture features that are non-local and that persist in time.
%%%%%%%%%%%%%%%%%%%%%%%%%%%%%%%%%%%%%%%%%%

% back sections
%%%%%%%%%%%%%%%%%%%%%%%%%%%%%%%%%%%%%%%%%%
\backsection[Supplementary data]{\label{SupMat}Supplementary movies are available at \href{https://doi.org/10.1017/jfm.2024.90}{https://doi.org/10.1017/jfm.2024.90}.}

\backsection[Acknowledgements]{The authors gratefully acknowledge B. Dorschner and K. Yu for their extensive guidance on the computations. }

\backsection[Funding]{R.A. was supported by the Department of Defense (DoD) through the National Defense Science \& Engineering Graduate (NDSEG) Fellowship Program. This work used the Extreme Science and Engineering Discovery Environment (XSEDE, \citet{Tow2014}), which is supported by National Science Foundation grant number ACI-1548562. Specifically, it used the Bridges-2 system, which is supported by NSF award number ACI-1928147, at the Pittsburgh Supercomputing Center (PSC).}

\backsection[Declaration of interests]{The authors report no conflict of interest.}

%\backsection[Data availability statement]{The data that support the findings of this study are available from the authors upon reasonable request.}

\backsection[Author ORCIDs]{

\noindent \orcidlink{0000-0002-5942-169X} Rahul Arun \href{https://orcid.org/0000-0002-5942-169X}{https://orcid.org/0000-0002-5942-169X}

\noindent \orcidlink{0000-0003-0326-3909} Tim Colonius \href{https://orcid.org/0000-0003-0326-3909}{https://orcid.org/0000-0003-0326-3909}}

%\backsection[Author contributions]{Authors may include details of the contributions made by each author to the manuscript.}
%%%%%%%%%%%%%%%%%%%%%%%%%%%%%%%%%%%%%%%%%%

%%%%%%%%%%%%%%%%%%%%%%%%%%%%%%%%%%%%%%%%%%
% appendices
%%%%%%%%%%%%%%%%%%%%%%%%%%%%%%%%%%%%%%%%%%

\appendix

%%%%%%%%%%%%%%%%%%%%%%%%%%%%%%%%%%%%%%%%%%
\section{Computational formulation}\label{sec:app:comp_meth}
%%%%%%%%%%%%%%%%%%%%%%%%%%%%%%%%%%%%%%%%%%

Here, we briefly document the adaptive computational framework. We refer to \citet{Yu2022} for a detailed description of the framework and a discussion of its novel aspects.

The non-dimensional, incompressible \copyedit{NSE} are given by
\begin{equation}\label{eq:app:NSE}
   \partial_t \boldsymbol{u} + (\boldsymbol{u} \boldsymbol{\cdot} \nabla)\boldsymbol{u} = -\nabla p + \frac{1}{Re} \nabla^2 \boldsymbol{u}, \quad \nabla \boldsymbol{\cdot} \boldsymbol{u} = 0,
\end{equation}
where $\boldsymbol{u} = (u, v, w)$ is the velocity, $p$ is the pressure, $t$ denotes \copyedit{time} and $Re$ is the Reynolds number. We focus particularly on the class of unbounded flows obeying the following far-field boundary conditions: $\boldsymbol{u}(\boldsymbol{x},t) \rightarrow \boldsymbol{0}$, $p(\boldsymbol{x},t) \rightarrow p_\infty$ \copyedit{and} $\boldsymbol{\omega}(\boldsymbol{x},t) \rightarrow \boldsymbol{0}$ (exponentially) as $\lvert \boldsymbol{x} \rvert \rightarrow \infty$. These boundary conditions differ slightly from the more generic (time-varying) \copyedit{free-stream} conditions considered by \citet{Lis2016}. For the present simulations, variables are non-dimensionalized using the initial radius and circulation of each vortex ring ($R_0$ and $\mathit{\Gamma}_0$, respectively) and $Re$ is given by the initial circulation Reynolds number ($Re_{\mathit{\Gamma}_0}$).

\begin{figure}
    \centering
    \includegraphics[width=\textwidth,trim=10 130 90 110,clip]{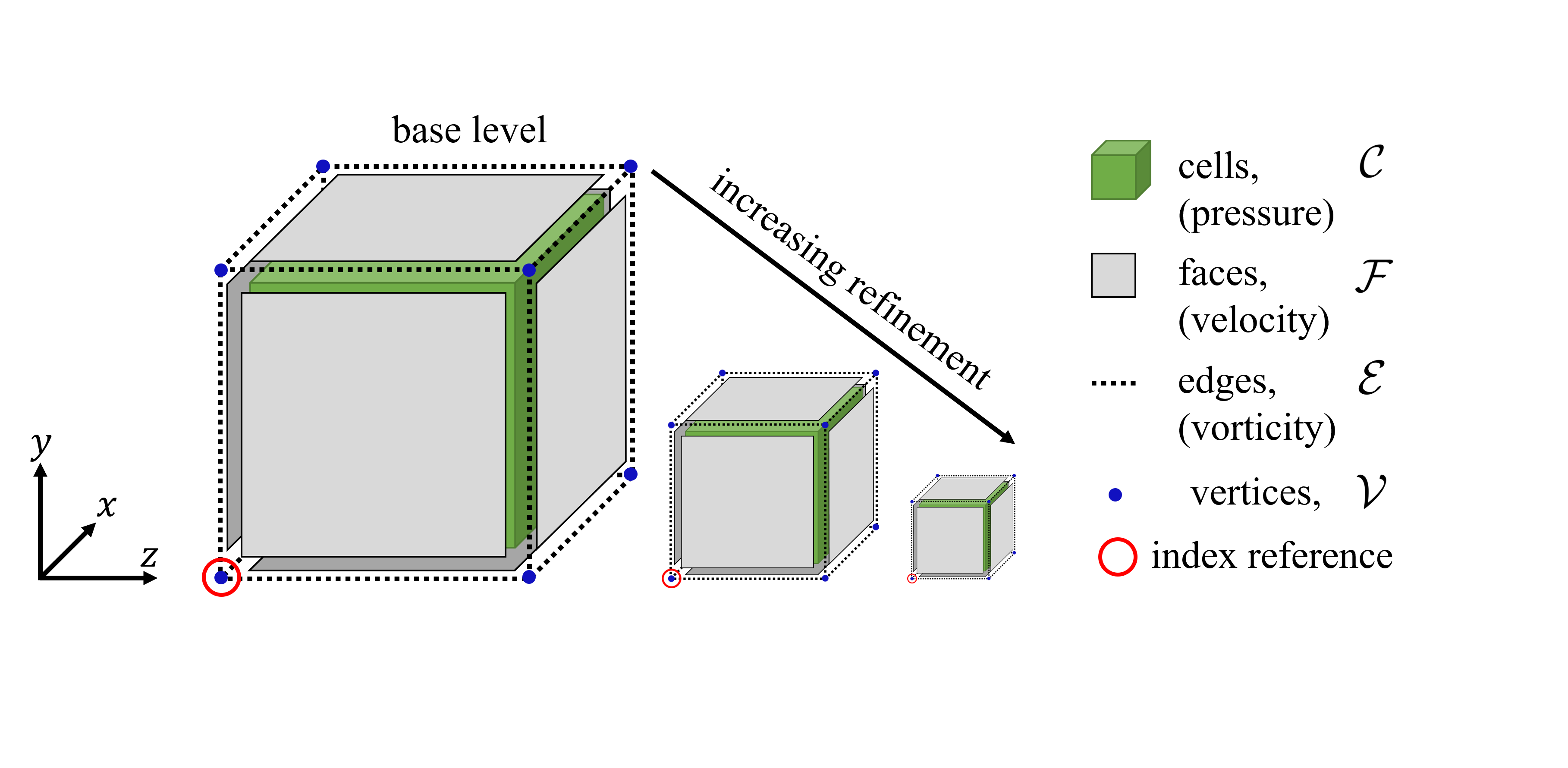}
    \caption{Unit cells of the staggered Cartesian grid at the base and refinement levels of the mesh, showing the locations of relevant flow variables.}
    \label{fig:app:cell}
\end{figure}

The NSE are spatially discretized on the composite grid, which contains a series of uniform staggered Cartesian meshes with increasing resolution. \autoref{fig:app:cell} depicts the locations of various vector and scalar flow variables on the cells of these meshes. We use $\mathcal{Q} \in \left\{ \mathcal{C}, \mathcal{F}, \mathcal{E}, \mathcal{V} \right\}$ to denote operations that are constrained to the corresponding locations on the cells. The semi-discrete NSE on the composite grid are given by
\begin{equation}\label{eq:app:semi}
\frac{d \mathsfbi{u}}{dt} - \mathsfbi{N}(\mathsfbi{u}) = -\mathsfbi{G} \mathsfi{p}_{\mathsf{tot}} + \frac{1}{Re} \mathsfbi{L}_\mathcal{F}^{} \mathsfbi{u}, \quad \mathsfbi{D} \mathsfbi{u} = 0,
\end{equation}
where we represent discretized variables and operators using sans-serif symbols, which are bold in vector settings and non-bold in scalar settings. Here, $\mathsfbi{G}$, $\mathsfbi{D}$, $\mathsfbi{L}$ \copyedit{and} $\mathsfbi{N}$ represent the discrete forms of the gradient, divergence, \copyedit{Laplace} and nonlinear operators, respectively. We have used the rotational form of the convective term in (\ref{eq:app:semi}) such that $\mathsfi{p}_{\mathsf{tot}} = \mathsfi{p} + \frac{1}{2}|\mathsfbi{u}|^2$ discretely represents the total pressure perturbation and $\mathsfbi{N}(\mathsfbi{u}) = \mathsfbi{r}$ discretely represents the Lamb vector, $\boldsymbol{r} = \boldsymbol{u} \times \boldsymbol{\omega}$.

The semi-discrete momentum equations, subject to the continuity constraint, are integrated in time using the IF-HERK method \citep{Lis2016, Yu2022}. This method combines an integrating factor (IF) technique for the viscous term with a half-explicit Runge-Kutta (HERK) technique for the convective term. In the HERK time-stepping scheme \citep{Bra1993, Lis2016, Yu2022}, the task of integrating (\ref{eq:app:semi}) at each time step is subdivided into $N_{\rm stage}$ stages. Using a block \rahul{lower--upper (LU)} decomposition and the mimesis and commutativity properties of the relevant operators, the subproblem associated with stage $i$ of the HERK scheme is formulated on the composite grid as
\begin{equation}\label{eq:app:herk}
\mathsfi{L}_\mathcal{C}^{} \mathsfi{p}_{\mathsf{tot}}^i = \mathsfi{f}^i = \mathsfbi{D} \mathsfbi{r}^i, \quad \mathsfbi{u}^i = \mathsfbi{H}_\mathcal{F}^i \left( \mathsfbi{r}^i - \mathsfbi{G} \mathsfi{p}_{\mathsf{tot}}^i \right).
\end{equation}
Here, $\mathsfbi{H}$ represents the IF operator and $\mathsfi{f} = \mathsfbi{D} \mathsfbi{r}$ approximately represents the divergence of \copyedit{the} Lamb vector. For brevity, we omit the exact dependencies of $\mathsfbi{r}^i$ on various flow variables from stages 1 to $i$ of the HERK scheme. We refer to the formulation in \textsection 2.4 of \citet{Yu2022} for these details and for the corresponding Butcher tableau.

While the discrete operators in (\ref{eq:app:semi}) and (\ref{eq:app:herk}) are formally defined on the unbounded composite grid, they are practically applied to the finite subset representing the AMR grid. The operator $\mathsfi{R}_\mathcal{Q}$ restricts variables from the composite grid \copyedit{$\left( \, \cdot \, \right)$} to the AMR grid as \copyedit{$\widehat{ \left( \, \cdot \, \right) } = \mathsfi{R}_\mathcal{Q} \left( \, \cdot \, \right)$}. In the other direction, the operator $\mathsfi{P}_\mathcal{Q}$ approximates variables on the composite grid using the values on the AMR grid as \copyedit{$\left( \, \cdot \, \right) \approx \mathsfi{P}_\mathcal{Q} \widehat{ \left( \, \cdot \, \right)}$}. Using these operators, solutions to the subproblems associated with each stage of the HERK formulation in (\ref{eq:app:herk}) can be approximated on the AMR grid.

The two steps of each subproblem in (\ref{eq:app:herk}) involve (i) solving the discrete pressure Poisson equation and (ii) applying the IF to recover the velocity. The solution to the pressure Poisson equation on the AMR grid can be expressed as
\begin{equation}\label{eq:app:lgf}
    \widehat{\mathsfi{p}_{\mathsf{tot}}^i} = \mathsfi{R}_\mathcal{C}^{} \mathcal{G}_\mathcal{C}^{} * \mathsfi{f}^i =  \mathsfi{R}_\mathcal{C}^{} \mathsfi{L}_\mathcal{C}^{-1} \mathsfi{f}^i \approx  \mathsfi{R}_\mathcal{C}^{} \mathsfi{L}_\mathcal{C}^{-1} \mathsfi{P}_\mathcal{C}^{} \widehat{\mathsfi{f}_{\vphantom{\mathcal{C}}}^i},
\end{equation}
where $\mathcal{G}_\mathcal{C}^{}$ is the LGF and $*$ represents the discrete convolution. We efficiently evaluate (\ref{eq:app:lgf}) using a fast multipole method \citep{Lis2014, Dor2020} that accelerates solutions by incorporating summation techniques based on the fast Fourier transform. This method is key to enabling the linear algorithmic complexity and high parallel efficiency of the flow solver. The application of the IF operator can similarly be expressed as
\begin{equation}\label{eq:app:if}
    \widehat{\mathsfbi{u\vphantom{f}}^i} = \mathsfbi{R}_\mathcal{F}^{} \mathsfbi{H}_\mathcal{F}^i \left( \mathsfbi{r}^i - \mathsfbi{G} \mathsfi{p}_{\mathsf{tot}}^i \right) \approx \mathsfbi{R}_\mathcal{F}^{} \mathsfbi{H}_\mathcal{F}^i \mathsfbi{P}_\mathcal{F}^{} \left( \widehat{\mathsfbi{r}_{\vphantom{\mathcal{C}}}^i} - \widehat{ \mathsfbi{G} \mathsfi{p}_{\mathsf{tot}}^i } \right),
\end{equation}
where $\widehat{ \mathsfbi{G} \mathsfi{p}_{\mathsf{tot}}^i } = \mathsfbi{R}_\mathcal{C}^{} \mathsfbi{G} \mathsfi{p}_{\mathsf{tot}}^i$. The application of the IF operator represents a convolution with an exponentially decaying kernel and it can also be evaluated using fast LGF techniques \citep{Lis2016, Yu2022}. 

At each time step, the simulation adapts the extent of the AMR grid and adaptively refines regions within the AMR grid according to the spatial adaptivity and mesh refinement criteria, respectively. The spatial adaptivity criterion sets the boundaries of the AMR grid to capture regions where the source of the pressure Poisson equation exceeds a threshold, $\varepsilon_{\rm adapt}$, relative to its maximum value in the domain. In other words, the AMR grid is adaptively truncated to capture the subset of the unbounded domain satisfying
\begin{equation}\label{eq:app:adapt}
    \lvert \mathsfi{f} (\mathsfbi{x}, \mathsfi{t}) \rvert < \varepsilon_{\rm adapt} \lVert \mathsfi{f} \rVert_\infty (\mathsfi{t}),
\end{equation}
where $\varepsilon_{\rm adapt} = 10^{-6}$ for the present simulation. One caveat is that the IF convolution involves a velocity source that decays slower than vorticity. Correspondingly, its evaluation requires the velocity field in a slightly extended domain based on a cutoff distance that is selected to capture the IF kernel with high accuracy. For the present simulation, the initial rectangular domain is large enough to contain as a subset the domain satisfying the adaptivity criterion (\ref{eq:app:adapt}).

At each level $k$ of the AMR grid, the mesh refinement criteria are formulated in terms of a combined source, $\mathsfi{f}_k(\mathsfi{t})$, which includes the source of the pressure Poisson equation and a correction term. The correction term accounts for the differences between the partial solutions on the coarse and fine grids and it is evaluated using an extended region that can overlap with \copyedit{neighbouring} levels. We refer to \citet{Yu2022} for the details of its formulation and implementation, which we omit for brevity. Using the combined source, a region is refined or coarsened when
\begin{equation}\label{eq:app:refine}
   \mathsfi{f}_k (\mathsfbi{x}, \mathsfi{t}) > \alpha^{N_{\rm level}-k}  \mathsfi{f}_{\mathsf{max}}(\mathsfi{t}) \; \quad \; {\rm or} \; \quad \;  \mathsfi{f}_k (\mathsfbi{x}, \mathsfi{t}) < \beta \alpha^{N_{\rm level}-k}  \mathsfi{f}_{\mathsf{max}}(\mathsfi{t}),
\end{equation}
respectively, where $\alpha \in (0,1)$ and $\beta \in (0,1)$ and we select $\alpha = 0.125$ and $\beta = 0.875$ for the present simulation. In these criteria, the combined source is evaluated relative to its maximum blockwise root-mean-square (BRMS) value computed over all blocks and previous times, which is expressed as
\begin{equation}\label{eq:app:brms}
    \mathsfi{f}_{\mathsf{max}}(\mathsfi{t}) = \max_{\mathsfi{\tau} < \mathsfi{t}}{\rm BRMS}\left(\mathsfi{f}_k(\mathsfbi{x}, \mathsfi{\tau})\right).
\end{equation}

%%%%%%%%%%%%%%%%%%%%%%%%%%%%%%%%%%%%%%%%%%
\section{Instability development during transition}\label{sec:app:elliptic}
%%%%%%%%%%%%%%%%%%%%%%%%%%%%%%%%%%%%%%%%%%

\begin{figure}
    \centering
    \includegraphics[width=\textwidth,trim=60 60 60 20,clip]{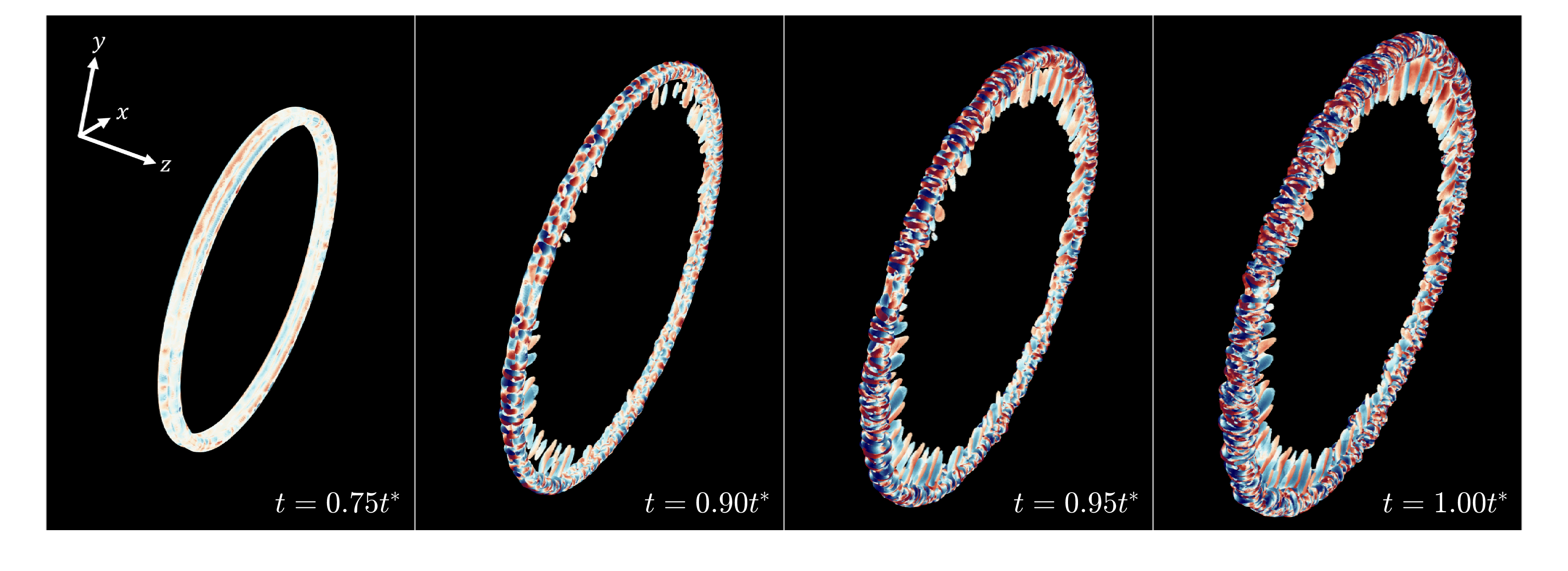}
    \caption{Magnified versions of the auxiliary viewpoints in \cref{fig:tim_try_lvec2} at the reference times associated with instability growth and transition.}
    \label{fig:rev_magnified}
\end{figure}

Whereas the \copyedit{short-wave} elliptic instability has a wavelength \copyedit{of} the order of the vortex core radius, $a(t)$, the \copyedit{long-wave} Crow instability occurs at wavelengths much larger than $a(t)$ \citep{Lew2016,McK2020,Mis2021}. Here, we track the development of \copyedit{wave-like} instabilities around the azimuth of the ring at the reference times from \cref{tab:res:regimes} associated with the transition to turbulence. \autoref{fig:rev_magnified} shows closeups of the vortical flow structures at these times from the auxiliary viewpoint in \cref{fig:tim_try_lvec2}.

The dominant scales of the \copyedit{wave-like} perturbations are identified by decomposing the flow into azimuthal Fourier modes, which are denoted using \rahul{$\hat{( \, \cdot \, )}$} and have corresponding wavenumbers $m$. To obtain these Fourier modes, we linearly interpolate the flow at the finest level of the AMR grid to a uniform cylindrical grid. This uniform grid is discretized into $N_\theta = 1001$ points in the azimuthal direction and it has spacings, $\Delta r_{\rm uni} = \Delta z_{\rm uni} = 0.01$, that are consistent with that of the finest level, $\Delta x_{\rm fine} = 0.01$. In the following analysis, we limit our consideration to $(r,z)$ pairs for which $\langle \omega_z^2 \rangle_\theta$ exceeds 10\% of $\max_{r,z} \langle \omega_z^2 \rangle_\theta$, where \copyedit{$\langle \, \cdot \, \rangle_\theta$} denotes azimuthal averaging.

\begin{figure}
    \centering
    \includegraphics[width=\textwidth,trim=0 60 0 0,clip]{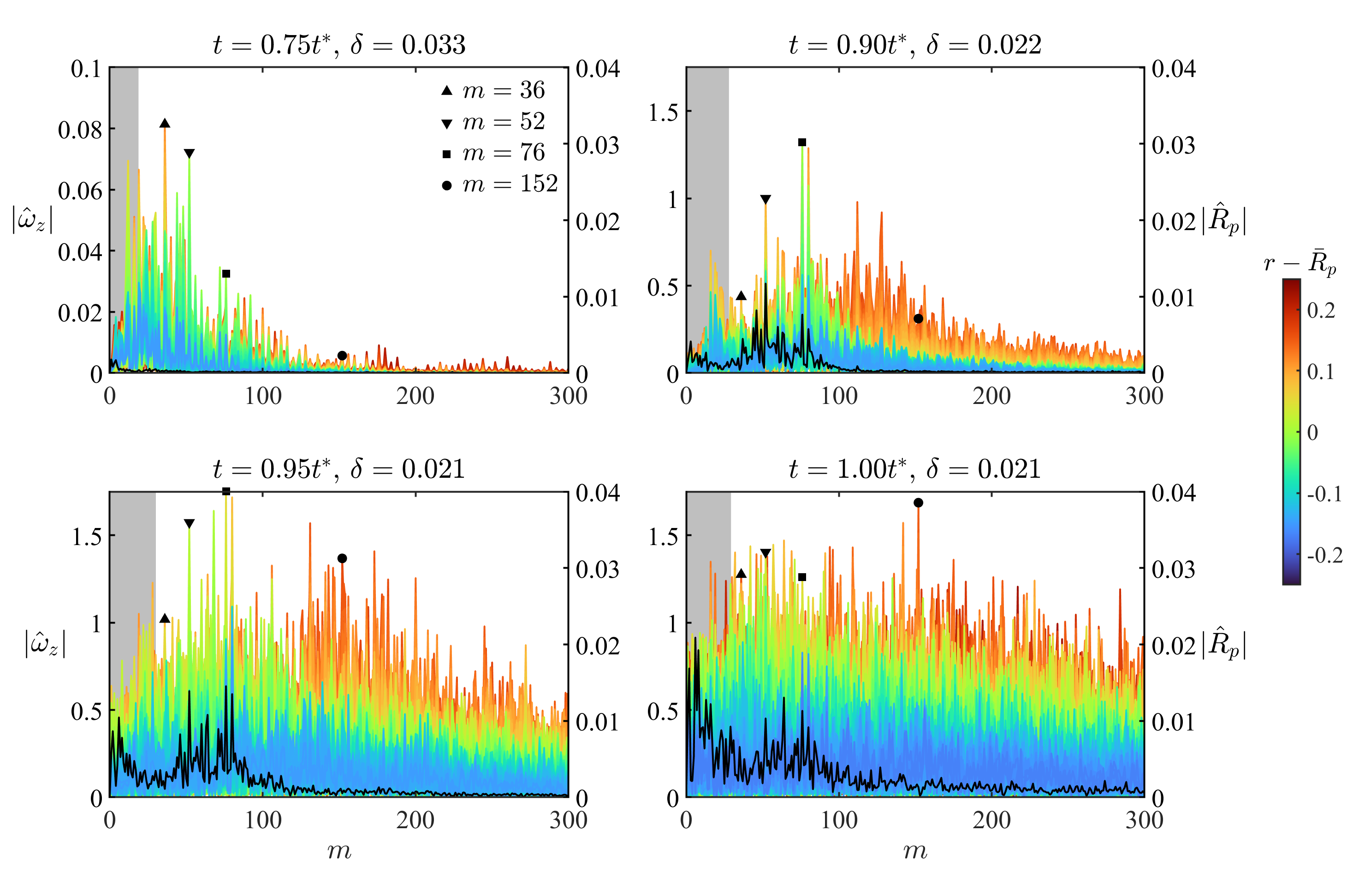}
    \caption{Fourier coefficient amplitudes: $| \hat{\omega}_z |$ (\copyedit{coloured} by radial displacement from $\bar{R}_p$) and $| \hat{R}_p |$ (black). At each of the reference times shown, $\delta$ represents the average slenderness ratio of the vortex rings. The values of $| \hat{\omega}_z |$ at wavenumbers pertinent to instability development are marked by symbols and the shaded regions represent wavenumbers for which $C > 10$.}
    \label{fig:rev_elliptic}
\end{figure}

As depicted in \cref{fig:rev_elliptic}, we characterize instability development using two flow variables. First, following previous studies \citep{McK2020,Mis2021}, we approximate the positions, $\boldsymbol{X}^+$ and $\boldsymbol{X}^-$, of the vortex cores by identifying the locations of the pressure minima for $z > 0$ and $z < 0$, respectively. The radial perturbations of these cores, $R_p^+(\theta)$ and $R_p^-(\theta)$, about their mean radial positions, $\bar{R}_p^+$ and $\bar{R}_p^-$, characterize the interactions between the rings. We measure the strength of the vortex core perturbations using $| \hat{R}_p | = [( | \hat{R}_p^+ |^2 + | \hat{R}_p^- |^2 )/2]^{1/2}$. Second, we consider $\omega_z$ since it captures the antiparallel vortex structures that develop around the ring during transition. The Fourier coefficient amplitudes of the corresponding perturbations, $| \hat{\omega}_z |$, are \copyedit{coloured} according to their displacement from average vortex ring radius, $\bar{R}_p = (\bar{R}_p^+ + \bar{R}_p^-)/2$.

To characterize the wavelengths of the dominant perturbations in terms of the elliptic and Crow instabilities, we estimate the core radii, $a^+$ and $a^-$, of the vortex rings at each reference time. Following \citet{McK2020}, we fit a \copyedit{two-dimensional} Gaussian function to the vortex core profiles, $\langle \omega_\theta \rangle_\theta$, for each ring. However, whereas \citet{McK2020} estimated the core radii by averaging the standard deviations of the Gaussian fits, ${\sigma^\pm}_{\mkern-12mu \rm min}$ and ${\sigma^\pm}_{\mkern-12mu \rm max}$, we instead estimate the core radii as $a^\pm = [({\sigma^\pm}_{\mkern-12mu \rm min})^2 + ({\sigma^\pm}_{\mkern-12mu \rm max})^2]^{1/2}$. Our estimates produce core radii that are consistent with the definition in (\ref{eq:intro:Gaussian}) for Gaussian vortex rings. Hence, we correctly identify $a_0 = 0.2$ for the initial condition, which is larger than the alternate core radius definition \citep{McK2020,Ost2021} by a factor of $\sqrt{2}$. We compute the average slenderness ratio of the vortex rings as $\delta = (\delta^+ + \delta^-)/2$, where $\delta^\pm = a^\pm/\bar{R}_p^\pm$. This definition is consistent with the Gaussian fits we consider since the radial locations of their centroids coincide with $\bar{R}_p^\pm$ to within 0.38\%. The average ratio of the perturbation wavelength to the core thickness can be expressed as $C = (C^+ + C^-)/2$, where $C^\pm = 2\pi/m \delta^\pm$. Here, we loosely associate $C > 10$ and $C < 10$ with the Crow and elliptic instabilities, respectively.

At $t = 0.75 t^*$, the wavenumber ($m = 36$) of the largest $\hat{\omega}_z$ perturbation corresponds to $C = 5.08$. A similar value, $C = 5.38$, is obtained at $t = 0.90 t^*$ for the $m = 52$ perturbation. However, $m = 76$ represents the largest perturbation at both $t = 0.90 t^*$ and $t = 0.95 t^*$, for which $C = 3.68$ and $3.97$, respectively. At each of these times, the dominant \copyedit{short-wave perturbations are of} the order of the core thickness. This result suggests that the corresponding development of secondary antiparallel vortex filaments (see \cref{fig:rev_magnified}) can be associated with the development of the elliptic instability.

The vortex core perturbations further support the notion that the elliptic instability is the dominant mechanism in the transitional regime. The perturbations at $t = 0.75 t^*$ are too small, relative to $\Delta r_{\rm uni} = 0.01$, to resolve. The dominant core perturbations are resolved for $t = 0.90 t^*$ and $t = 0.95 t^*$ and, consistent with the vorticity perturbations, they are largest at $m = 52$ and $m = 76$. For both $\hat{\omega}_z$ and $\hat{R}_p$, we also observe prominent perturbations at $m = 80$, but we do not speculate on their source.

For a similar vortex ring collision with $Re_{\mathit{\Gamma}_0} = 3500$ and $\delta_0 = 0.1$, \citet{Mis2021} attributed the growth of the $m = 40$ mode to the elliptic instability in a regime where $2 \pi R_p(t) / m \approx 0.2 - 0.4$. Employing a crude volume-conserving approximation for the vortex cores \citep{McK2020,Mis2021}, the core radii can be \copyedit{modelled} as $a^\pm = a_0 \sqrt{R_0/R_p^\pm}$, which suggests that $C \approx 2.26 - 6.38$ in that regime. Hence, the scales of the dominant perturbations relative to the core thickness in the present case are consistent with those previously attributed to the elliptic instability in a similar collision.

At $t = t^*$, we remarkably observe that the dominant $\hat{\omega}_z$ perturbation occurs at $m = 152$, which is the second harmonic of the $m = 76$ perturbation that governed the generation of secondary vorticity. This observation suggests that the elliptic instability retains an important role in mediating the production of subsequent generations of vortical structures at progressively smaller scales. It thus qualitatively supports the initial stages of iterative elliptic instability scenario leading to the generation of turbulence \citep{McK2020}. Identifying the later stages of this pathway would require a more refined analysis of the orientation of each generation of vortices relative to previous generations.

Although perturbations for which $C > 10$ are non-negligible, their signatures in $\hat{\omega}_z$ are not as prominent as those for which $C < 10$. This observation supports the notion that the Crow instability plays a secondary role to the elliptic instability in the transitional regime, consistent with previous studies in similar configurations \citep{McK2020,Mis2021}. Nevertheless, especially as the flow becomes turbulent, the broadening range of active scales obscures the interplay between these mechanisms. As this occurs, the vortex core perturbations gain significant energy at lower wavenumbers ($C > 10$), indicating that \copyedit{long-wave} mechanisms like the Crow instability may become important. Altogether, while relatively limited, the present analysis of instability development confirms the \copyedit{pre-eminence} of the elliptic instability during transition and supports our interpretation of the corresponding velocity gradients. 

%%%%%%%%%%%%%%%%%%%%%%%%%%%%%%%%%%%%%%%%%%
\section{\copyedit{Shear--rotation} correlations and vortical flow structures}\label{sec:app:src_vfs}
%%%%%%%%%%%%%%%%%%%%%%%%%%%%%%%%%%%%%%%%%%

\begin{figure}
    \centering
    \includegraphics[width=0.5\textwidth,trim=320 0 320 0,clip]{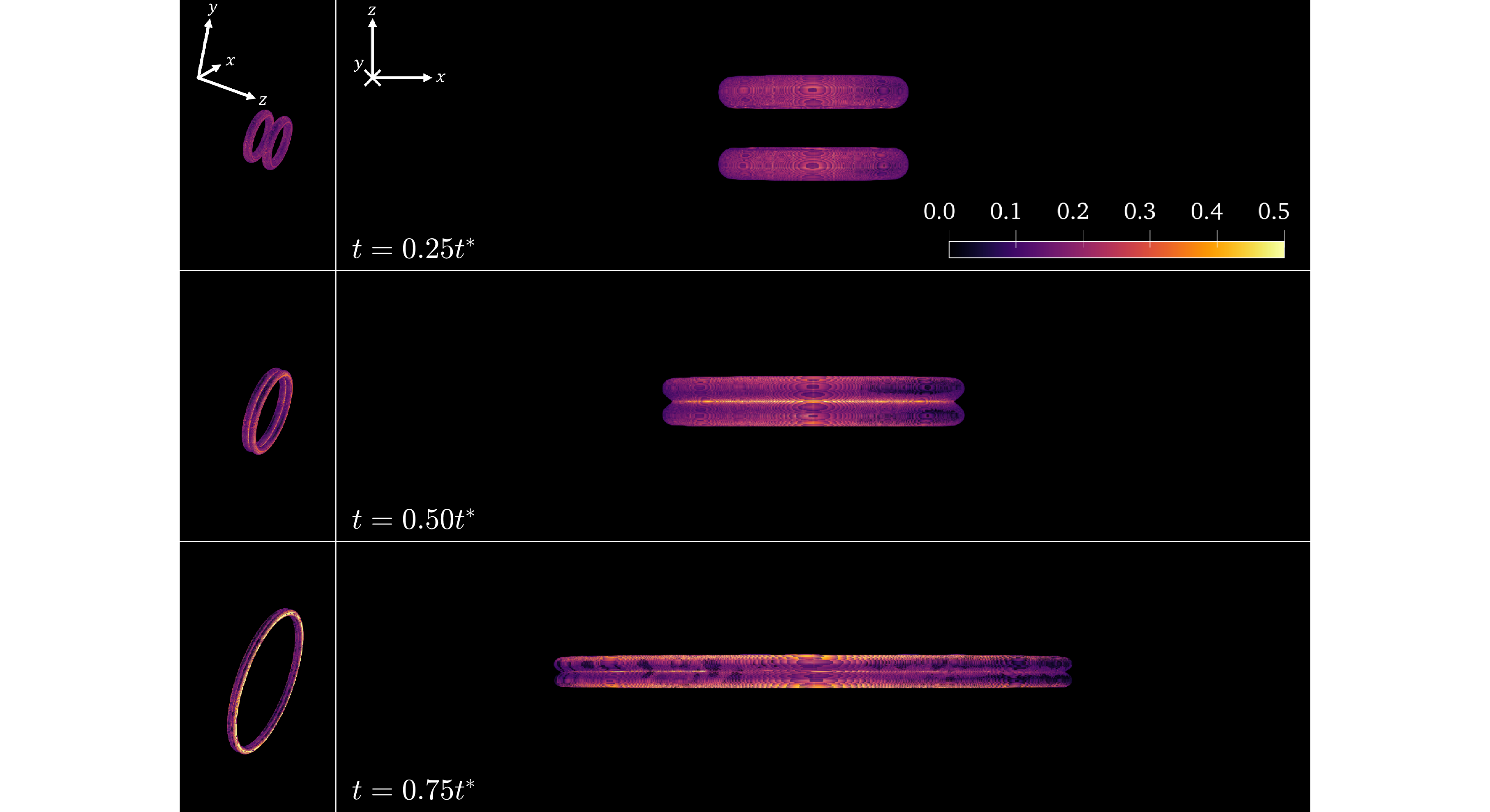}\includegraphics[width=0.5\textwidth,trim=320 0 320 0,clip]{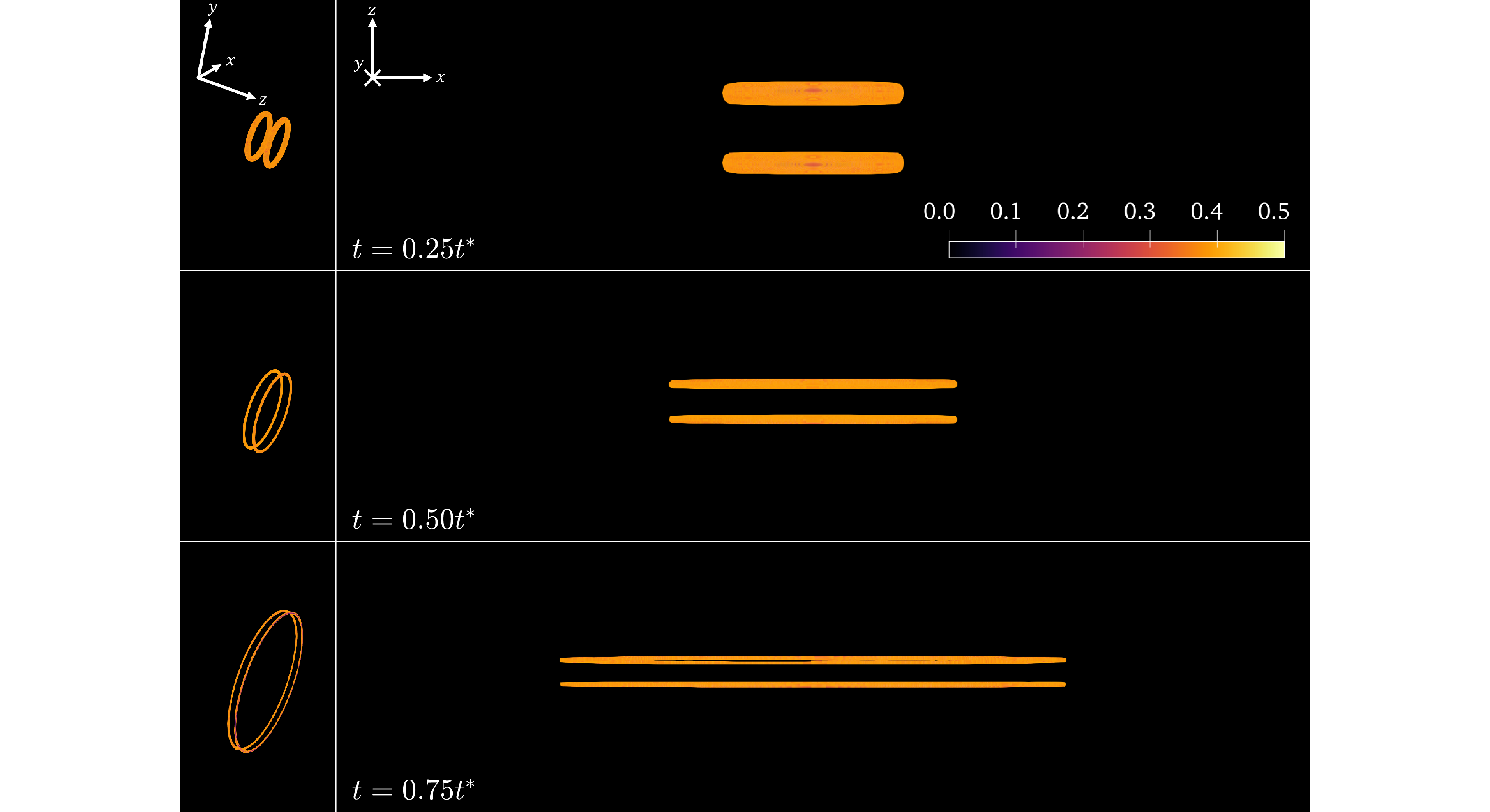}
    \includegraphics[width=0.5\textwidth,trim=320 0 320 0,clip]{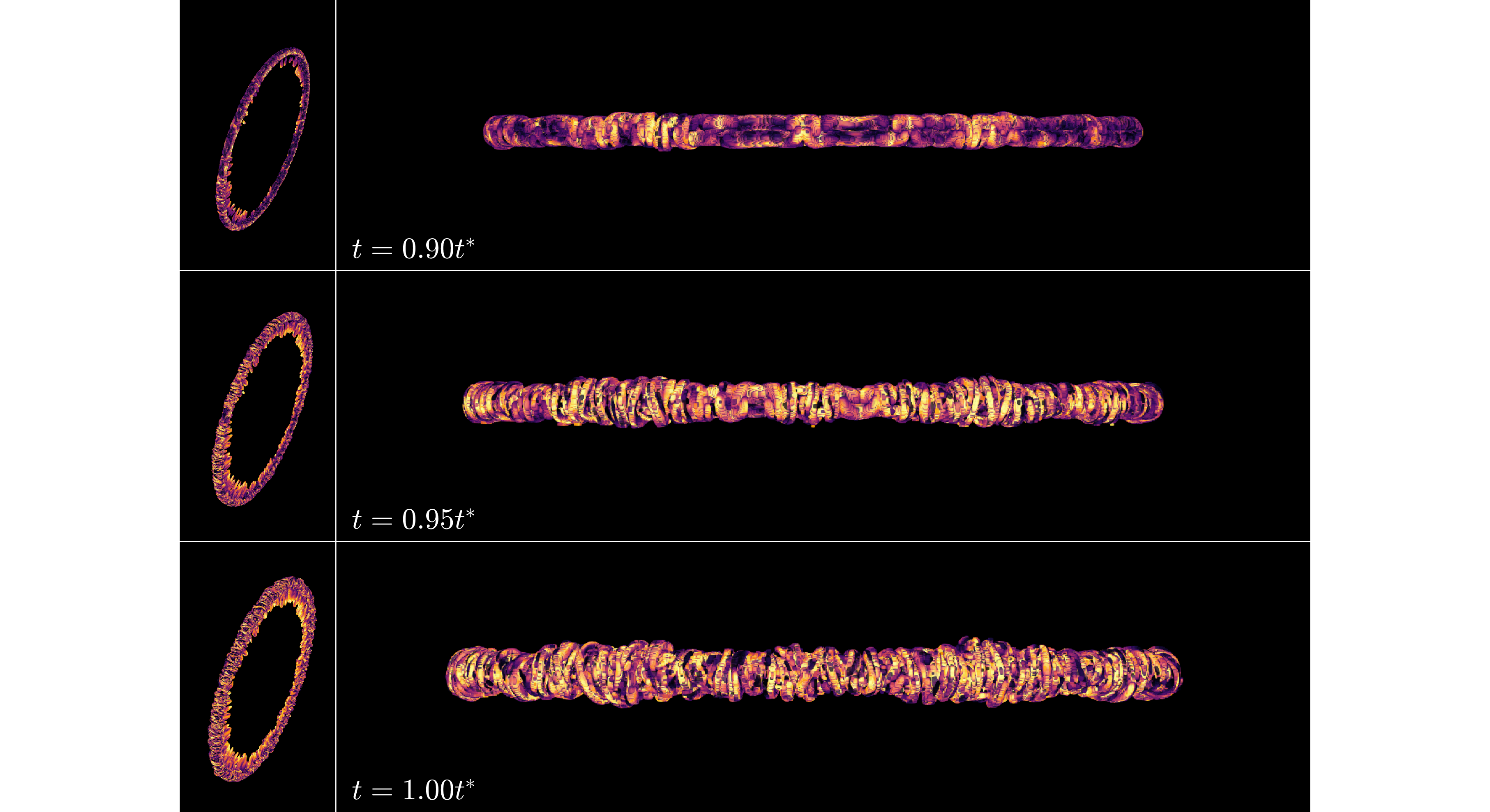}\includegraphics[width=0.5\textwidth,trim=320 0 320 0,clip]{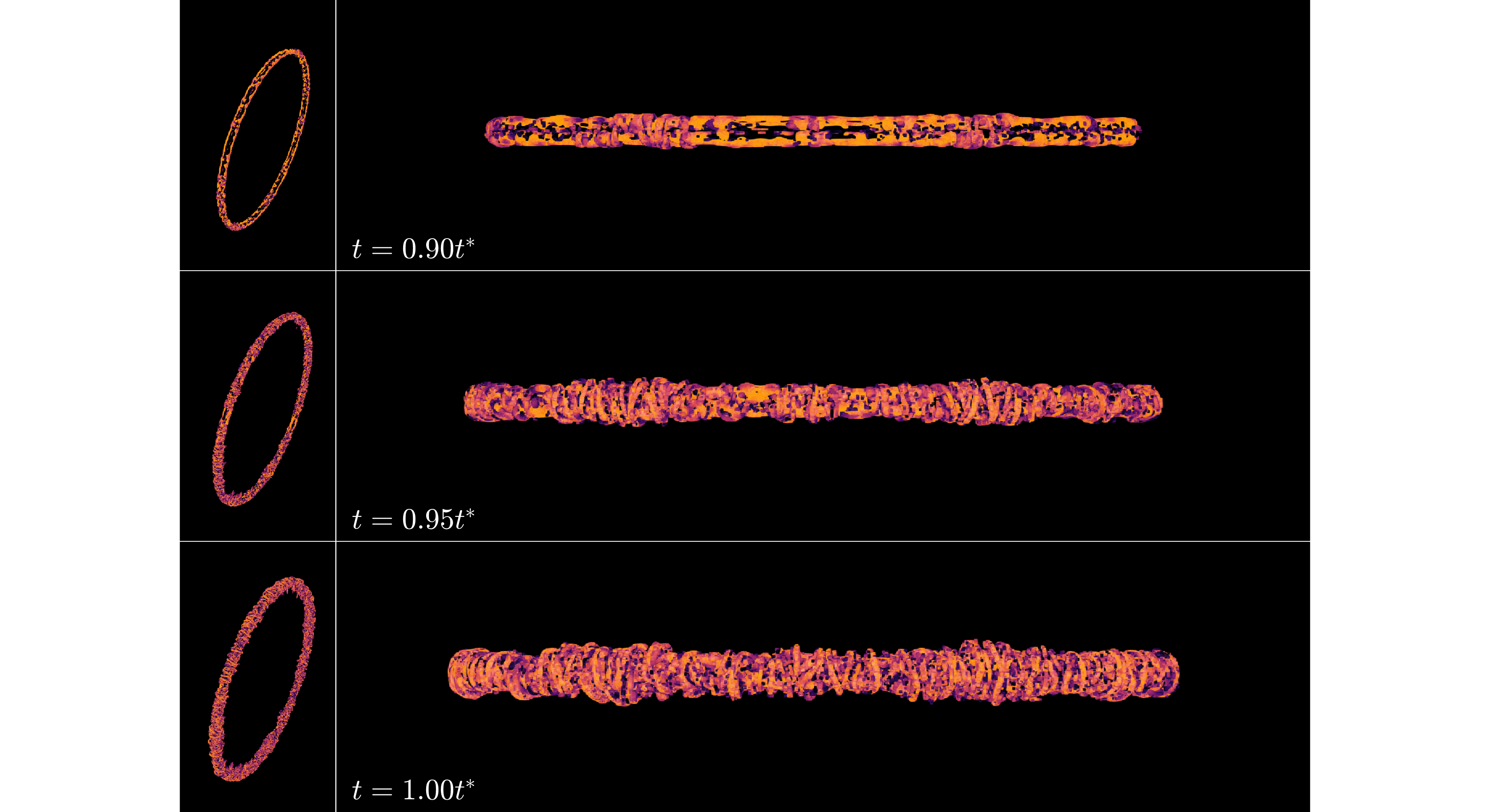}
    \includegraphics[width=0.5\textwidth,trim=320 0 320 0,clip]{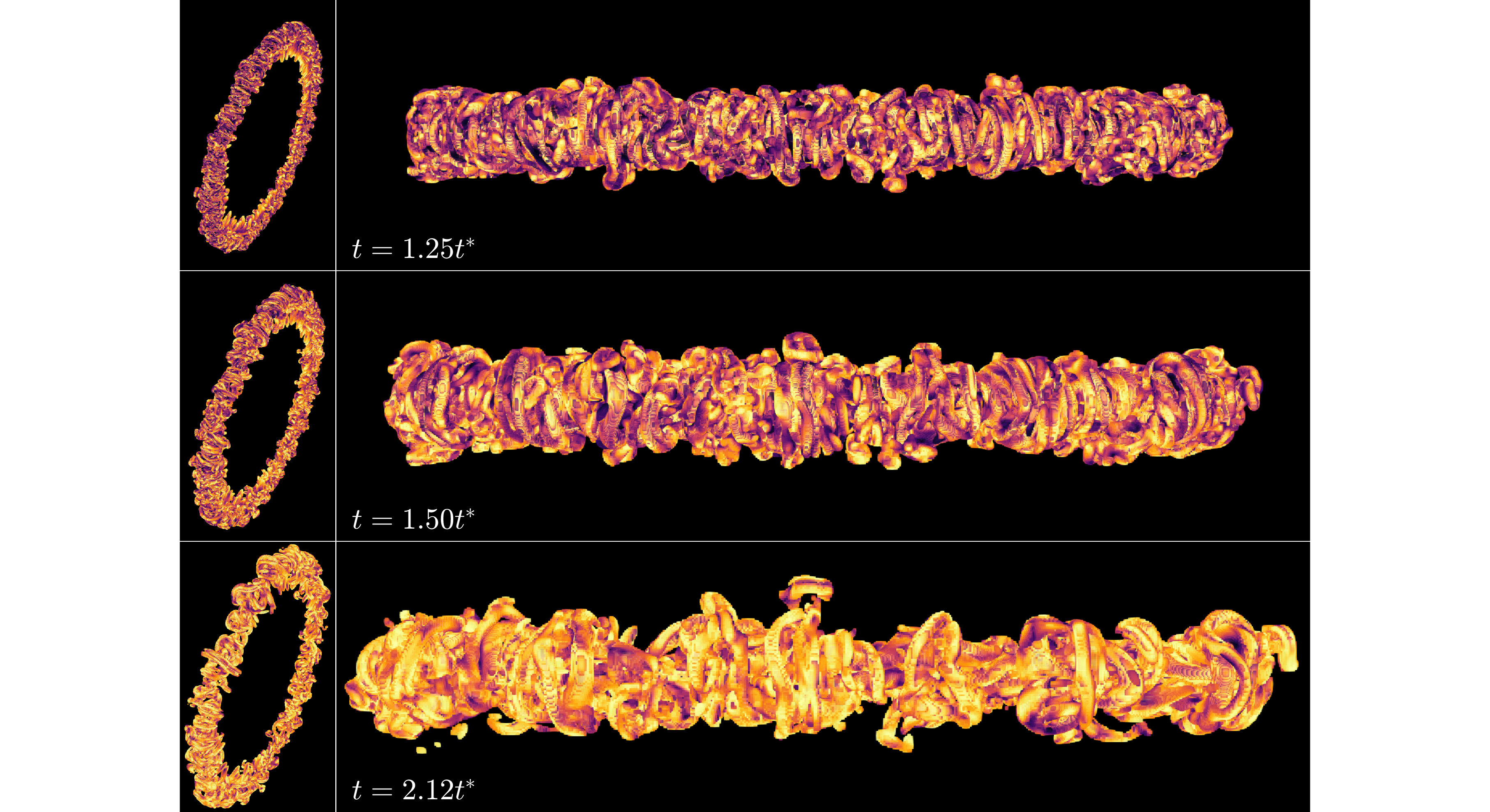}\includegraphics[width=0.5\textwidth,trim=320 0 320 0,clip]{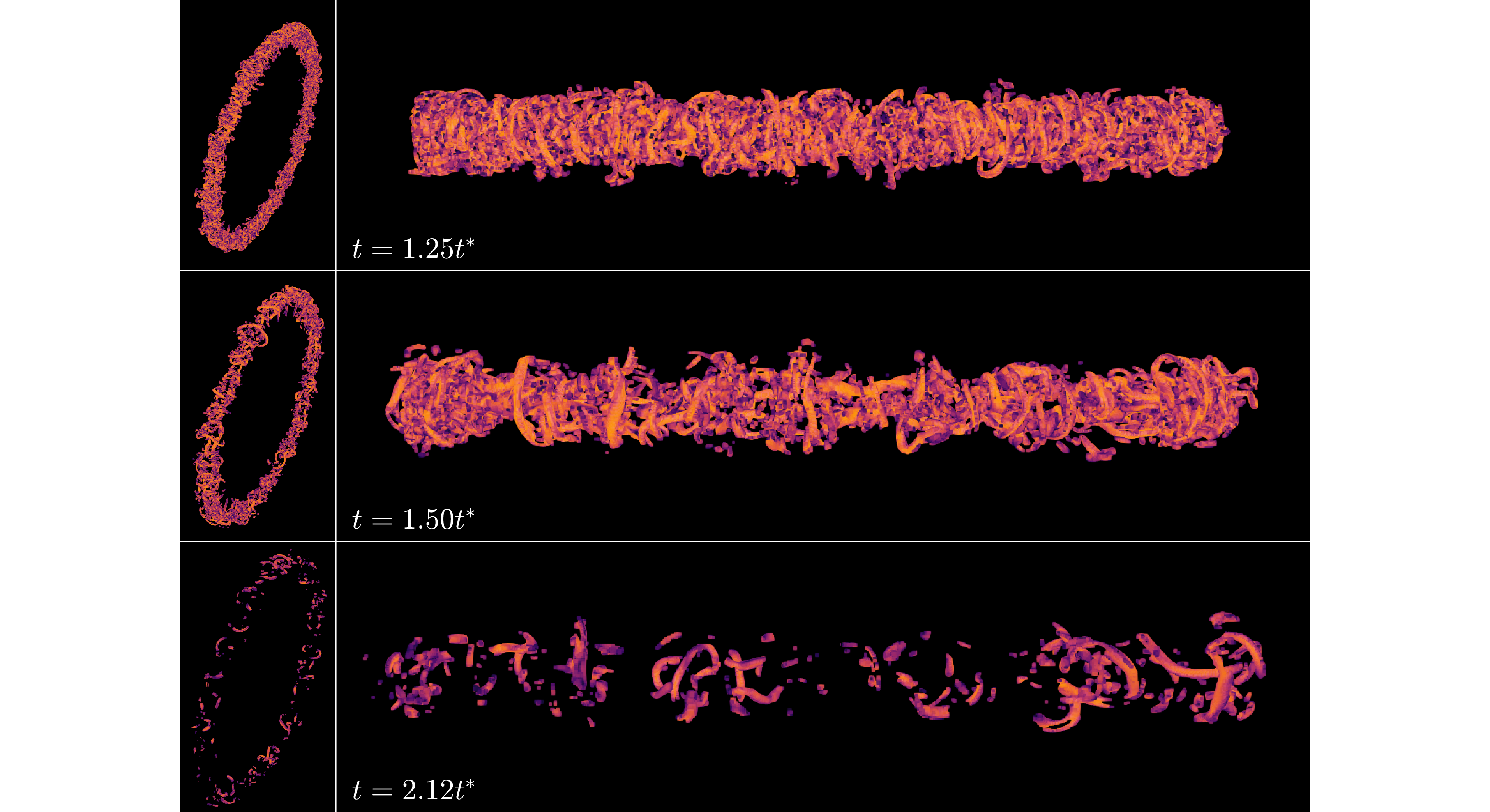}
    \caption{Visualizations of the vortex boundaries ($\Omega_r = 0.52$, left side) and vortex cores ($\Omega_r = 0.93$, right side), \copyedit{coloured} by $2 \src / \tilde{W}^2$, for each reference time from \cref{tab:res:regimes}. A movie depicting the evolution of the vortex boundaries from the auxiliary viewpoint (leftmost column) is provided as supplementary material.}
    \label{fig:tim_try_corr2}
\end{figure}

The visualizations in \cref{fig:tim_try_lvec2} help identify antiparallel vortex filaments and interactions between vortices, but the comparisons between the vortex boundary ($\Omega_r = 0.52$) and core ($\Omega_r = 0.93$) structures provide relatively little information. In \cref{fig:tim_try_corr2}, we visualize the same vortex structures but instead \copyedit{colour} them using $2 \src / \tilde{W}^2$ to probe how conditions conducive to the elliptic instability are structured throughout the vortices in the flow.

As the vortex boundaries merge and expand radially, the \copyedit{shear--rotation} correlations are relatively large at the collision plane and the outer boundaries in $z$ and they are relatively small at the inner and outer boundaries in the radial direction. This structuring illustrates how \copyedit{shear--rotation} correlations are especially enhanced in regions where the vortex boundaries become thinner, corresponding to the shift from a rigid-rotation-dominated regime to a shearing-dominated regime. During transition, the secondary vortex filaments are initially associated with relatively high and low \copyedit{shear--rotation} correlations near their boundaries and cores, respectively. As the turbulence develops, this structuring of $2 \src / \tilde{W}^2$ within the vortices remains similar to that of the secondary vortices mediating transition.

This persistent partitioning opens up an interesting possibility of \copyedit{analysing} the action of various mechanisms (e.g. the elliptic and Crow instabilities) in turbulent flows based on their proximity to vortex cores. For example, the phase space transformations in \textsection \ref{sec:res:phase} can be used to characterize local streamline geometries throughout vortices using the structure of the \copyedit{shear--rotation} correlations. Consistent with the transformations depicted in \cref{fig:eccentricity2}, our results suggest that local streamlines are more elliptic near vortex boundaries and more circular near vortex cores. This conceptual picture is consistent with the notion that the breakup and displacement of vortex core structures can be loosely associated with the elliptic and Crow instabilities, respectively. 

%%%%%%%%%%%%%%%%%%%%%%%%%%%%%%%%%%%%%%%%%%
\section{Effect of \copyedit{shear--rotation} alignment}\label{sec:app:shear_rot}
%%%%%%%%%%%%%%%%%%%%%%%%%%%%%%%%%%%%%%%%%%

\begin{figure}
    \centering
    \includegraphics[width=\textwidth,trim=80 60 60 0,clip]{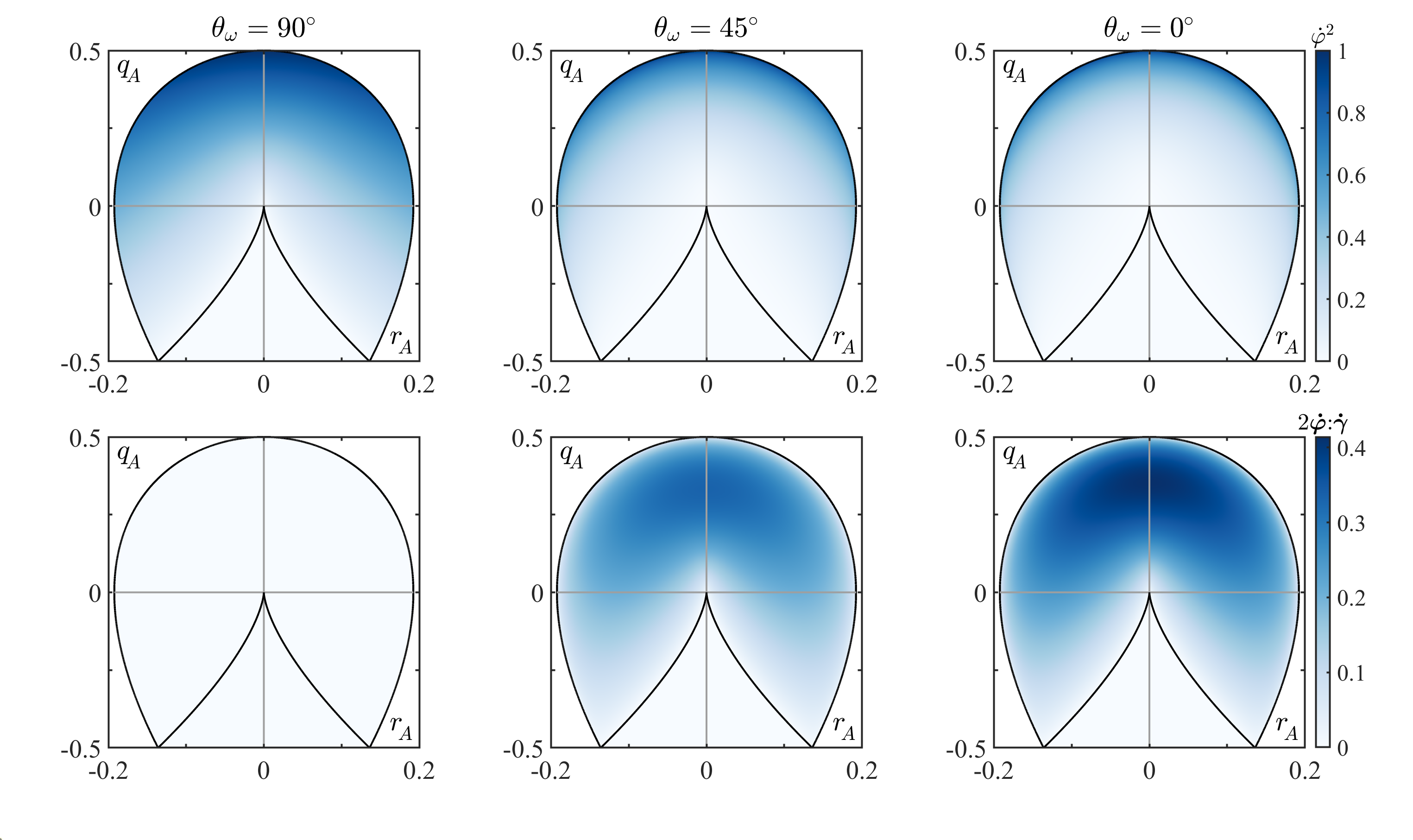}
    \includegraphics[width=\textwidth,trim=80 60 60 40,clip]{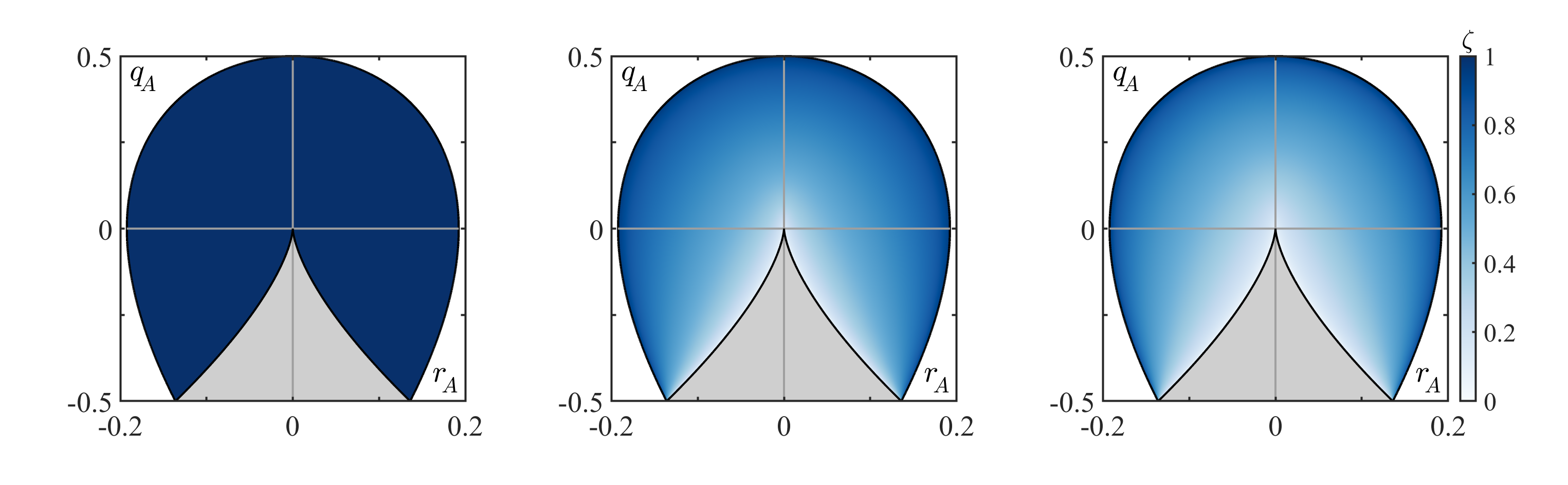}
    \caption{Transformations to $\psq$, $2 \src$ \copyedit{and} $\zeta$ from the $\qA -\, \rA$ phase space for various alignment angles, $\theta_\omega$. The plots are in the same style as those in \cref{fig:qr_and_nrh_v1}.}
    \label{fig:shear_rotation_angle_v1}
\end{figure}

Here, we characterize the effect of the alignment between shearing and rigid rotation, as measured by $\theta_\omega$, on the phase space transformations associated with rotational local streamline geometries. \autoref{fig:shear_rotation_angle_v1} depicts how the corresponding transformations vary with $\theta_\omega$ in the $\qA -\, \rA$ phase space. When $\theta_\omega = 90^\circ$, shearing and rigid rotation occur in orthogonal planes. In this case, $2 \src = 0$, $\zeta = 1$, and the region where $\psq$ dominates $A^2$ extends the furthest from the external boundary of the phase space. When $\theta_\omega = 45^\circ$, the regions where $\psq$ and $\zeta$ are large concentrate more sharply near the external boundary and $2 \src$ grows in the intermediate region between the boundaries of the rotational geometries. The concentration of $\psq$ and $\zeta$ and the amplification of $2 \src$ are most extreme when $\theta_\omega = 0^\circ$. In this case, the peak contribution of $2 \src$ is $( \sqrt{2} + 1 )^{-1}$ \citep{Das2020} and, for all $\theta_\omega < 90^\circ$, it occurs when $\rA = 0$. The location of this maximum approaches $\qA \rightarrow \frac{1}{4}$ as $\theta_\omega \rightarrow 90^\circ$ and $\qA \rightarrow \frac{1}{2\sqrt{2}}$ as $\theta_\omega \rightarrow 0^\circ$. The qualitative features of the distributions vary more significantly from $\theta_\omega = 90^\circ$ to $\theta_\omega = 45^\circ$ than they do from $\theta_\omega = 45^\circ$ to $\theta_\omega = 0^\circ$. 

%%%%%%%%%%%%%%%%%%%%%%%%%%%%%%%%%%%%%%%%%%

% ensure bib. starts after appendices
%%%%%%%%%%%%%%%%%%%%%%%%%%%%%%%%%%%%%%%%%%
% \newpage
%%%%%%%%%%%%%%%%%%%%%%%%%%%%%%%%%%%%%%%%%%

% bibliography
%%%%%%%%%%%%%%%%%%%%%%%%%%%%%%%%%%%%%%%%%%
\bibliographystyle{jfm}
\bibliography{references}
%%%%%%%%%%%%%%%%%%%%%%%%%%%%%%%%%%%%%%%%%%

% end document
%%%%%%%%%%%%%%%%%%%%%%%%%%%%%%%%%%%%%%%%%%
\end{document}